\documentclass{biometrika}

\makeatletter
\let\originalpsplain\ps@plain
\def\ps@plain{%
  \originalpsplain
  \let\@oddhead\@empty
  \let\@evenhead\@empty
}
\makeatother

\usepackage{amsmath}
\usepackage{adjustbox}
\usepackage{booktabs}
\usepackage{multirow}
\usepackage{float}
\usepackage{xcolor}
\usepackage{newtxtext}
\usepackage[subscriptcorrection]{newtxmath}

\graphicspath{{./art/}}

\usepackage[plain,noend]{algorithm2e}

\makeatletter
\renewcommand{\algocf@captiontext}[2]{#1\algocf@typo. \AlCapFnt{}#2} % text of caption
\def\@algocf@capt@plain{top}
\renewcommand{\algocf@makecaption}[2]{%
  \addtolength{\hsize}{\algomargin}%
  \sbox\@tempboxa{\algocf@captiontext{#1}{#2}}%
  \ifdim\wd\@tempboxa >\hsize%     % if caption is longer than a line
    \hskip .5\algomargin%
    \parbox[t]{\hsize}{\algocf@captiontext{#1}{#2}}% then caption is not centered
  \else%
    \global\@minipagefalse%
    \hbox to\hsize{\box\@tempboxa}% else caption is centered
  \fi%
  \addtolength{\hsize}{-\algomargin}%
}
\makeatother

\newcommand{\E}{\mathbb{E}}
\newcommand{\Var}{\mathrm{Var}}

\newcommand{\Prb}{\mathbb{P}}

\newcommand{\toP}{\xrightarrow{p}}
\newcommand{\toD}{\xrightarrow{d}}
\newcommand{\norm}[1]{\left\lVert #1 \right\rVert}

\begin{document}

% \jname{}
% %% The year, volume, and number are determined on publication
% \jyear{}
% \jvol{}
% \jnum{}
% \cyear{}
%% The \doi{...} and \accessdate commands are used by the production team
%\doi{10.1093/biomet/asm023}
% \accessdate{}

%% These dates are usually set by the production team
% \received{}
% \revised{}

%% The left and right page headers are defined here:
%\markboth{Y. Wang et~al.}{Biometrika style}

%% Here are the title, author names and addresses
\title{Semiparametric inference for a Target Outcome Leveraging Heterogeneous External Summaries of Auxiliary Outcomes}

\author{Yifei Wang}
\affil{Public Health Sciences Division, Fred Hutchinson Cancer Center,\\ Seattle, WA 98109, USA
\email{ywang6@fredhutch.org}}

\author{Jiayin Zheng}
\affil{Department of Biostatistics, Epidemiology and Informatics, Perelman
School of Medicine, University of Pennsylvania, Philadelphia, PA 19104, USA \email{Jiayin.Zheng1@Pennmedicine.upenn.edu}}

\author{Christina C. Newton}
\affil{Department of Population Science, American Cancer Society, \\Atlanta, GA 30303, USA \email{Christina.Newton@cancer.org}}

\author{Li Hsu}
\affil{Public Health Sciences Division, Fred Hutchinson Cancer Center,\\ Seattle, WA 98109, USA
\email{lih@fredhutch.org}}

\maketitle

\begin{abstract}
The integration of external information can substantially improve the efficiency of statistical inference in an internal study, but is often complicated by between-study heterogeneity. Additional challenges arise when external studies examine auxiliary outcomes rather than on the target outcomes and provide only summary statistics. To address these challenges, we propose a semiparametric framework that incorporates external summary statistics into the internal analysis via the generalized method of moments. To accommodate between-study heterogeneity, the external summaries are formulated within a random-effects model centered on the internal parameter estimates and subsequently embedded into the internal estimating equations. We establish the asymptotic properties of the proposed estimator. While uncertainty from the first-stage estimation vanishes as the number of external studies increases, it remains non-negligible where the number of external studies is moderate. In this setting, we show that a hybrid bootstrap procedure correctly propagates this uncertainty and consistently estimates the asymptotic covariance of the proposed estimator, thereby ensuring valid inference. We further derive theoretical efficiency bounds and demonstrate robust finite-sample performance through simulation studies. In an application to colorectal cancer tumor sequencing data, we show that integrating external information of auxiliary outcomes reveals molecular subtype-specific etiologic associations that would not be detected using the internal data alone.
\end{abstract}


\section{Introduction} \label{sec:intro}

% motivation and basic setting

Leveraging related auxiliary outcomes from external studies can substantially improve the inference for the target outcomes in an internal study. However, differences in underlying populations between internal and external studies can introduce between-study heterogeneity and lead to biased inference. Moreover, individual-level data from external studies are often inaccessible due to logistical, ethical or privacy constraints. Consequently, the available external information is often limited to summary statistics, such as association estimates and their standard errors. 

These challenges arise frequently in biomedical research. For example, in cancer epidemiology, advances in sequencing technologies have enabled increasingly detailed characterization of tumors, creating new opportunities to investigate associations between novel molecular tumor features and etiologic risk factors \citep{Lawrence2014, Limsui2010, Kandoth2013}. However, sequencing remains costly and is not routinely available for all tumors. As a result, an internal study that contains sequencing data on novel tumor markers has the limited sample size, while usually much larger external studies may provide summary association estimates for established markers that are correlated with the novel markers and may therefore serve as auxiliary outcomes \citep{Botteri2021}. In applied settings, the internal and external studies may differ in their underlying populations, giving rise to heterogeneity across studies. This motivates a general framework for integrating potentially heterogeneous external summary information on auxiliary outcomes to improve inference for novel outcomes observed in only a small internal study.

% lit review organized around the gaps. para

 A number of methods have been developed to incorporate summary-level external information, but important limitations remain. \cite{Chatterjee2016} proposed a constrained maximum likelihood framework, closely related to empirical likelihood \citep{Qin1994}, for incorporating such information. \cite{Zhang2020} extended this framework to account for estimation uncertainty in the external summary statistics. \cite{Kundu2019} developed a generalized method of moments (GMM) approach that combines external estimates from different sets of covariates. These methods all assume common covariate effects between the internal and external studies. In practice, effects often differ across studies. First, even when the same covariates are measured, the true relationship between the outcome and those covariates may genuinely differ across study populations, for example, due to differences in unmeasured factors associated with both the covariates and the outcome. Second, studies may adjust for different sets of covariates; when covariate distributions also differ across studies, this mismatch can yield the different estimates. Such between-study heterogeneity is difficult to identify or quantify with summary-level information alone, yet ignoring it can lead to substantial bias.

Recently, several methods have been developed to account for between-study heterogeneity, each requiring additional assumptions. \cite{Shao2023} proposed a GMM approach that adjusts for differing covariate distributions through a density-ratio model, which requires both a correctly specified ratio model and additional external summary statistics on the covariates. \cite{Zhai2022} and \cite{Fang2025} proposed penalized estimating-equation approaches that adaptively select the equations with zero expectation under the internal-study sample distribution. By framing the problem as valid moment selection, these approaches rely on binary inclusion decisions and penalty-parameter tuning. \cite{Gu2023} proposed a synthetic-data approach that explicitly models differences in covariate effects across studies, but does not establish theoretical efficiency guarantees. More fundamentally, however, all of these methods require the internal and external studies to share a common outcome, a major limitation when the target outcome is a novel outcome available only in the internal study.

A related line of work uses prediction-powered inference to combine gold-standard labeled observations with individual-level covariates and black-box predictions from a typically much larger unlabeled sample \citep{Angelopoulos2023,Angelopoulos2023b}. Subsequent extensions improve efficiency through recalibration \citep{Ji2025}, accommodating conditional inferential targets \citep{Sui2026}, or aggregate predictions from multiple black-box models \citep{Shan2026}. Despite these advances, prediction-powered inference methods generally require external individual-level covariates and predictions for the unlabeled sample, and therefore cannot directly leverage external data available only in aggregate form. In contrast, our approach requires individual-level data only from the internal study and incorporates external information solely through the summary statistics of parameter estimates and their standard errors. 

% your contributions. 

%  In this article, we develop a semiparametric framework that overcomes these limitations by integrating robust meta-analytic summaries of auxilary outcomes into internal estimating equations. The proposed approach employs a two-stage estimation procedure. In the first stage, we formulate the summary statistics for the auxiliary outcomes using a random-effects model that centers around the internal parameter estimates, yielding  an unbiased and efficient estimator that accounts for between-study heterogeneity and limits the influence of outlying external studies. In the second stage, we incorporate this estimator into a GMM framework to improve estimation of the primary parameter of interest. %We establish the consistency, asymptotic normality, and efficiency of the proposed estimator, and demonstrate efficiency gains over the internal-study-only estimator through simulation studies and a real-data analysis.

% {\color{blue}
% AI's suggestion with Li's input:

We establish the asymptotic properties of the proposed estimator while incorporating estimation uncertainty from the external studies. In particular, when the number of external studies increases, this uncertainty becomes asymptotically negligible. However, in more practically relevant settings where the external information is limited, either because the number of external studies is small or because their sample sizes are small, this uncertainty remains non-negligible and contributes to the first-order asymptotic distribution. To address this, we propose a hybrid bootstrap procedure that properly propagates this uncertainty and consistently reproduces the resulting asymptotic distribution, thereby ensuring valid inference. We further derive theoretical efficiency bounds and show that incorporating external information improves efficiency over the internal study only when the variance reduction from auxiliary outcomes outweighs the added variance due to uncertainty in external-study estimates.

% Biometrika papers do not contain a `contents of the paper' paragraph.

% The rest of the article is organized as follows. Section~\ref{sec:method} describes the proposed method and its asymptotic properties. Section~\ref{sec:sim} evaluates the proposed approach through simulation studies. Section~\ref{sec:app} applies the method to an association study of smoking with novel colorectal cancer (CRC) tumor markers using individual-level data from the Cancer Prevention Study-II (CPS-II) Nutrition Cohort, while leveraging external information on established markers. Section~\ref{sec:disc} concludes with a discussion. Technical details and additional numerical results are provided in the supplementary material.

\section{Method} \label{sec:method}

\subsection{Notation and setup} \label{sec:notation}

Consider an internal study with individual-level observations of a covariate $X$, a target outcome $Y_1$, and an auxiliary outcome $Y_2$. The target outcome is of direct interest, whereas additional information on the association between $X$ and $Y_2$ is available from external studies only through summary statistics. Our goal is to integrate these external summaries with the internal data to improve inference on the association between $X$ and $Y_1$, while allowing the association involving $Y_2$ to vary across studies. We now formalize the setup.

Let $Y_1\in\{0,\ldots,J\}$ denote the target outcome and let $Y_2\in\{0,\ldots,L\}$ denote the auxiliary outcome, where $J$ and $L$ are the numbers of non-reference categories for the target and auxiliary outcomes, respectively. The two outcomes share a common reference category, represented by $Y_1=0$ and $Y_2=0$. For example, in the colorectal cancer application in Section~\ref{sec:app}, $Y_1$ and $Y_2$ characterize disease subtypes defined by the novel APC and established BRAF tumor markers, respectively. Individuals without colorectal cancer belong to the common reference group and are therefore assigned $Y_1=Y_2=0$, whereas cases are classified into the nonzero categories according to their marker status. We model the conditional distributions given the covariate $X$ using separate polytomous logistic regressions \citep{Zabor2017}:
\begin{equation}\label{eq:plr}
\begin{aligned}
\Pr(Y_1=j \mid X)
&=
\frac{\exp\left(\beta_{0j}+\beta_{1j}X\right)}
{1+\sum_{h=1}^{J}\exp\left(\beta_{0h}+\beta_{1h}X\right)},
\quad j=1,\ldots,J,\\
\Pr(Y_2=l \mid X)
&=
\frac{\exp\left(\theta_{0l}+\theta_{1l}X\right)}
{1+\sum_{m=1}^{L}\exp\left(\theta_{0m}+\theta_{1m}X\right)},
\quad l=1,\ldots,L.
\end{aligned}
\end{equation}

The primary parameter of interest, $\beta_{1j}$, represents the log odds ratio  characterizing the association between covariate $X$ and disease risk specific for the $j$th subtype defined by the target outcome $Y_1$, relative to the reference category, for $j = 1, \dots, J$. Similarly, $\theta_{1l}$ characterizes the association between $X$ and the $l$th subtype disease defined by the auxiliary outcome $Y_2$, relative to the reference category, for $l = 1, \ldots, L$.

Suppose we observe $n_0$ independent and identically distributed individuals from the internal study, with individual-level data denoted by $\{(X_i,Y_{i1},Y_{i2}):i=1,\ldots,n_0\}$. In addition, suppose summary information are available from $K$ independent external studies, indexed by $k=1,\ldots,K$, on the association between $X$ and the auxiliary outcome $Y_2$. Specifically, for the $k$th external study, the available information consists of a vector of log odds ratio estimates $\widehat \theta^{(k)}=\{\widehat\theta^{(k)}_{0l}, \widehat\theta^{(k)}_{1l}, l=1,\ldots,L\}$ for the association between $X$ and $Y_2$, and the corresponding covariance matrix $S^{(k)}$, obtained by fitting the polytomous logistic regression model \eqref{eq:plr} for $Y_2$ to the $k$th external dataset.

We propose a two-stage estimation framework that uses external summary information on the auxiliary outcome to improve estimation of the association analysis for 
the target outcome. In the first stage, we combine internal and
external information on the auxiliary outcome using a robust
random-effects meta-analysis that accounts for between-study
heterogeneity. In the second stage, we incorporate the resulting
meta-analytic estimates into a generalized method of moments
framework to improve estimation of the association between the
covariate and the target outcome.

\subsection{Meta-analysis for integrating internal and external studies on the auxiliary outcome} \label{sec:meta}

We begin with a homogeneity assumption, whereby the association between the covariate $X$ and the auxiliary outcome $Y_2$ is assumed to be common across internal and all external studies. Let $\theta^{(0)}$ denote the vector of true log odds ratios of $X$ and $Y_2$ for the internal study. Under the homogeneity assumption, the estimator from each external study is approximately distributed as
\begin{equation}
\sqrt{n_k}\left(\widehat{\theta}^{(k)} - \theta^{(0)}\right) \sim \operatorname{MVN}\left(0, n_kS^{(k)}\right),
\qquad k=1,\ldots,K,
\label{eq:fe}
\end{equation}
where $S^{(k)}$ can be different across studies. Under the fixed-effects model, the meta-analytic estimator combines information across studies using inverse-variance weighting. Specifically,  the fixed-effects estimator is given by
\begin{equation}
\tilde{\theta}_{FE}
=
\left\{
S^{(0)^{-1}} + \sum_{k=1}^{K}S^{(k)^{-1}}
\right\}^{-1}
\left\{
S^{(0)^{-1}}\hat{\theta}^{(0)} + \sum_{k=1}^{K}S^{(k)^{-1}}\hat{\theta}^{(k)}
\right\},
\label{eq:fe_tilde_theta}
\end{equation}
with corresponding covariance
\begin{equation}
\operatorname{Var}(\tilde{\theta}_{FE})
=
\left\{
S^{(0)^{-1}} + \sum_{k=1}^{K}S^{(k)^{-1}}
\right\}^{-1}.
\label{eq:fe_tilde_theta_var}
\end{equation}

When the homogeneity assumption fails, the fixed-effects estimator in \eqref{eq:fe_tilde_theta} is no longer consistent for the internal-study  parameter $\theta^{(0)}$. While a standard random-effects meta analysis is a natural alternative, it typically targets an overall mean effect and treats all studies exchangeably \citep{Hsu2022}. Consequently, heterogeneous external evidence can bias the estimation away from the target $\theta^{(0)}$.  To overcome this limitation, we propose a tailored random-effects framework that explicitly models between-study heterogeneity as random deviations centered around the internal parameter true value $\theta^{(0)}$. Specifically, for the internal study, we have 
$$
\hat{\theta}^{(0)} \sim \operatorname{MVN}\left(\theta^{(0)}, S^{(0)}\right),
$$
whereas for external studies $k=1,\ldots,K$,
\begin{equation}
\begin{aligned}
        \gamma^{(k)} &\sim \operatorname{MVN}\left(0, \Omega\right), \\
\theta^{(k)} &= \theta^{(0)} + \gamma^{(k)}, \\
\hat{\theta}^{(k)} &\sim \operatorname{MVN}\left(\theta^{(k)}, S^{(k)}\right)
\end{aligned}
\label{eq:re}
\end{equation}
Under this formulation, external study $\theta^{(k)}$ is allowed to differ from the target parameter $\theta^{(0)}$ through study-specific random deviations $\gamma^{(k)}$. The random deviations $\gamma^{(k)}$ are independent across studies and independent of the within-study estimation terms. The between-study covariance $\Omega$ characterizes the magnitude of this heterogeneity and adaptively regularizes the integration of  external evidence, thereby preventing over-borrowing from highly precise but potentially incompatible studies. 

While the random-effects model accommodates random heterogeneity, certain external studies may exhibit systematic departures from the internal study. To ensure a robust estimation against such extreme outliers, we therefore seek a principled approach to limit the influence of outlying external studies. To this end, we construct a power prior \citep{Ibrahim2015} for $\theta^{(0)}$ using a weighted external-study likelihood function:
$$
\pi\!\left(\theta^{(0)} \mid \hat{\theta}^{(1:K)}, S^{(1:K)}, \alpha_{1:K}, \Omega\right)
\propto
\prod_{k=1}^K
\phi\!\left(\hat{\theta}^{(k)};\theta^{(0)},\Omega+S^{(k)}\right)^{\alpha_k}
\pi_0(\theta^{(0)})
$$
where $\pi_0(\theta^{(0)})\propto 1$ is a non-informative prior, $\phi(\cdot;\mu,\Sigma)$ denotes the multivariate normal density with mean $\mu$ and covariance $\Sigma$, and $0 \le \alpha_k \le 1$ is a study-specific weight that controls the extent of borrowing from external study $k$. Smaller values of  $\alpha_k$ downweight studies that are less compatible with the internal study, thereby limiting the influence of outlying external evidence.  

Combining this power prior with the internal-study likelihood function yields the posterior distribution
$$
\pi\!\left(\theta^{(0)} \mid \hat{\theta}^{(0:K)}, S^{(0:K)}, \alpha_{1:K}, \Omega\right)
\sim
\mathcal{N}(\theta^{\star}, V^{\star}),
$$
where
$$
V^{\star}
=
\left[
S^{(0)^{-1}}
+
\sum_{k=1}^K \alpha_k(\Omega + S^{(k)})^{-1}
\right]^{-1},
\quad
\theta^{\star}
=
V^{\star}
\left[
S^{(0)^{-1}}\hat{\theta}^{(0)}
+
\sum_{k=1}^K \alpha_k(\Omega + S^{(k)})^{-1}\hat{\theta}^{(k)}
\right].
$$
The posterior mean $\theta^{\star}$ combines the internal-study estimator $\hat{\theta}^{(0)}$ and the external-study estimators $\hat{\theta}^{(k)}$ through the precision-weighted average. The contribution of each external study is governed by its sampling covariance $S^{(k)}$, the borrowing weight $\alpha_k$, and the between-study covariance $\Omega$.

To fit this model, we first estimate the study-specific borrowing weights $\alpha_k$ using a Huber-type rule based on the squared Mahalanobis distance \citep{Huber1981}:
$$
\hat{\alpha}_k
=
\min\left\{1,\sqrt{\frac{c_{K,n_0}}{d_k^2}}\right\},
$$
where
$$
c_{K,n_0}=\chi^2_{p,\,1-(Kn_0)^{-1/2}}
\qquad
d_k^2
=
\left(\hat{\theta}^{(k)}-\hat{\theta}^{(0)}\right)^\top
\left(S^{(k)}+S^{(0)}\right)^{-1}
\left(\hat{\theta}^{(k)}-\hat{\theta}^{(0)}\right),
$$
and $p=\dim(\theta^{(0)})$. The distance $d_k^2$ measures the discrepancy between external study $k$ and the internal study relative to their estimation variability. 
We use an adaptive critical value $c_{K,n_0}$ to ensure that, under the local-heterogeneity regime, $\Pr(\hat{\alpha}_k=1)\to 1$ as $n_0\to\infty$. Under the fixed-heterogeneity regime, by contrast, $\hat{\alpha}_k \rightarrow 0$ in probability as $n_0\to\infty$. These two regimes are formally defined in Section~\ref{sec: asy_prop}. The cap at $\hat{\alpha}_k=1$ follows the standard Huber-type robust weighting rule \citep{Huber1964,HuberRonchetti2009,VenablesRipley2002}.

Given these robust weights, we estimate the between-study covariance $\Omega$ and the posterior parameters $\theta^{\star}$ and $V^{\star}$ through an iterative weighted method-of-moments (MM) method. Specifically, given the posterior mean $\tilde{\theta}_{RE}$ and the corresponding covariance $\tilde V_{RE}$, the between-study covariance is estimated by
\begin{equation}
\hat{\Omega}
=
\frac{1}{A}\left[
\sum_{k=1}^{K}\hat{\alpha}_k
\left(\hat{\theta}^{(k)}-\tilde{\theta}_{RE}\right)
\left(\hat{\theta}^{(k)}-\tilde{\theta}_{RE}\right)^{\top}
-
\sum_{k=1}^{K}\hat{\alpha}_k S^{(k)}
\right]
+
\tilde V_{RE},
\label{eq:omega_hat}
\end{equation}
where $A=\sum_{k=1}^{K}\hat{\alpha}_k$. Given $\hat{\Omega}$, the posterior mean and covariance are estimated by
\begin{equation}
\tilde V_{RE}
=
\left[
(S^{(0)})^{-1}
+
\sum_{k=1}^K \hat{\alpha}_k(\hat{\Omega} + S^{(k)})^{-1}
\right]^{-1},
\quad
\tilde{\theta}_{RE}
=
\tilde V_{RE}
\left[
(S^{(0)})^{-1}\hat{\theta}^{(0)}
+
\sum_{k=1}^K \hat{\alpha}_k(\hat{\Omega} + S^{(k)})^{-1}\hat{\theta}^{(k)}
\right].
\label{eq:posterior_param}
\end{equation}

Since $\hat{\Omega}$, $\tilde{\theta}_{RE}$, and $\tilde V_{RE}$ depend on one another, these quantities are obtained iteratively until convergence.  The algorithm is summarized below.
\begin{algo}
Iterative method-of-moments (MM) algorithm
\begin{tabbing}
\qquad \enspace Input $(\hat{\theta}^{(0)},S^{(0)})$ and $\{(\hat{\theta}^{(k)},S^{(k)}): k=1,\dots,K\}$ \\
\qquad \enspace Compute $\hat{\alpha}_k$ for $k=1,\dots,K$ and set $A=\sum_{k=1}^K \hat{\alpha}_k$ \\
\qquad \enspace Initialize
$
\hat{\Omega}^{(0)}
=
\frac{1}{A}\left[
\sum_{k=1}^{K}\hat{\alpha}_k
(\hat{\theta}^{(k)}-\hat{\theta}^{(0)})(\hat{\theta}^{(k)}-\hat{\theta}^{(0)})^{\top}
-
\sum_{k=1}^{K}\hat{\alpha}_k(S^{(k)}+S^{(0)})
\right]
$ \\
\qquad \enspace Set $t=0$ \\
\qquad \enspace Repeat \\
\qquad \qquad Estimate $\tilde{\theta}_{RE}^{(t)}$ and $\tilde V_{RE}^{(t)}$ from \eqref{eq:posterior_param} given $\hat{\Omega}^{(t)}$ \\
\qquad \qquad Update $\hat{\Omega}^{(t+1)}$ from \eqref{eq:omega_hat} using $\tilde{\theta}_{RE}^{(t)}$ and $\tilde V_{RE}^{(t)}$ \\
\qquad \qquad $t \leftarrow t+1$ \\
\qquad \enspace Until convergence \\
\qquad \enspace Output $\tilde{\theta}_{RE}^{(t)}$ and $\tilde V_{RE}^{(t)}$
\end{tabbing}
\end{algo}

This iterative MM algorithm jointly estimates $\Omega$, $\theta^{\star}$, and $V^{\star}$ through a sequence of closed-form updates. By using the pooled estimator $\tilde{\theta}_{RE}$ rather than the often small-sample internal-study estimator $\hat{\theta}^{(0)}$, the algorithm improves precision robustly. It is also computationally highly efficient, since each update has a closed-form expression and avoids numerical optimization. In contrast, likelihood-based approaches require iterative optimization over the between-study covariance parameters. 

\subsection{Generalized method of moments for association analysis of the target outcome} \label{sec:gmm}

Building on the first-stage meta-analysis estimates $(\tilde{\theta}_{RE}, \tilde V_{RE})$ for the auxiliary outcome $Y_2$, we develop a generalized method of moments (GMM) approach to improve estimation of the parameter of interest, $\beta$, that characterizes the association between $X$ and the target outcome $Y_1$. Because the joint distribution of the novel and auxiliary outcomes, $(Y_1, Y_2)$ conditional on $X$, is observed in the internal study, information for the auxiliary outcome can be incorporated into inference on $\beta$ through a joint system of estimating equations. Following the GMM framework \citep{Hansen1982, Imbens2002}, we construct the moment conditions implied by the score equations of the polytomous logistic regression model \eqref{eq:plr}:
$$
\operatorname{E}\!\left\{\mathbf{g}(Y_1,Y_2,X;\beta,\theta)\right\} = \mathbf{0},
\qquad
\mathbf{g}(Y_1,Y_2,X;\beta,\theta)
=
\begin{bmatrix}
\mathbf{g}_1(Y_1,X;\beta) \\
\mathbf{g}_2(Y_2,X;\theta)
\end{bmatrix},
$$
where

$$
\begin{aligned}
\mathbf{g}_1(Y_1,X;\beta)
&=
\begin{bmatrix}
\begin{array}{@{}l@{}}
\mathrm{1}_{(Y_1=1)}-\Pr(Y_1=1\mid X;\beta)\\
\multicolumn{1}{c}{\vdots}\\
\mathrm{1}_{(Y_1=J)}-\Pr(Y_1=J\mid X;\beta)\\
\left[\mathrm{1}_{(Y_1=1)}-\Pr(Y_1=1\mid X;\beta)\right]X\\
\multicolumn{1}{c}{\vdots}\\
\left[\mathrm{1}_{(Y_1=J)}-\Pr(Y_1=J\mid X;\beta)\right]X
\end{array}
\end{bmatrix}, \\[1ex]
\mathbf{g}_2(Y_2,X;\theta)
&=
\begin{bmatrix}
\begin{array}{@{}l@{}}
\mathrm{1}_{(Y_2=1)}-\Pr(Y_2=1\mid X;\theta)\\
\multicolumn{1}{c}{\vdots}\\
\mathrm{1}_{(Y_2=L)}-\Pr(Y_2=L\mid X;\theta)\\
\left[\mathrm{1}_{(Y_2=1)}-\Pr(Y_2=1\mid X;\theta)\right]X\\
\multicolumn{1}{c}{\vdots}\\
\left[\mathrm{1}_{(Y_2=L)}-\Pr(Y_2=L\mid X;\theta)\right]X
\end{array}
\end{bmatrix}.
\end{aligned}
$$

and
$$
\beta = \left[\beta_{01},\ldots,\beta_{0J},\beta_{11},\ldots,\beta_{1J}\right],
\qquad
\theta = \left[\theta_{01},\ldots,\theta_{0L},\theta_{11},\ldots,\theta_{1L}\right].
$$
The corresponding sample estimating functions based on the internal study are given by 
\begin{equation}
\bar{\mathbf{g}}(\beta,\theta)
=
\frac{1}{n_0}\sum_{i=1}^{n_0}
\mathbf{g}(Y_{i1},Y_{i2},X_i;\beta,\theta).
\label{eq:est_eqns}
\end{equation}

External information on the auxiliary outcomes can be incorporated into \eqref{eq:est_eqns} in two ways. The first is a plug-in formulation, in which $\theta$ is replaced directly by the posterior mean estimator $\tilde{\theta}_{RE}$. We denote the corresponding estimating functions by $\mathbf{g}_{2}(Y_{i2}, X_i; \tilde{\theta}_{RE})$, and refer to the resulting GMM estimator as $\widehat\beta_{\text{rGMM}}$ .
\iffalse========================
. Under this plug-in formulation, external information enters through the estimating-equation block
$$
\mathbf{g}_{2}(Y_{i2}, X_i; \tilde{\theta}_{RE})
=
\begin{bmatrix}
\left[\mathrm{1}_{(Y_{i2}=1)}-\Pr\!\left(Y_{i2}=1\mid X_i;\tilde{\theta}_{RE}\right)\right]X_i \\
\vdots \\
\left[\mathrm{1}_{(Y_{i2}=L)}-\Pr\!\left(Y_{i2}=L\mid X_i;\tilde{\theta}_{RE}\right)\right]X_i
\end{bmatrix}.
$$
=================================\fi
The second approach accounts for uncertainty from the meta-analysis by integrating over the posterior uncertainty in $\theta$. Specifically, rather than treating $\theta$ as fixed, we assume that $\theta$ follows the posterior distribution $\mathcal{N}(\theta^{*}, V^{*})$, replacing $\Pr(Y_2=l \mid X;\theta)$ in the estimating functions with the corresponding marginal probability 
$$
\mu_l\!\left(\theta^{*}, V^{*}\right)
=
\int \Pr\!\left(Y_2=l \mid X,\theta\right)\,
\phi\!\left(\theta;\theta^{*},V^{*}\right)\,
d\theta,
\qquad l=1,\ldots,L
$$
where $\phi(\cdot;\theta^{*},V^{*})$ denotes the multivariate normal density with mean $\theta^{*}$ and covariance $V^{*}$. We denote the resulting estimating functions by $\mathbf{g}_{2}(Y_{i2}, X_i; \theta^{*}, V^{*})$, and refer to the corresponding GMM estimator as $\widehat\beta_{\text{mGMM}}$.
\iffalse==============================
The corresponding estimating-equation block is
$$
\mathbf{g}_{2}(Y_{i2}, X_i; \theta^{\star}, V^{\star})
=
\begin{bmatrix}
\left[\mathrm{1}_{(Y_{i2}=1)}-\mu_1\!\left(\theta^\star, V^\star\right)\right]X_i \\
\vdots \\
\left[\mathrm{1}_{(Y_{i2}=L)}-\mu_L\!\left(\theta^\star, V^\star\right)\right]X_i
\end{bmatrix}.
$$
======================================\fi
For implementation, we replace $(\theta^{*},V^{*})$ by their estimators $(\tilde{\theta}_{RE},\tilde V_{RE})$ and approximate marginal probability $\mu_l(\tilde{\theta}_{RE},\tilde V_{RE})$ by Monte Carlo integration:
$$
\hat{\mu}_l\!\left(\tilde{\theta}_{RE}, \tilde V_{RE}\right)
=
\frac{1}{B}\sum_{b=1}^B
\Pr\!\left(Y_2=l \mid X,\theta_b\right),
\qquad
\theta_b \stackrel{\text{i.i.d.}}{\sim} \operatorname{MVN}(\tilde{\theta}_{RE},\tilde V_{RE}).
$$

For notational convenience, let $\mathbf{g}_i(\beta)$ generically denote the resulting estimating function $\mathbf{g}(Y_1,Y_2,X;\beta,\theta)$ after replacing the $Y_2$ component by either $\mathbf g_2(Y_{i2},X_i;\tilde\theta_{RE})$ under the plug-in formulation or $\mathbf g_2(Y_{i2},X_i;\tilde\theta_{RE},\tilde V_{RE})$ under the uncertainty-propagating formulation and $\bar{\mathbf{g}}(\beta,\tilde\theta_{RE},\tilde V_{RE})$ denote the sample estimating functions $\bar{\mathbf{g}}(\beta,\tilde\theta_{RE},\tilde V_{RE}) = \frac{1}{n_0}\sum_{i=1}^{n_0}\mathbf{g}_i(\beta)$. Correspondingly, let $\widehat\beta_{\mathrm{GMM}}$ generically denote either $\widehat\beta_{\mathrm{rGMM}}$ or $\widehat\beta_{\mathrm{mGMM}}$, depending on the formulation used.  We estimate $\beta$ by the following two-step GMM procedure:
\begin{algo}
Two-step GMM algorithm for estimating $\beta$
\begin{tabbing}
\qquad \enspace Input internal data $\{(Y_{i1},Y_{i2},X_i): i=1,\dots,n_0\}$ and meta-analysis result $(\tilde\theta_{RE},\tilde V_{RE})$ \\

\qquad \enspace \textbf{Step 1: Internal MLE and fixed weight matrix} \\

\qquad \qquad Obtain the internal-study maximum likelihood estimator $\hat{\beta}_{\mathrm{MLE}}$ and compute \\
\qquad \qquad the empirical covariance matrix of the estimating equations: \\
\qquad \qquad
\parbox{0.90\linewidth}{
$$
\hat{W}_{\mathrm{MLE}}
=
\frac{1}{n_0}\sum_{i=1}^{n_0}
\left\{
\mathbf{g}_i(\hat{\beta}_{\mathrm{MLE}})
-
\bar{\mathbf{g}}(\hat{\beta}_{\mathrm{MLE}},\tilde\theta_{RE},\tilde V_{RE})
\right\}
\left\{
\mathbf{g}_i(\hat{\beta}_{\mathrm{MLE}})
-
\bar{\mathbf{g}}(\hat{\beta}_{\mathrm{MLE}},\tilde\theta_{RE},\tilde V_{RE})
\right\}^{\top}.
$$
}
\\

\qquad \enspace \textbf{Step 2: GMM estimation} \\

\qquad \qquad Estimate $\beta$ by minimizing the fixed-weight GMM criterion: \\

\qquad \qquad
\parbox{0.90\linewidth}{
$$
\hat{\beta}_{\mathrm{GMM}}
=
\operatorname*{argmin}_{\beta}\;
\bar{\mathbf{g}}(\beta,\tilde\theta_{RE},\tilde V_{RE})^{\top}
\left\{\hat{W}_{\mathrm{MLE}}\right\}^{-1}
\bar{\mathbf{g}}(\beta,\tilde\theta_{RE},\tilde V_{RE}).
$$
}
\\

\qquad \enspace Output $\hat{\beta}_{\mathrm{GMM}}$
\end{tabbing}
\end{algo}

Compared with constrained empirical likelihood estimation \citep{Zheng2022}, the proposed GMM approach is assumption-lean without specifying an explicit joint distribution of $(Y_1,Y_2)$ among diseased individuals, thereby more robust against mis-specification of dependence between $Y_1$ and $Y_2$. The plug-in formulation provides a simple implementation, whereas the uncertainty-propagating formulation additionally accounts for  uncertainty from the meta-analysis. Both formulations are scalable and flexible to accommodate multiple external studies.

\subsection{Asymptotic properties}
\label{sec: asy_prop}

We consider an asymptotic setting in which the internal-study sample size $n_0$ and the external-study sample sizes $n_k (k=1,\ldots,K)$ grow proportionally. Let $n=\operatorname{median}\{n_1,\ldots,n_K\}$ and define the sample-size ratio $r=n/n_0$. We assume that $n_k \asymp n$ and $r\to r_0\in(0,\infty)$, so that the external-study sample size is of the same order as the internal-study sample size. The detailed regularity conditions and technical proofs are provided in the Supplementary Material.

The proposed meta-estimator is designed to adapt the degree of external borrowing to the magnitude of between-study heterogeneity relative to within-study estimation uncertainty. To formalize this, we first investigate the asymptotic behavior under a local heterogeneity regime where the between-study covariance $\Omega=n^{-1}\Psi=(rn_0)^{-1}\Psi$ for a fixed positive-definite matrix $\Psi$. Under this regime, the between-study covariance is of the same order as the within-study sampling covariance. Let $\hat{\Psi}=n\hat{\Omega}=rn_0\hat{\Omega}$, where $\hat{\Omega}$ is the iterative method-of-moments estimator, and let $(\tilde{\theta}_{RE},\tilde V_{RE})$ denote the corresponding posterior mean and covariance estimators.

\begin{theorem}
\label{theorem1}
Under the regularity conditions stated in the Supplementary Material, as $n_0\to\infty$, $K\to\infty$, and $K(rn_0)^{-1/2}\to0$,
$$
\|\hat{\Psi}-\Psi\|
=
O_p\!\left(K^{-1/2}\right),
\qquad
\|\tilde{\theta}_{RE}-\theta^{(0)}\|
=
O_p\!\left((Krn_0)^{-1/2}\right),
\qquad
\|\tilde V_{RE}\|
=
O_p\!\left((Krn_0)^{-1}\right).
$$
\end{theorem}

Theorem~\ref{theorem1} establishes that under large $K$, the estimation error in $\tilde{\theta}_{RE}$ is asymptotically negligible relative to the variation in the GMM estimator $\hat{\beta}_{\mathrm{GMM}}$. Because $r\to r_0\in(0,\infty)$, $K\to\infty$, and $(Kr)^{-1/2}\to 0$, we have
$$
\|\tilde{\theta}_{RE}-\theta^{(0)}\|
=
O_p\!\left((Krn_0)^{-1/2}\right)
=
o_p(n_0^{-1/2}).
$$
Therefore, substituting $\tilde{\theta}_{RE}$ for $\theta^{(0)}$ in the estimating equations does not affect the first-order asymptotic distribution of $\hat{\beta}_{\mathrm{GMM}}$. Consequently, the first-order behavior of $\hat{\beta}_{\mathrm{GMM}}$ is governed by the internal-study sample size $n_0$.

\begin{theorem}
\label{theorem2}
Under the regularity conditions stated in the Supplementary Material and  Theorem~\ref{theorem1}, as $n_0\to\infty$:
\begin{enumerate}
\item[(a)] \textbf{Consistency}
$$
\hat{\beta}_{\mathrm{GMM}}\xrightarrow{P}\beta_0.
$$

\item[(b)] \textbf{Asymptotic normality}
$$
\sqrt{n_0}\,(\hat{\beta}_{\mathrm{GMM}}-\beta_0)
\xrightarrow{d}
N\!\left(
0,
\left(D_\beta^{\top}\Gamma^{-1}D_\beta\right)^{-1}
\right),
$$
where
$$
D_\beta
=
\left.
\frac{\partial}{\partial\beta}
\operatorname{E}
\{\mathbf{g}(Y_1,Y_2,X;\beta,\theta^{(0)})\}
\right|_{\beta=\beta_0}
$$
and
$$
\Gamma
=
\operatorname{E}
\left[
\mathbf{g}(Y_1,Y_2,X;\beta_0,\theta^{(0)})
\mathbf{g}(Y_1,Y_2,X;\beta_0,\theta^{(0)})^\top
\right].
$$
\end{enumerate}
\end{theorem}

We next characterize the efficiency gain. Partition
$$
\mathbf g
=
\begin{pmatrix}
\mathbf g_1\\
\mathbf g_2
\end{pmatrix},
\qquad
\Gamma
=
\begin{pmatrix}
\Gamma_{11} & \Gamma_{12}\\
\Gamma_{21} & \Gamma_{22}
\end{pmatrix},
$$
where $\mathbf g_1$ is the internal-study score for $\beta$ and
$\mathbf g_2$ is the auxiliary estimating function for $\theta$. Define the
Schur complement
$$
\Gamma_{22\cdot1}
=
\Gamma_{22}
-
\Gamma_{21}\Gamma_{11}^{-1}\Gamma_{12},
$$
and let $\Sigma_{\mathrm{MLE}}$ denote the asymptotic covariance of the
internal-study-only maximum likelihood estimator.

\begin{corollary}[Optimal GMM efficiency bound]
\label{cor:efficiency_gain}
Under Theorem~\ref{theorem2}, the asymptotic covariance of
the proposed GMM estimator satisfies
$$
\Sigma_{\mathrm{GMM}}
=
\left\{
\Sigma_{\mathrm{MLE}}^{-1}
+
\Gamma_{21}^{\top}
\Gamma_{22\cdot1}^{-1}
\Gamma_{21}
\right\}^{-1}
\preceq
\Sigma_{\mathrm{MLE}}.
$$
Thus, the proposed estimator is asymptotically at least as efficient
as the internal-study-only maximum likelihood estimator. If
$\Gamma_{21}\neq0$, the efficiency gain is strict in at least one direction.
If $\Gamma_{21}$ has full column rank, then
$$
\Sigma_{\mathrm{GMM}}
\prec
\Sigma_{\mathrm{MLE}},
$$
so the efficiency gain is strict in every nonzero direction.
\end{corollary}

Under the fixed-heterogeneity regime $\Omega=\Psi$, where $\Psi$ is positive definite, the differences between the external and internal studies persist as their sample sizes increase. The external contribution becomes asymptotically negligible relative to the internal contribution. The following result establishes that the proposed estimator becomes asymptotically equivalent to the internal-study-only estimator.

\begin{theorem}
\label{theorem:automatic_debiasing}
Under the the fixed-heterogeneity regime $\Omega=\Psi$ and the regularity conditions stated in the Supplementary Material, as $n_0\to\infty$, $n_k \asymp n$, $K\to\infty$, and $K(rn_0)^{-1/2}\to0$,
$$
\sqrt{n_0}
\left(\tilde{\theta}_{RE}-\hat{\theta}^{(0)}\right)
\xrightarrow{P}0,
\qquad
n_0\left\|\tilde V_{RE}-S^{(0)}\right\|
\xrightarrow{P}0.
$$
Moreover
$$
\sqrt{n_0}
\left(
\hat{\beta}_{\mathrm{GMM}}
-
\hat{\beta}_{\mathrm{MLE}}
\right)
\xrightarrow{P}0.
$$
\end{theorem}

The preceding results consider regimes in which the first-stage meta-analysis error is either asymptotically negligible at the \(n_0^{1/2}\) scale or external borrowing is asymptotically negligible. In practice, however, first-stage uncertainty often remains first-order relevant when the number of the external studies is moderate. As a result $\sqrt{n_0}\left(\tilde{\theta}_{RE}-\theta^{(0)}\right)$ has a nondegenerate limiting distribution. To account for this uncertainty and ensure proper inference, we propose a hybrid bootstrap that resamples the internal observations nonparametrically \citep{Efron1979,Efron1981} and independently generates the external summary estimates from the fitted random-effects model in \eqref{eq:re} \citep{VanDenNoortgate2005}. The full two-stage procedure is then refitted to obtain $\hat{\beta}_{\mathrm{GMM}}^{*}$, and the bootstrap standard error is estimated by the empirical standard deviation of the bootstrap replicates. Define
$$
D_\theta
=
\left.
\frac{\partial}{\partial\theta^\top}
\operatorname{E}
\left\{
\mathbf g(Y_1,Y_2,X;\beta_0,\theta)
\right\}
\right|_{\theta=\theta^{(0)}},
\qquad
J=D_\beta^\top\Gamma^{-1}D_\beta.
$$
The following theorem establishes the validity of the hybrid bootstrap under this regime.

\begin{theorem}
\label{thm:boot}
Consider an asymptotic regime in which,
as $n_0\to\infty$ and $K\to\infty$,
$$
\left(
\sqrt{n_0}
\bar{\mathbf g}_n(\beta_0,\theta^{(0)}),
\sqrt{n_0}
(\tilde\theta_{RE}-\theta^{(0)})
\right)
\xrightarrow{d}
(\mathbb G,\mathbb T),
$$
where $\mathbb G$ is a mean-zero Gaussian random
vector with covariance matrix $\Gamma$, and
$\mathbb T$ is a mean-zero, nondegenerate random
vector with finite covariance matrix. Under the bootstrap regularity conditions stated in the Supplementary Material,
$$
\sqrt{n_0}
(\hat\beta_{\mathrm{GMM}}-\beta_0)
\xrightarrow{d}\mathbb B,
$$
where
$$
\mathbb B
=
-J^{-1}D_\beta^\top\Gamma^{-1}
(\mathbb G+D_\theta\mathbb T).
$$

Moreover, the hybrid bootstrap consistently
estimates the asymptotic covariance of the
proposed estimator:
$$
n_0\operatorname{Var}^*
(\hat\beta_{\mathrm{GMM}}^*)
\xrightarrow{p}
\operatorname{Var}(\mathbb B).
$$
\end{theorem}

We next characterize how the uncertainty from the first-stage meta-analysis affects the efficiency gain. Let $\Sigma_0=J^{-1}$ and $\mathcal G = (\Sigma_{\mathrm{MLE}}-\Sigma_0)$ where $\Sigma_0$ is the optimal GMM covariance in
Corollary~\ref{cor:efficiency_gain}. Define
$$
V_T
=
\operatorname{Var}(\mathbb T),
\qquad
C_{GT}
=
\operatorname{Cov}(\mathbb G,\mathbb T),
$$
and
$$
\Delta_T
=
D_\theta V_TD_\theta^\top
+
C_{GT}D_\theta^\top
+
D_\theta C_{GT}^\top,\quad \mathcal H_T
=
J^{-1}D_\beta^\top\Gamma^{-1}
\Delta_T
\Gamma^{-1}D_\beta J^{-1}
$$

\begin{corollary}[Net efficiency with first-stage uncertainty]
\label{cor:efficiency_firststage}
Under Theorem~\ref{thm:boot}, the proposed estimator is
asymptotically at least as efficient as the internal-study-only maximum
likelihood estimator if and only if
$$
\mathcal H_T
\preceq
\mathcal G.
$$
\end{corollary}
Corollary~\ref{cor:efficiency_firststage} reveals a simple trade-off underlying the proposed method. Dependence between the target and auxiliary outcomes improves efficiency through leveraging external information on the auxiliary outcomes, whereas uncertainty in the first-stage meta-analysis can offset part of this gain. The proposed estimator would remain more efficient than the internal-study-only estimator if the reduction in variance achieved by incorporating the auxiliary information exceeded the additional variance introduced by uncertainty in the first-stage meta-analysis.

\subsection{Practical Consideration}

In practice, external studies often report only point estimates and standard errors, which makes adopting a diagonal structure for the between-study covariance $\Omega$ appealing for simplicity. However, this simplification is appropriate only when heterogeneity acts independently across parameters. In many situations, differences between the external studies and the internal study, such as unobserved study-level confounding or systemic differences in populations, may induce simultaneous shifts across multiple parameters, which yields cross-parameter dependence. Imposing a diagonal structure then forces these correlated shifts to be treated as independent variations. This misspecification can distort the discrepancy measures $d_k^2$ and weights and, in turn, introduce bias. To ensure robust information borrowing in the presence of potential systematic structural shifts, we adopt an unrestricted structure for $\Omega$ to fully capture the between-study heterogeneity.

\section{Simulation} \label{sec:sim}

\subsection{Simulation setting}

We conducted simulation studies to evaluate the finite-sample performance of the proposed methods under varying degrees of between-study heterogeneity. We consider two distinct scenarios, a baseline setting with no location shift and a contaminated setting with systematic location shift. In both scenarios, we generated one internal study and $K=15$ external studies. The internal-study sample size was fixed at $n_0=1000$, and each external study had sample size $n_k=1500$. For each study, we generated covariates $X_i=(X_{i1},X_{i2})^\top$, where $X_{i1}\sim \operatorname{Bernoulli}(0.5)$ and $X_{i2}\mid X_{i1}\sim \operatorname{Bernoulli}(0.5+0.2X_{i1})$. Each simulation setting was evaluated using 2000 replicates. Performance was evaluated using percentage relative bias (Bias), empirical standard deviation (ESD), average bootstrap standard error (SE), and the coverage probability (CP) of the $95\%$ confidence interval.

For the auxiliary outcome, the internal-study parameter was set to $\theta^{(0)}=(\theta_{01},\theta_{11},\theta_{21},\theta_{02},\theta_{12},\theta_{22})^\top$, with $(\theta_{01},\theta_{11},\theta_{21})^\top = (-1,\,0.5,\,0.5)^\top$ and $(\theta_{02},\theta_{12},\theta_{22})^\top = (-1,\,0.5,\,-0.5)^\top$. Under the no-location-shift scenario, the external-study parameters were generated independently as $\theta^{(k)} \sim \operatorname{N}(\theta^{(0)},\Sigma_\tau)$, where $\Sigma_\tau$ is a covariance matrix with diagonal entries equal to $\tau^2$ and off-diagonal entries equal to $\rho\tau^2$, with $\tau^2 \in \{0, 0.01,0.03,0.05\}$ and $\rho=0.5$. Under the location-shift scenario, the first five external studies were generated from $\operatorname{N}(\theta^{(0)}+0.5,\Sigma_\tau)$, whereas the remaining ten studies were generated from $\operatorname{N}(\theta^{(0)},\Sigma_\tau)$. Given the corresponding study-specific parameters and covariates,  the auxiliary outcome $Y_{i2}$ was generated according to  the polytomous logistic regression model \eqref{eq:plr}.

For the target outcome, we set $Y_{i1}=0$ whenever $Y_{i2}=0$. Among individuals with $Y_{i2}>0$ (i.e. diseased individuals), $Y_{i1}$ was generated according to the conditional model $\log \{\Pr(Y_{i1}=2)/\Pr(Y_{i1}=1)\} = \delta_0 + \delta_1 \mathbb{I}(Y_{i2}=2) + \delta_2^\top X_i$, where $\delta_0=-2$, $\delta_1=3$, and $\delta_2=(1,0.5, -0.5)^\top$. Because this conditional specification does not directly yield the marginal parameter \(\beta_0\), we obtained its true value by integrating out \(Y_2\) under the data-generating mechanism to obtain \(P(Y_1=j\mid X)\), \(j=0,1,2\), and solving the corresponding score equations for the polytomous logistic regression model \eqref{eq:plr}.

\subsection{Simulation results for meta-analysis of auxiliary outcomes}

For the first-stage meta-analysis of the auxiliary outcome, we evaluated five estimators: the internal-sample-only maximum likelihood estimator $\hat{\theta}_{\text{MLE}}$, the fixed-effects estimator $\hat{\theta}_{\text{FE}}$ assigning full weight to all external studies, a fixed-effect estimator  $\hat{\theta}_{\text{wFE}}$ using the Huber-type weight in Section \ref{sec:meta}, the random-effects estimator $\hat{\theta}_{\text{RE}}$ with full weights, and the proposed robust random-effects estimator with the power prior $\hat{\theta}_{\text{wRE}}$ using the Huber-type weight.

Table \ref{tab:compare_meta_models_no_ls_K15} presents simulation results under the no-location-shift scenario. When $\tau^2=0$, i.e., there was no between-study heterogeneity, all estimators were approximately unbiased and $\hat{\theta}_{\text{FE}}$ achieved the smallest ESD, as expected. $\hat{\theta}_{\text{RE}}$ and $\hat{\theta}_{\text{wRE}}$ incurred little efficiency loss despite estimating an additional variance component. As heterogeneity increased from $\tau^2=0.01$ to $\tau^2=0.05$, $\hat{\theta}_{\text{FE}}$ showed growing bias and increasing underestimation of uncertainty, with CPs deteriorating from approximately 87\%--91\% to 66\%--74\%. The robust weighting $\hat{\theta}_{\text{wFE}}$ partially mitigated the bias, raising CPs by roughly 2\%--6\% over $\hat{\theta}_{\text{FE}}$, yet both fixed effects estimators failed to maintain nominal coverage. In contrast, $\hat{\theta}_{\text{RE}}$ and $\hat{\theta}_{\text{wRE}}$ were nearly unbiased across all heterogeneity levels, with CPs consistently close to the nominal level (92\% - 96\%). Compared with the internal-study-only estimator $\hat{\theta}_{\text{MLE}}$, $\hat{\theta}_{\text{RE}}$ and $\hat{\theta}_{\text{wRE}}$ achieved substantially smaller ESDs, demonstrating efficiency gains extracted from incorporating external study information in the presence of properly modeled heterogeneity. 

\begin{table}[htbp]
\centering
\caption{Simulation results for $\theta$ under no location shift with $K = 15$ external studies.}
\label{tab:compare_meta_models_no_ls_K15}
\scriptsize
\setlength{\tabcolsep}{1.5pt}
\renewcommand{\arraystretch}{0.95}
\begin{adjustbox}{width=\textwidth}
\begin{tabular}{llcccccccccccccccccc}
\toprule
 &  & \multicolumn{2}{c}{MLE(Int)} & \multicolumn{4}{c}{FE} & \multicolumn{4}{c}{$\text{wFE}$} & \multicolumn{4}{c}{RE} & \multicolumn{4}{c}{$\text{wRE}$} \\
\cmidrule(lr){3-4} \cmidrule(lr){5-8} \cmidrule(lr){9-12} \cmidrule(lr){13-16} \cmidrule(lr){17-20}
$\tau^2$ & Parameter & Bias & ESD & Bias & ESD & SE & CP & Bias & ESD & SE & CP & Bias & ESD & SE & CP & Bias & ESD & SE & CP \\
\midrule
\multirow{6}{*}{0} & $\theta_{11}$ & 0.08 & 14.99 & 0.00 & 3.07 & 3.04 & 94.75 & 0.00 & 3.07 & 3.06 & 94.80 & 0.04 & 3.08 & 3.39 & 96.55 & 0.04 & 3.08 & 3.50 & 97.05 \\
 & $\theta_{21}$ & 0.35 & 18.06 & 0.02 & 3.72 & 3.71 & 94.95 & 0.02 & 3.72 & 3.74 & 95.05 & 0.06 & 3.74 & 4.09 & 96.65 & 0.07 & 3.74 & 4.21 & 96.70 \\
 & $\theta_{21} - \theta_{11}$ & 0.43 & 19.39 & 0.02 & 3.97 & 4.00 & 95.20 & 0.02 & 3.97 & 4.02 & 95.20 & 0.02 & 4.00 & 4.50 & 96.90 & 0.02 & 4.01 & 4.63 & 97.05 \\
 & $\theta_{12}$ & 0.25 & 15.64 & 0.13 & 3.08 & 3.21 & 95.75 & 0.13 & 3.08 & 3.23 & 96.10 & 0.06 & 3.11 & 3.55 & 96.70 & 0.06 & 3.11 & 3.66 & 97.15 \\
 & $\theta_{22}$ & 0.55 & 17.96 & 0.08 & 3.60 & 3.71 & 95.45 & 0.08 & 3.60 & 3.73 & 95.70 & 0.00 & 3.62 & 4.09 & 96.60 & 0.01 & 3.62 & 4.21 & 97.00 \\
 & $\theta_{22} - \theta_{12}$ & 0.31 & 19.68 & 0.21 & 4.01 & 4.07 & 95.20 & 0.21 & 4.01 & 4.10 & 95.35 & 0.06 & 4.02 & 4.55 & 96.90 & 0.05 & 4.03 & 4.69 & 97.20 \\
\midrule
\multirow{6}{*}{0.01} & $\theta_{11}$ & 0.08 & 14.99 & 0.59 & 3.87 & 3.06 & 87.30 & 0.54 & 3.83 & 3.12 & 88.55 & 0.04 & 3.79 & 3.89 & 95.90 & 0.05 & 3.83 & 4.03 & 96.10 \\
 & $\theta_{21}$ & 0.35 & 18.06 & 0.37 & 4.42 & 3.72 & 90.85 & 0.33 & 4.39 & 3.79 & 91.55 & 0.01 & 4.42 & 4.53 & 94.90 & 0.00 & 4.46 & 4.70 & 95.80 \\
 & $\theta_{21} - \theta_{11}$ & 0.43 & 19.39 & 0.22 & 4.70 & 4.01 & 90.40 & 0.21 & 4.68 & 4.08 & 91.20 & 0.04 & 4.73 & 4.97 & 95.90 & 0.05 & 4.76 & 5.19 & 96.45 \\
 & $\theta_{12}$ & 0.25 & 15.64 & 0.33 & 3.83 & 3.22 & 90.00 & 0.34 & 3.78 & 3.28 & 91.05 & 0.09 & 3.77 & 4.01 & 95.90 & 0.09 & 3.81 & 4.16 & 96.20 \\
 & $\theta_{22}$ & 0.55 & 17.96 & 0.40 & 4.44 & 3.72 & 89.90 & 0.42 & 4.41 & 3.79 & 90.25 & 0.07 & 4.37 & 4.54 & 94.65 & 0.06 & 4.40 & 4.70 & 94.90 \\
 & $\theta_{22} - \theta_{12}$ & 0.31 & 19.68 & 0.74 & 4.77 & 4.08 & 90.00 & 0.76 & 4.77 & 4.15 & 90.15 & 0.16 & 4.77 & 5.03 & 95.25 & 0.15 & 4.82 & 5.25 & 95.90 \\
\midrule
\multirow{6}{*}{0.03} & $\theta_{11}$ & 0.08 & 14.99 & 1.70 & 4.94 & 3.08 & 75.60 & 1.35 & 4.78 & 3.23 & 79.60 & 0.01 & 4.70 & 4.43 & 92.70 & 0.02 & 4.84 & 4.69 & 93.30 \\
 & $\theta_{21}$ & 0.35 & 18.06 & 1.03 & 5.55 & 3.73 & 82.30 & 0.76 & 5.40 & 3.90 & 84.40 & 0.09 & 5.34 & 5.06 & 93.50 & 0.11 & 5.52 & 5.34 & 94.20 \\
 & $\theta_{21} - \theta_{11}$ & 0.43 & 19.39 & 0.68 & 5.80 & 4.01 & 82.75 & 0.59 & 5.78 & 4.19 & 84.15 & 0.10 & 5.86 & 5.75 & 94.10 & 0.13 & 6.04 & 6.14 & 94.70 \\
 & $\theta_{12}$ & 0.25 & 15.64 & 0.74 & 5.00 & 3.24 & 79.60 & 0.71 & 4.80 & 3.39 & 83.00 & 0.12 & 4.66 & 4.49 & 93.80 & 0.14 & 4.76 & 4.73 & 94.45 \\
 & $\theta_{22}$ & 0.55 & 17.96 & 1.08 & 5.58 & 3.73 & 79.85 & 1.10 & 5.39 & 3.90 & 83.65 & 0.03 & 5.14 & 5.09 & 94.00 & 0.00 & 5.31 & 5.38 & 94.70 \\
 & $\theta_{22} - \theta_{12}$ & 0.31 & 19.68 & 1.82 & 6.05 & 4.08 & 79.05 & 1.81 & 6.01 & 4.26 & 81.25 & 0.15 & 5.94 & 5.83 & 93.45 & 0.14 & 6.11 & 6.22 & 94.60 \\
\midrule
\multirow{6}{*}{0.05} & $\theta_{11}$ & 0.08 & 14.99 & 2.74 & 5.85 & 3.10 & 65.70 & 2.06 & 5.53 & 3.33 & 72.90 & 0.01 & 5.30 & 4.89 & 91.85 & 0.01 & 5.54 & 5.29 & 93.00 \\
 & $\theta_{21}$ & 0.35 & 18.06 & 1.63 & 6.33 & 3.74 & 73.70 & 1.11 & 6.08 & 3.98 & 79.05 & 0.23 & 5.93 & 5.49 & 91.90 & 0.29 & 6.21 & 5.91 & 93.10 \\
 & $\theta_{21} - \theta_{11}$ & 0.43 & 19.39 & 1.10 & 6.88 & 4.02 & 72.50 & 0.95 & 6.80 & 4.28 & 77.10 & 0.25 & 6.75 & 6.46 & 93.05 & 0.27 & 7.06 & 7.00 & 93.90 \\
 & $\theta_{12}$ & 0.25 & 15.64 & 1.12 & 5.76 & 3.25 & 72.25 & 1.01 & 5.47 & 3.49 & 78.45 & 0.16 & 5.25 & 4.90 & 92.15 & 0.14 & 5.48 & 5.27 & 93.65 \\
 & $\theta_{22}$ & 0.55 & 17.96 & 1.89 & 6.83 & 3.74 & 68.80 & 1.80 & 6.39 & 3.98 & 75.05 & 0.15 & 5.95 & 5.50 & 92.55 & 0.07 & 6.20 & 5.92 & 93.00 \\
 & $\theta_{22} - \theta_{12}$ & 0.31 & 19.68 & 3.02 & 7.26 & 4.08 & 69.75 & 2.82 & 7.07 & 4.36 & 74.05 & 0.30 & 6.82 & 6.52 & 92.85 & 0.21 & 7.15 & 7.07 & 94.10 \\
\bottomrule
\end{tabular}
\end{adjustbox}
\vspace{2pt}
\begin{minipage}{\textwidth}\footnotesize
Note: Bias (x100), mean estimation error multiplied by 100; Note: Bias, relative bias ($\%$); ESD (x100), empirical standard deviation multiplied by 100; SE (x100), mean estimated standard error multiplied by 100; CP, coverage probability of the 95\% Wald confidence interval (\%). MLE(Int) is based only on the internal sample.SE (x100), mean of bootstrap standard errors multiplied by 100; CP, coverage probability of 95\% confidence ($\%$). MLE(Int) is based only on the internal sample, so its results are the same across values of $\tau^2$.
\end{minipage}
\end{table}

Table~\ref{tab:compare_meta_models_ls_K15} presents simulation results under the contaminated location-shift scenario. $\hat{\theta}_{\text{FE}}$ exhibited substantial bias and well-below nominal level CP when $\tau^2 = 0$, as it borrowed indiscriminately from all shifted external studies. The unweighted $\hat{\theta}_{\text{RE}}$ partially mitigated bias by absorbing a portion of the structural shift into the estimated between-study variance component. Nevertheless, the CPs remained somewhat below the nominal level (87\%--92\%). The robust $\hat{\theta}_{\text{wFE}}$ downweighted shifted studies and achieved near-nominal CPs for some parameters, but showed noticeable residual bias for others (e.g., $\theta_{12}$ and $\theta_{22}$). $\hat{\theta}_{\text{wRE}}$, combining study-specific downweighting with random-effects adjustment, achieved the smallest bias and CPs closest to the nominal level among all integrative estimators. As heterogeneity increased from $\tau^2 = 0.01$ to $\tau^2 = 0.05$, $\hat{\theta}_{\text{FE}}$ and $\hat{\theta}_{\text{wFE}}$ suffer from substantial bias and undercoverage, while $\hat{\theta}_{\text{RE}}$ and especially $\hat{\theta}_{\text{wRE}}$ maintained more stable performance, with $\hat{\theta}_{\text{wRE}}$ CPs generally in the range of 91\%--93\%. The improvement for $\hat{\theta}_{\text{wRE}}$ came at only a modest increase in ESD relative to $\hat{\theta}_{\text{RE}}$. Compared with the internal data MLE, $\hat{\theta}_{\text{wRE}}$ achieved substantially smaller ESDs, demonstrating efficiency gains while maintaining robustness against systematic structural contamination.

\begin{table}[htbp]
\centering
\caption{Simulation results for $\theta$ under location shift with $K = 15$ external studies.}
\label{tab:compare_meta_models_ls_K15}
\scriptsize
\setlength{\tabcolsep}{1.5pt}
\renewcommand{\arraystretch}{0.95}
\begin{adjustbox}{width=\textwidth}
\begin{tabular}{llcccccccccccccccccc}
\toprule
 &  & \multicolumn{2}{c}{MLE(Int)} & \multicolumn{4}{c}{FE} & \multicolumn{4}{c}{$\text{wFE}$} & \multicolumn{4}{c}{RE} & \multicolumn{4}{c}{$\text{wRE}$} \\
\cmidrule(lr){3-4} \cmidrule(lr){5-8} \cmidrule(lr){9-12} \cmidrule(lr){13-16} \cmidrule(lr){17-20}
$\tau^2$ & Parameter & Bias & ESD & Bias & ESD & SE & CP & Bias & ESD & SE & CP & Bias & ESD & SE & CP & Bias & ESD & SE & CP \\
\midrule
\multirow{6}{*}{0} & $\theta_{11}$ & 0.08 & 14.99 & 3.00 & 3.01 & 3.11 & 84.50 & 0.49 & 3.30 & 3.33 & 94.85 & 2.20 & 4.35 & 3.81 & 87.00 & 1.49 & 4.24 & 4.14 & 92.45 \\
 & $\theta_{21}$ & 0.35 & 18.06 & 4.01 & 3.56 & 3.68 & 82.05 & 1.01 & 3.87 & 3.97 & 94.60 & 2.20 & 5.03 & 4.40 & 88.30 & 1.50 & 4.97 & 4.79 & 93.00 \\
 & $\theta_{21} - \theta_{11}$ & 0.43 & 19.39 & 1.01 & 3.77 & 3.74 & 93.75 & 0.52 & 3.99 & 4.11 & 95.50 & 0.01 & 4.70 & 4.33 & 93.05 & 0.01 & 4.79 & 4.79 & 94.75 \\
 & $\theta_{12}$ & 0.25 & 15.64 & 8.33 & 3.03 & 3.24 & 25.55 & 2.98 & 3.35 & 3.49 & 88.20 & 2.08 & 4.52 & 3.92 & 87.65 & 1.38 & 4.43 & 4.26 & 92.70 \\
 & $\theta_{22}$ & 0.55 & 17.96 & 8.39 & 3.48 & 3.67 & 36.70 & 3.08 & 3.80 & 3.97 & 89.10 & 2.31 & 4.84 & 4.38 & 88.65 & 1.45 & 4.78 & 4.77 & 93.60 \\
 & $\theta_{22} - \theta_{12}$ & 0.31 & 19.68 & 0.06 & 3.77 & 3.81 & 95.10 & 0.10 & 3.99 & 4.18 & 96.20 & 0.23 & 4.68 & 4.39 & 92.75 & 0.07 & 4.77 & 4.87 & 95.00 \\
\midrule
\multirow{6}{*}{0.01} & $\theta_{11}$ & 0.08 & 14.99 & 2.46 & 4.17 & 3.13 & 79.35 & 0.16 & 4.30 & 3.37 & 87.10 & 2.14 & 4.68 & 4.03 & 87.10 & 1.42 & 4.69 & 4.45 & 91.45 \\
 & $\theta_{21}$ & 0.35 & 18.06 & 3.67 & 4.59 & 3.69 & 76.80 & 0.87 & 4.77 & 4.01 & 89.65 & 2.14 & 5.47 & 4.64 & 88.00 & 1.39 & 5.53 & 5.11 & 92.10 \\
 & $\theta_{21} - \theta_{11}$ & 0.43 & 19.39 & 1.22 & 4.54 & 3.75 & 88.85 & 0.70 & 4.71 & 4.13 & 91.10 & 0.00 & 5.52 & 4.80 & 90.95 & 0.04 & 5.67 & 5.37 & 93.10 \\
 & $\theta_{12}$ & 0.25 & 15.64 & 7.95 & 3.78 & 3.25 & 34.05 & 2.84 & 4.10 & 3.52 & 82.95 & 1.90 & 4.74 & 4.13 & 88.65 & 1.19 & 4.72 & 4.54 & 92.65 \\
 & $\theta_{22}$ & 0.55 & 17.96 & 8.65 & 4.55 & 3.68 & 37.20 & 3.54 & 4.76 & 4.00 & 80.95 & 2.28 & 5.33 & 4.64 & 87.95 & 1.36 & 5.37 & 5.13 & 91.95 \\
 & $\theta_{22} - \theta_{12}$ & 0.31 & 19.68 & 0.70 & 4.61 & 3.82 & 88.15 & 0.70 & 4.73 & 4.20 & 91.45 & 0.38 & 5.48 & 4.87 & 91.35 & 0.18 & 5.64 & 5.47 & 93.40 \\
\midrule
\multirow{6}{*}{0.03} & $\theta_{11}$ & 0.08 & 14.99 & 1.48 & 5.61 & 3.14 & 71.45 & 0.30 & 5.40 & 3.45 & 78.65 & 1.85 & 5.42 & 4.48 & 86.65 & 1.13 & 5.59 & 5.06 & 91.00 \\
 & $\theta_{21}$ & 0.35 & 18.06 & 3.10 & 6.16 & 3.70 & 69.50 & 0.86 & 5.99 & 4.07 & 82.05 & 1.84 & 6.18 & 5.09 & 87.40 & 1.06 & 6.44 & 5.72 & 90.60 \\
 & $\theta_{21} - \theta_{11}$ & 0.43 & 19.39 & 1.63 & 5.78 & 3.76 & 78.00 & 1.16 & 5.81 & 4.16 & 83.05 & 0.01 & 6.54 & 5.64 & 88.90 & 0.07 & 6.89 & 6.38 & 92.10 \\
 & $\theta_{12}$ & 0.25 & 15.64 & 7.34 & 5.11 & 3.26 & 42.00 & 2.85 & 5.24 & 3.59 & 76.05 & 1.61 & 5.32 & 4.51 & 88.30 & 0.88 & 5.49 & 5.08 & 91.35 \\
 & $\theta_{22}$ & 0.55 & 17.96 & 9.40 & 5.76 & 3.69 & 35.65 & 4.75 & 5.74 & 4.06 & 70.05 & 2.10 & 6.00 & 5.11 & 87.65 & 1.13 & 6.25 & 5.75 & 91.85 \\
 & $\theta_{22} - \theta_{12}$ & 0.31 & 19.68 & 2.06 & 6.01 & 3.82 & 76.00 & 1.90 & 6.01 & 4.24 & 80.25 & 0.49 & 6.74 & 5.72 & 89.20 & 0.25 & 7.05 & 6.47 & 91.60 \\
\midrule
\multirow{6}{*}{0.05} & $\theta_{11}$ & 0.08 & 14.99 & 0.48 & 6.80 & 3.16 & 63.05 & 0.68 & 6.28 & 3.52 & 71.65 & 1.79 & 5.86 & 4.94 & 87.80 & 1.02 & 6.27 & 5.64 & 90.50 \\
 & $\theta_{21}$ & 0.35 & 18.06 & 2.54 & 7.09 & 3.71 & 66.30 & 0.83 & 6.70 & 4.11 & 76.75 & 1.56 & 6.65 & 5.50 & 86.85 & 0.73 & 7.04 & 6.27 & 90.95 \\
 & $\theta_{21} - \theta_{11}$ & 0.43 & 19.39 & 2.06 & 6.91 & 3.76 & 68.15 & 1.52 & 6.75 & 4.20 & 76.15 & 0.23 & 7.45 & 6.39 & 90.05 & 0.29 & 7.97 & 7.27 & 92.20 \\
 & $\theta_{12}$ & 0.25 & 15.64 & 6.63 & 5.96 & 3.28 & 47.90 & 2.66 & 6.03 & 3.65 & 72.55 & 1.45 & 5.95 & 4.90 & 87.05 & 0.75 & 6.28 & 5.60 & 90.50 \\
 & $\theta_{22}$ & 0.55 & 17.96 & 10.06 & 7.06 & 3.70 & 34.35 & 5.82 & 6.65 & 4.10 & 61.25 & 2.21 & 6.58 & 5.51 & 86.90 & 1.19 & 6.99 & 6.26 & 90.95 \\
 & $\theta_{22} - \theta_{12}$ & 0.31 & 19.68 & 3.43 & 7.27 & 3.83 & 64.80 & 3.17 & 7.05 & 4.27 & 71.55 & 0.76 & 7.55 & 6.44 & 89.75 & 0.44 & 8.09 & 7.33 & 92.05 \\
\bottomrule
\end{tabular}
\end{adjustbox}
\vspace{2pt}
\begin{minipage}{\textwidth}\footnotesize
Note: Bias (x100), mean estimation error multiplied by 100; Note: Bias, relative bias ($\%$); ESD (x100), empirical standard deviation multiplied by 100; SE (x100), mean estimated standard error multiplied by 100; CP, coverage probability of the 95\% Wald confidence interval (\%). MLE(Int) is based only on the internal sample.SE (x100), mean of bootstrap standard errors multiplied by 100; CP, coverage probability of 95\% confidence ($\%$). MLE(Int) is based only on the internal sample, so its results are the same across values of $\tau^2$.
\end{minipage}
\end{table}

Additional results are reported in the Supplementary Material. Table S1 evaluates settings with fewer external studies $K \in \left\{3,5,7,10\right\}$, whereas Table S2 considers moderate external-study sample sizes $n_k \in \left\{100,300,500,1000\right\}$ with $n_0=1000$. Across these settings, $\hat{\theta}_{\text{wRE}}$ remained nearly unbiased and had smaller ESDs than the internal-study estimator $\hat{\theta}_{\mathrm{MLE}}$, indicating stable performance even when external information is limited.

\subsection{Simulation results for generalized method of moments of target outcomes}

For combining the meta-analysis results and estimating the association between risk factor $X$ and the target outcome $Y_1$, we compared six estimators: the internal-study-only estimator $\hat{\beta}_{\text{MLE}}$, the empirical likelihood estimator $\hat{\beta}_{\text{EL}}$ \citep{Zheng2022}, and four variations of the proposed GMM estimators that arise from crossing two choices. The first choice is how the meta-analysis result enters the estimating equations: $\hat{\beta}_{\text{mGMM}}$ averages over its posterior distribution, whereas $\hat{\beta}_{\text{rGMM}}$ substitutes the posterior mean as a plug-in. The second choice is the borrowing weight in the power prior: the base estimators use the full weight $\hat{\alpha}_k = 1$, whereas the \texttt{W} variants use the Huber-type weight $\hat{\alpha}_k = \min\left\{1,\sqrt{\chi^2_{p,\,1-(Kn_0)^{-1/2}}/d_k^2}\right\}$. Crossing the two choices yields the four estimators $\hat{\beta}_{\text{mGMM}}$, $\hat{\beta}_{\text{mGMM}_{\text{W}}}$, $\hat{\beta}_{\text{rGMM}}$, and $\hat{\beta}_{\text{rGMM}_{\text{W}}}$. For the estimator $\hat{\beta}_{\text{EL}}$, the external information is based on a fixed-effects model assuming homogeneity across all studies. 

Table \ref{tab:sim_beta_no_ls} presents the results under the no-location-shift scenario. When there was no between-study heterogeneity ($\tau^2=0$), all integrative estimators had low bias below $5\%$ and exhibited substantially smaller ESDs than the internal-study-only estimator $\hat{\beta}_{\text{MLE}}$. Notably, the GMM estimators had ESDs close to those of $\hat{\beta}_{\text{EL}}$, suggesting that the random-effects-based GMM retained efficiency even when the true data-generating model was the fixed-effects model. As between-study heterogeneity increased ($\tau^2>0$), $\hat{\beta}_{\text{EL}}$ showed increasing bias and undercovarage of CPs  due to its ignorance of between-study heterogeneity. In contrast, the GMM estimators showed little bias and maintained nominal coverage across the range of $\tau^2$. Both $\hat{\beta}_{\text{mGMM}}$ and $\hat{\beta}_{\text{rGMM}}$ yielded nearly indistinguishable results in terms of bias, ESD, and CP across all settings. Compared to equal-weighted $\hat{\beta}_{\text{mGMM}}$ and $\hat{\beta}_{\text{rGMM}}$, the power-prior estimators $\hat{\beta}_{\text{mGMM}_{\text{W}}}$ and $\hat{\beta}_{\text{rGMM}_{\text{W}}}$ had smaller ESDs, indicating that the Huber-type adaptive weights were able to isolate and prioritize the most compatible external information. 

\begin{table}[htbp]
\centering
\scriptsize
\setlength{\tabcolsep}{1.8pt}
\renewcommand{\arraystretch}{0.90}
\caption{Simulation results for $\beta$ under no location shift with $K = 15$ external studies.}
\label{tab:sim_beta_no_ls}
\begin{adjustbox}{max width=\textwidth}
\begin{tabular}{llcccccccccccccccccccccc}
\toprule
 &  & \multicolumn{2}{c}{MLE} & \multicolumn{4}{c}{EL} & \multicolumn{4}{c}{mGMM} & \multicolumn{4}{c}{$\text{mGMM}_{\text{W}}$} & \multicolumn{4}{c}{rGMM} & \multicolumn{4}{c}{$\text{rGMM}_{\text{W}}$} \\
\cmidrule(lr){3-4} \cmidrule(lr){5-8} \cmidrule(lr){9-12} \cmidrule(lr){13-16} \cmidrule(lr){17-20} \cmidrule(lr){21-24}
$\tau^2$ & Parameter & Bias & ESD & Bias & ESD & SE & CP & Bias & ESD & SE & CP & Bias & ESD & SE & CP & Bias & ESD & SE & CP & Bias & ESD & SE & CP \\
\midrule
\multirow{6}{*}{0} & $\beta_{11}$ & 1.37 & 15.35 & 0.69 & 7.91 & 7.73 & 95.10 & 0.36 & 9.18 & 9.97 & 96.47 & 0.13 & 7.84 & 8.06 & 95.37 & 0.29 & 9.17 & 9.97 & 96.41 & 0.04 & 7.82 & 8.07 & 95.36 \\
 & $\beta_{21}$ & 0.47 & 16.40 & 0.78 & 8.66 & 8.40 & 94.35 & 0.24 & 9.91 & 10.59 & 96.52 & 0.36 & 8.52 & 8.76 & 95.17 & 0.27 & 9.88 & 10.61 & 96.51 & 0.40 & 8.49 & 8.78 & 95.11 \\
 & $\beta_{21} - \beta_{11}$ & 0.49 & 18.71 & 2.36 & 15.59 & 15.17 & 94.20 & 0.87 & 15.82 & 16.20 & 94.55 & 0.90 & 15.34 & 15.60 & 95.02 & 0.87 & 15.78 & 16.23 & 94.99 & 0.88 & 15.30 & 15.63 & 95.41 \\
 & $\beta_{12}$ & 0.04 & 17.05 & 0.16 & 8.93 & 8.86 & 95.55 & 0.04 & 10.13 & 11.12 & 96.97 & 0.23 & 8.90 & 9.24 & 95.53 & 0.07 & 10.18 & 11.10 & 97.11 & 0.26 & 8.91 & 9.24 & 96.17 \\
 & $\beta_{22}$ & 0.88 & 16.25 & 1.35 & 8.33 & 8.22 & 94.45 & 1.20 & 9.83 & 10.42 & 96.62 & 0.85 & 8.37 & 8.57 & 94.77 & 1.09 & 9.76 & 10.43 & 96.81 & 0.73 & 8.38 & 8.58 & 95.31 \\
 & $\beta_{22} - \beta_{12}$ & 0.34 & 19.06 & 0.60 & 15.86 & 15.83 & 95.35 & 0.46 & 16.46 & 16.91 & 95.81 & 0.46 & 16.03 & 16.29 & 94.82 & 0.44 & 16.46 & 16.91 & 96.15 & 0.43 & 16.04 & 16.29 & 95.36 \\
\midrule
\multirow{6}{*}{0.01} & $\beta_{11}$ & 1.37 & 15.35 & 1.13 & 8.15 & 7.74 & 94.50 & 0.29 & 9.42 & 10.13 & 96.67 & 0.55 & 10.27 & 8.28 & 94.87 & 0.68 & 10.61 & 10.12 & 96.63 & 0.05 & 8.07 & 8.31 & 95.30 \\
 & $\beta_{21}$ & 0.47 & 16.40 & 1.48 & 8.96 & 8.40 & 93.40 & 0.13 & 10.06 & 10.71 & 96.26 & 0.37 & 9.17 & 8.94 & 94.97 & 0.23 & 10.12 & 10.74 & 96.28 & 0.34 & 8.76 & 8.96 & 94.90 \\
 & $\beta_{21} - \beta_{11}$ & 0.49 & 18.71 & 1.86 & 15.64 & 15.17 & 94.10 & 0.58 & 15.90 & 16.26 & 95.05 & 1.35 & 17.81 & 15.66 & 95.07 & 1.20 & 17.01 & 16.29 & 94.67 & 0.76 & 15.34 & 15.70 & 95.25 \\
 & $\beta_{12}$ & 0.04 & 17.05 & 0.49 & 9.28 & 8.86 & 94.30 & 0.01 & 10.45 & 11.26 & 96.77 & 0.14 & 9.21 & 9.44 & 95.52 & 0.09 & 10.65 & 11.25 & 96.93 & 0.13 & 9.24 & 9.45 & 96.16 \\
 & $\beta_{22}$ & 0.88 & 16.25 & 1.82 & 8.45 & 8.22 & 93.15 & 1.04 & 9.90 & 10.56 & 96.67 & 0.62 & 8.48 & 8.76 & 95.37 & 0.96 & 10.08 & 10.56 & 96.33 & 0.66 & 8.51 & 8.76 & 94.95 \\
 & $\beta_{22} - \beta_{12}$ & 0.34 & 19.06 & 0.98 & 15.87 & 15.83 & 95.45 & 0.37 & 16.58 & 16.96 & 95.56 & 0.32 & 16.00 & 16.35 & 95.17 & 0.29 & 17.00 & 16.95 & 95.62 & 0.33 & 16.06 & 16.36 & 95.50 \\
\midrule
\multirow{6}{*}{0.03} & $\beta_{11}$ & 1.37 & 15.35 & 3.57 & 8.95 & 7.74 & 90.85 & 0.39 & 9.66 & 10.31 & 96.32 & 0.26 & 8.51 & 8.58 & 95.36 & 0.31 & 9.70 & 10.29 & 96.06 & 0.19 & 8.51 & 8.59 & 94.80 \\
 & $\beta_{21}$ & 0.47 & 16.40 & 2.42 & 9.32 & 8.41 & 92.50 & 0.06 & 10.42 & 10.86 & 95.67 & 0.24 & 9.18 & 9.15 & 94.20 & 0.08 & 10.40 & 10.87 & 96.17 & 0.25 & 9.25 & 9.15 & 94.40 \\
 & $\beta_{21} - \beta_{11}$ & 0.49 & 18.71 & 1.19 & 15.84 & 15.17 & 94.15 & 0.53 & 16.04 & 16.33 & 94.76 & 0.78 & 15.50 & 15.77 & 94.60 & 0.49 & 16.05 & 16.34 & 95.16 & 0.71 & 15.58 & 15.78 & 94.95 \\
 & $\beta_{12}$ & 0.04 & 17.05 & 0.85 & 9.64 & 8.86 & 92.90 & 0.03 & 10.64 & 11.42 & 96.57 & 0.03 & 9.60 & 9.67 & 94.91 & 0.16 & 10.64 & 11.42 & 96.67 & 0.17 & 9.60 & 9.67 & 94.85 \\
 & $\beta_{22}$ & 0.88 & 16.25 & 3.00 & 9.18 & 8.23 & 91.80 & 0.85 & 10.20 & 10.70 & 96.02 & 0.59 & 8.92 & 8.93 & 94.76 & 1.10 & 10.22 & 10.70 & 96.32 & 0.91 & 8.85 & 8.96 & 95.06 \\
 & $\beta_{22} - \beta_{12}$ & 0.34 & 19.06 & 1.63 & 16.13 & 15.82 & 94.45 & 0.29 & 16.68 & 17.03 & 95.42 & 0.24 & 16.33 & 16.43 & 94.96 & 0.50 & 16.69 & 17.03 & 95.76 & 0.44 & 16.25 & 16.44 & 95.11 \\
\midrule
\multirow{6}{*}{0.05} & $\beta_{11}$ & 1.37 & 15.35 & 6.60 & 9.41 & 7.75 & 88.40 & 0.63 & 9.88 & 10.46 & 96.02 & 0.46 & 8.87 & 8.89 & 94.25 & 0.46 & 9.85 & 10.45 & 96.27 & 0.38 & 8.94 & 8.89 & 94.86 \\
 & $\beta_{21}$ & 0.47 & 16.40 & 3.41 & 9.61 & 8.41 & 90.75 & 0.06 & 10.57 & 10.95 & 96.02 & 0.04 & 9.52 & 9.33 & 94.35 & 0.10 & 10.54 & 10.99 & 96.02 & 0.07 & 9.69 & 9.35 & 94.76 \\
 & $\beta_{21} - \beta_{11}$ & 0.49 & 18.71 & 0.01 & 15.90 & 15.17 & 93.50 & 0.80 & 16.08 & 16.40 & 95.12 & 0.59 & 15.72 & 15.88 & 95.21 & 0.70 & 16.01 & 16.41 & 94.60 & 0.56 & 15.99 & 15.90 & 95.06 \\
 & $\beta_{12}$ & 0.04 & 17.05 & 1.11 & 10.18 & 8.85 & 91.20 & 0.19 & 10.74 & 11.54 & 96.63 & 0.01 & 9.75 & 9.90 & 95.11 & 0.01 & 10.74 & 11.53 & 96.62 & 0.07 & 9.74 & 9.90 & 95.11 \\
 & $\beta_{22}$ & 0.88 & 16.25 & 4.14 & 9.60 & 8.23 & 89.75 & 0.88 & 10.44 & 10.78 & 95.97 & 0.65 & 9.25 & 9.13 & 94.30 & 0.98 & 10.46 & 10.81 & 95.81 & 0.57 & 9.23 & 9.13 & 94.81 \\
 & $\beta_{22} - \beta_{12}$ & 0.34 & 19.06 & 2.22 & 16.36 & 15.82 & 94.35 & 0.20 & 16.61 & 17.09 & 95.52 & 0.24 & 16.23 & 16.54 & 95.46 & 0.35 & 16.62 & 17.08 & 95.26 & 0.25 & 16.21 & 16.52 & 95.41 \\
\bottomrule
\end{tabular}
\end{adjustbox}
\vspace{2pt}
\begin{minipage}{\textwidth}
\footnotesize
Note: Bias, relative bias ($\%$); ESD (x100), empirical standard deviation displayed after multiplying by 100; SE (x100), mean of bootstrap standard errors displayed after multiplying by 100; CP, coverage probability of 95\% confidence intervals ($\%$). MLE is based only on the internal sample, so its results are the same across values of $\tau^2$.
\end{minipage}
\end{table}

Table \ref{tab:sim_beta_ls} presents the simulation results under the contaminated location-shift scenario. While all estimators had much smaller variances compared to the internal study estimator $\hat{\beta}_{\text{MLE}}$, the estimator $\hat{\beta}_{\text{EL}}$  showed noticeably larger bias and undercoverage, reflecting its sensitivity to outlying external studies. In contrast, the GMM estimators were more robust, showing little bias while preserving efficiency gains. Consistent with the no-location-shift scenario, $\hat{\beta}_{\text{mGMM}}$ and $\hat{\beta}_{\text{rGMM}}$ yielded nearly identical results across all settings. Importantly, the power-prior weight  $\hat{\beta}_{\text{mGMM}_{\text{W}}}$ and $\hat{\beta}_{\text{rGMM}_{\text{W}}}$ noticeably outperformed equal-weighted estimators $\hat{\beta}_{\text{mGMM}}$ and $\hat{\beta}_{\text{rGMM}}$ with smaller ESDs. It is interesting to note that the random-effects model remained robust even under location shift because it absorbed the outliers by essentially inflating the between-study covariance estimates, which indiscriminately downweights all external studies. The proposed Huber-type adaptive weights mitigated this issue by limiting the contribution of outlying studies, leading to more robust borrowing information from non-outlying studies. 

\begin{table}[htbp]
\centering
\scriptsize
\setlength{\tabcolsep}{1.8pt}
\renewcommand{\arraystretch}{0.90}
\caption{Simulation results for $\beta$ under location shift with $K = 15$ external studies.}
\label{tab:sim_beta_ls}
\begin{adjustbox}{max width=\textwidth}
\begin{tabular}{llcccccccccccccccccccccc}
\toprule
 &  & \multicolumn{2}{c}{MLE} & \multicolumn{4}{c}{EL} & \multicolumn{4}{c}{mGMM} & \multicolumn{4}{c}{$\text{mGMM}_{\text{W}}$} & \multicolumn{4}{c}{rGMM} & \multicolumn{4}{c}{$\text{rGMM}_{\text{W}}$} \\
\cmidrule(lr){3-4} \cmidrule(lr){5-8} \cmidrule(lr){9-12} \cmidrule(lr){13-16} \cmidrule(lr){17-20} \cmidrule(lr){21-24}
$\tau^2$ & Parameter & Bias & ESD & Bias & ESD & SE & CP & Bias & ESD & SE & CP & Bias & ESD & SE & CP & Bias & ESD & SE & CP & Bias & ESD & SE & CP \\
\midrule
\multirow{6}{*}{0} & $\beta_{11}$ & 1.37 & 15.35 & 10.07 & 7.88 & 7.78 & 92.25 & 0.36 & 9.18 & 9.97 & 96.47 & 0.13 & 7.84 & 8.06 & 95.37 & 0.29 & 9.17 & 9.97 & 96.41 & 0.04 & 7.82 & 8.07 & 95.36 \\
 & $\beta_{21}$ & 0.47 & 16.40 & 4.81 & 8.67 & 8.41 & 92.55 & 0.24 & 9.91 & 10.59 & 96.52 & 0.36 & 8.52 & 8.76 & 95.17 & 0.27 & 9.88 & 10.61 & 96.51 & 0.40 & 8.49 & 8.78 & 95.11 \\
 & $\beta_{21} - \beta_{11}$ & 0.49 & 18.71 & 0.84 & 15.56 & 15.14 & 94.15 & 0.87 & 15.82 & 16.20 & 94.55 & 0.90 & 15.34 & 15.60 & 95.02 & 0.87 & 15.78 & 16.23 & 94.99 & 0.88 & 15.30 & 15.63 & 95.41 \\
 & $\beta_{12}$ & 0.04 & 17.05 & 14.75 & 8.88 & 8.95 & 86.75 & 0.04 & 10.13 & 11.12 & 96.97 & 0.23 & 8.90 & 9.24 & 95.53 & 0.07 & 10.18 & 11.10 & 97.11 & 0.26 & 8.91 & 9.24 & 96.17 \\
 & $\beta_{22}$ & 0.88 & 16.25 & 26.62 & 8.27 & 8.22 & 81.50 & 1.20 & 9.83 & 10.42 & 96.62 & 0.85 & 8.37 & 8.57 & 94.77 & 1.09 & 9.76 & 10.43 & 96.81 & 0.73 & 8.38 & 8.58 & 95.31 \\
 & $\beta_{22} - \beta_{12}$ & 0.34 & 19.06 & 0.38 & 15.80 & 15.86 & 95.05 & 0.46 & 16.46 & 16.91 & 95.81 & 0.46 & 16.03 & 16.29 & 94.82 & 0.44 & 16.46 & 16.91 & 96.15 & 0.43 & 16.04 & 16.29 & 95.36 \\
\midrule
\multirow{6}{*}{0.01} & $\beta_{11}$ & 1.37 & 15.35 & 8.32 & 8.36 & 7.78 & 91.90 & 0.29 & 9.42 & 10.13 & 96.67 & 0.55 & 10.27 & 8.28 & 94.87 & 0.68 & 10.61 & 10.12 & 96.63 & 0.05 & 8.07 & 8.31 & 95.30 \\
 & $\beta_{21}$ & 0.47 & 16.40 & 4.17 & 8.99 & 8.41 & 92.65 & 0.13 & 10.06 & 10.71 & 96.26 & 0.37 & 9.17 & 8.94 & 94.97 & 0.23 & 10.12 & 10.74 & 96.28 & 0.34 & 8.76 & 8.96 & 94.90 \\
 & $\beta_{21} - \beta_{11}$ & 0.49 & 18.71 & 0.27 & 15.61 & 15.14 & 94.40 & 0.58 & 15.90 & 16.26 & 95.05 & 1.35 & 17.81 & 15.66 & 95.07 & 1.20 & 17.01 & 16.29 & 94.67 & 0.76 & 15.34 & 15.70 & 95.25 \\
 & $\beta_{12}$ & 0.04 & 17.05 & 14.22 & 9.20 & 8.94 & 85.80 & 0.01 & 10.45 & 11.26 & 96.77 & 0.14 & 9.21 & 9.44 & 95.52 & 0.09 & 10.65 & 11.25 & 96.93 & 0.13 & 9.24 & 9.45 & 96.16 \\
 & $\beta_{22}$ & 0.88 & 16.25 & 26.92 & 8.49 & 8.22 & 80.45 & 1.04 & 9.90 & 10.56 & 96.67 & 0.62 & 8.48 & 8.76 & 95.37 & 0.96 & 10.08 & 10.56 & 96.33 & 0.66 & 8.51 & 8.76 & 94.95 \\
 & $\beta_{22} - \beta_{12}$ & 0.34 & 19.06 & 0.82 & 15.82 & 15.86 & 95.35 & 0.37 & 16.58 & 16.96 & 95.56 & 0.32 & 16.00 & 16.35 & 95.17 & 0.29 & 17.00 & 16.95 & 95.62 & 0.33 & 16.06 & 16.36 & 95.50 \\
\midrule
\multirow{6}{*}{0.03} & $\beta_{11}$ & 1.37 & 15.35 & 6.06 & 9.08 & 7.78 & 90.35 & 0.39 & 9.66 & 10.31 & 96.32 & 0.26 & 8.51 & 8.58 & 95.36 & 0.31 & 9.70 & 10.29 & 96.06 & 0.19 & 8.51 & 8.59 & 94.80 \\
 & $\beta_{21}$ & 0.47 & 16.40 & 3.32 & 9.75 & 8.42 & 90.25 & 0.06 & 10.42 & 10.86 & 95.67 & 0.24 & 9.18 & 9.15 & 94.20 & 0.08 & 10.40 & 10.87 & 96.17 & 0.25 & 9.25 & 9.15 & 94.40 \\
 & $\beta_{21} - \beta_{11}$ & 0.49 & 18.71 & 0.39 & 15.84 & 15.15 & 94.25 & 0.53 & 16.04 & 16.33 & 94.76 & 0.78 & 15.50 & 15.77 & 94.60 & 0.49 & 16.05 & 16.34 & 95.16 & 0.71 & 15.58 & 15.78 & 94.95 \\
 & $\beta_{12}$ & 0.04 & 17.05 & 13.37 & 9.67 & 8.93 & 86.25 & 0.03 & 10.64 & 11.42 & 96.57 & 0.03 & 9.60 & 9.67 & 94.91 & 0.16 & 10.64 & 11.42 & 96.67 & 0.17 & 9.60 & 9.67 & 94.85 \\
 & $\beta_{22}$ & 0.88 & 16.25 & 27.85 & 9.21 & 8.23 & 77.05 & 0.85 & 10.20 & 10.70 & 96.02 & 0.59 & 8.92 & 8.93 & 94.76 & 1.10 & 10.22 & 10.70 & 96.32 & 0.91 & 8.85 & 8.96 & 95.06 \\
 & $\beta_{22} - \beta_{12}$ & 0.34 & 19.06 & 1.70 & 16.10 & 15.85 & 94.55 & 0.29 & 16.68 & 17.03 & 95.42 & 0.24 & 16.33 & 16.43 & 94.96 & 0.50 & 16.69 & 17.03 & 95.76 & 0.44 & 16.25 & 16.44 & 95.11 \\
\midrule
\multirow{6}{*}{0.05} & $\beta_{11}$ & 1.37 & 15.35 & 3.01 & 9.78 & 7.79 & 87.95 & 0.63 & 9.88 & 10.46 & 96.02 & 0.46 & 8.87 & 8.89 & 94.25 & 0.46 & 9.85 & 10.45 & 96.27 & 0.38 & 8.94 & 8.89 & 94.86 \\
 & $\beta_{21}$ & 0.47 & 16.40 & 2.31 & 10.14 & 8.42 & 89.60 & 0.06 & 10.57 & 10.95 & 96.02 & 0.04 & 9.52 & 9.33 & 94.35 & 0.10 & 10.54 & 10.99 & 96.02 & 0.07 & 9.69 & 9.35 & 94.76 \\
 & $\beta_{21} - \beta_{11}$ & 0.49 & 18.71 & 1.55 & 15.84 & 15.15 & 93.60 & 0.80 & 16.08 & 16.40 & 95.12 & 0.59 & 15.72 & 15.88 & 95.21 & 0.70 & 16.01 & 16.41 & 94.60 & 0.56 & 15.99 & 15.90 & 95.06 \\
 & $\beta_{12}$ & 0.04 & 17.05 & 12.53 & 10.17 & 8.92 & 85.05 & 0.19 & 10.74 & 11.54 & 96.63 & 0.01 & 9.75 & 9.90 & 95.11 & 0.01 & 10.74 & 11.53 & 96.62 & 0.07 & 9.74 & 9.90 & 95.11 \\
 & $\beta_{22}$ & 0.88 & 16.25 & 28.32 & 9.66 & 8.23 & 75.55 & 0.88 & 10.44 & 10.78 & 95.97 & 0.65 & 9.25 & 9.13 & 94.30 & 0.98 & 10.46 & 10.81 & 95.81 & 0.57 & 9.23 & 9.13 & 94.81 \\
 & $\beta_{22} - \beta_{12}$ & 0.34 & 19.06 & 2.40 & 16.35 & 15.85 & 94.05 & 0.20 & 16.61 & 17.09 & 95.52 & 0.24 & 16.23 & 16.54 & 95.46 & 0.35 & 16.62 & 17.08 & 95.26 & 0.25 & 16.21 & 16.52 & 95.41 \\
\bottomrule
\end{tabular}
\end{adjustbox}
\vspace{2pt}
\begin{minipage}{\textwidth}
\footnotesize
Note: Bias, relative bias ($\%$); ESD (x100), empirical standard deviation displayed after multiplying by 100; SE (x100), mean of bootstrap standard errors displayed after multiplying by 100; CP, coverage probability of 95\% confidence intervals ($\%$). MLE is based only on the internal sample, so its results are the same across values of $\tau^2$.
\end{minipage}
\end{table}

Additional simulation results are provided in the Supplementary Material. Table S3 considers fewer numbers of external studies, $K \in \{3,5,7,10\}$, and Table S4 considers moderate external-study sample sizes, $n_k \in \{100,300,500,1000\}$, with internal sample size $n_0=1000$. All integrative estimators had smaller ESDs than the internal-study-only estimator $\hat{\beta}_{\text{MLE}}$. When $K$ was small, $\hat{\beta}_{\text{EL}}$ underestimated the SEs and exhibited undercoverage, whereas small $n_k$ resulted in substantial bias. In contrast, $\hat{\beta}_{\text{mGMM}_{\text{W}}}$ and $\hat{\beta}_{\text{rGMM}_{\text{W}}}$ maintained little bias, bootstrap SEs close to the ESDs, and coverage near the nominal level across the considered settings. These results support the finite-sample robustness of the proposed estimators when external information is limited.

\section{Application} \label{sec:app}

\subsection{Tumor Marker Association Analysis}

Colorectal cancer (CRC) comprises molecular subtypes defined by genetic and epigenetic tumor features, including microsatellite instability (MSI), the CpG island methylator phenotype (CIMP), and mutations in BRAF and KRAS \citep{Jass2006,Ogino2010,Sung2021}. Associations with lifestyle risk factors may vary across these subtypes, but sequencing data for novel markers remain limited \citep{Zaidi2020}. We therefore combine individual-level data on novel markers as target outcomes with external summary information on established markers as auxiliary outcomes. 

We examined associations between smoking and 16 novel tumor markers in the Cancer Prevention Study-II Nutrition Cohort. The internal sample comprised 538 cases and 736 controls matched on age and sex, with ages ranging from 56 to 90 years; 53.6\% were female and 44.7\% were ever-smokers. External information consisted of smoking log-odds ratios for MSI, BRAF, CIMP, and KRAS from 11 publications, with 5,199--7,890 cases per marker; 4,867--9,811 controls per established marker \citep{Botteri2020, Botteri2021}. Among the 11 external studies, 9 reported log-ORs for MSI, whereas 4 or 5 reported log-ORs for these three established markers. 

% Model and result Setting

For each novel marker, we conducted separate analyses using each of the four established markers as the auxiliary outcome. And we computed the estimators $\hat{\beta}_{\text{mGMM}_{\text{W}}}$ and $\hat{\beta}_{\text{rGMM}_{\text{W}}}$ for the smoking log-ORs of the nonmutated and mutated subtypes, $\beta_{1,1}$ and $\beta_{2,1}$, and their heterogeneity contrast $\beta_{2,1}-\beta_{1,1}$, adjusting for age and sex. Both estimators used the Huber-type borrowing weights. Because the external studies reported only marginal smoking log-ORs, the age and sex effects were estimated from the internal data conditional on the external smoking estimates. We compared the proposed estimators with the internal-only estimator $\hat{\beta}_{\text{MLE}}$ and the EL estimator $\hat{\beta}_{\text{EL}}$ \citep{Zheng2022}. Standard errors for the proposed estimators were based on 500 hybrid bootstrap replications; model-based standard errors were used for the comparators. 

\begin{figure}
\centering
\includegraphics[width=\linewidth]{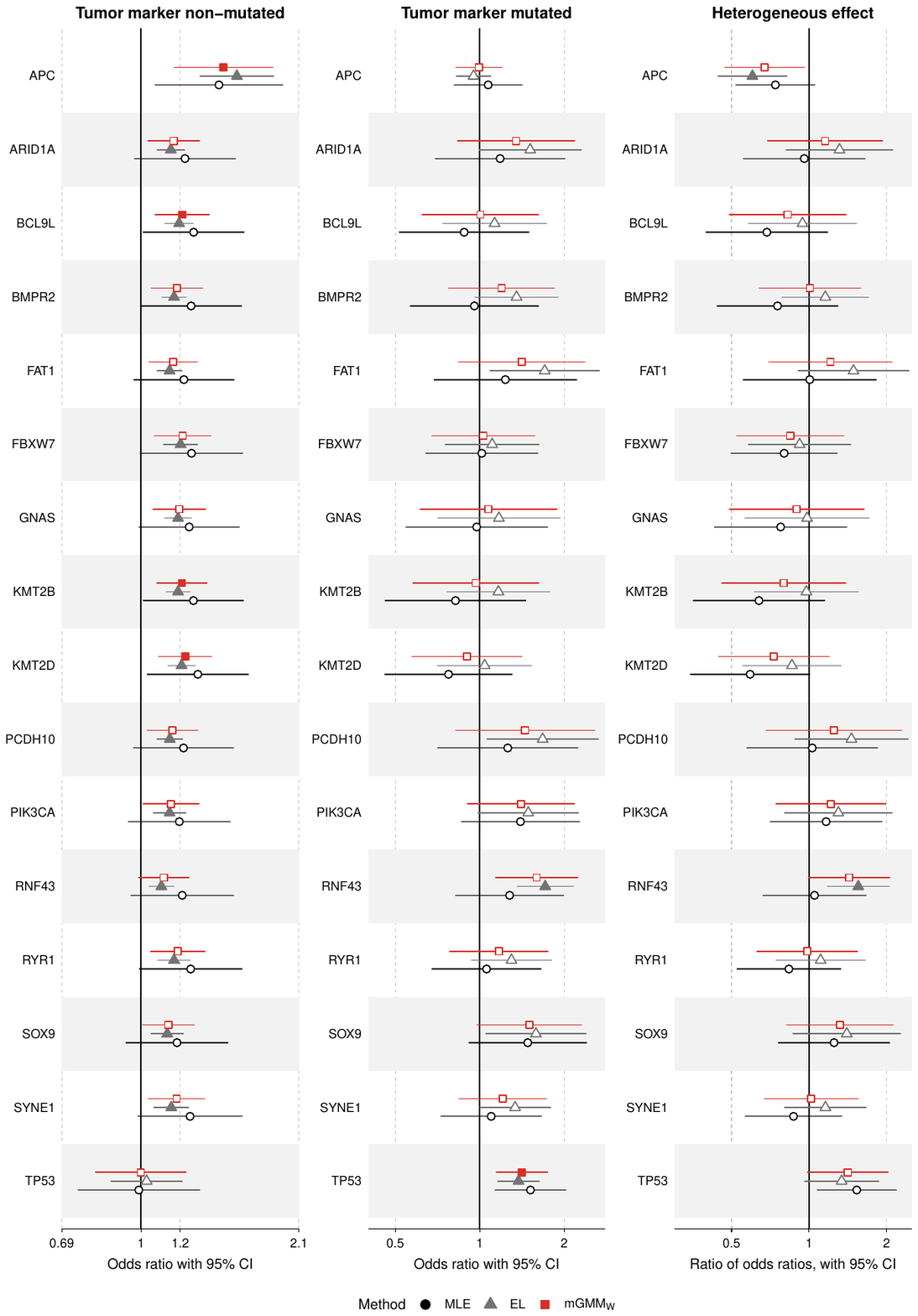}
\caption{Forest plot of ORs for the associations of smoking with tumor markers. Nonsignificant results are displayed as hollow points using a Bonferroni correction for multiple testing, with a significance threshold of $0.05/16 \approx 0.003$. The ratio of ORs compares the OR of smoking for CRC with the mutated marker with that for CRC with the non-mutated marker.}
\label{fig:forest_plot}
\end{figure}

Table~\ref{tab:smoking_novel_markers_mle_el_ppmgmm} and Figure~\ref{fig:forest_plot} summarize the results with external log-ORs on the established marker MSI. Because $\hat{\beta}_{\text{mGMM}_{\text{W}}}$ and $\hat{\beta}_{\text{rGMM}_{\text{W}}}$ gave nearly identical results, we report only $\hat{\beta}_{\text{mGMM}_{\text{W}}}$ in the main text; the complete results for $\hat{\beta}_{\text{rGMM}_{\text{W}}}$ are given in Table~S5 and Figure~S1 from the Supplementary Material. Relative to the internal-only estimator $\hat{\beta}_{\text{MLE}}$, $\hat{\beta}_{\text{mGMM}_{\text{W}}}$ reduced the SEs for the subtype specific associations $\beta_{1,1}$ and $\beta_{2,1}$ across all markers, with relative efficiency gains of $1.4\%$--$335.2\%$. Gains for the heterogeneity contrast were smaller, because this case-case contrast benefits from external case-control information only through the correlation between the novel and auxiliary markers. At the Bonferroni threshold of $0.05/16\approx0.003$, both $\hat{\beta}_{\text{mGMM}_{\text{W}}}$ and $\hat{\beta}_{\text{EL}}$ detected an association between smoking and mutated TP53, whereas $\hat{\beta}_{\text{MLE}}$ did not. No heterogeneity contrast was significant under $\hat{\beta}_{\text{mGMM}_{\text{W}}}$. $\hat{\beta}_{\text{EL}}$ identified significant heterogeneity for APC and RNF43. However, the estimates, particularly for RNF43, deviated substantially from the internal estimates, suggesting the possibility of the pact from heterogeneity in the external information.

\begin{table}[!htbp]
\centering
\caption{Log(OR) estimates, bootstrap SEs, and p-values for the association of smoking with the novel tumor markers.}
\label{tab:smoking_novel_markers_mle_el_ppmgmm}
\resizebox{\textwidth}{!}{%
\begin{tabular}{llrrrrrrrrr}
\toprule
 & & \multicolumn{3}{c}{MLE} & \multicolumn{3}{c}{EL} & \multicolumn{3}{c}{$\text{mGMM}_{\text{W}}$} \\
\cmidrule(lr){3-5} \cmidrule(lr){6-8} \cmidrule(lr){9-11}
Tumor marker & Parameter & log(OR) & SE & p-value & log(OR) & SE & p-value & log(OR) & SE & p-value \\
\midrule
APC & $\beta_{1,1}$ & 0.370 & 0.154 & 0.016 & 0.455 & 0.089 & <0.001 & 0.391 & 0.119 & 0.001 \\
 & $\beta_{2,1}$ & 0.069 & 0.142 & 0.626 & -0.051 & 0.072 & 0.480 & -0.007 & 0.096 & 0.945 \\
 & $\beta_{2,1}-\beta_{1,1}$ & -0.301 & 0.180 & 0.095 & -0.506 & 0.157 & 0.001 & -0.398 & 0.181 & 0.028 \\
\midrule
ARID1A & $\beta_{1,1}$ & 0.209 & 0.122 & 0.085 & 0.142 & 0.033 & <0.001 & 0.156 & 0.062 & 0.013 \\
 & $\beta_{2,1}$ & 0.167 & 0.271 & 0.538 & 0.414 & 0.214 & 0.053 & 0.298 & 0.245 & 0.223 \\
 & $\beta_{2,1}-\beta_{1,1}$ & -0.043 & 0.277 & 0.877 & 0.272 & 0.243 & 0.263 & 0.143 & 0.262 & 0.586 \\
\midrule
BCL9L & $\beta_{1,1}$ & 0.250 & 0.122 & 0.040 & 0.181 & 0.034 & <0.001 & 0.196 & 0.065 & 0.003 \\
 & $\beta_{2,1}$ & -0.128 & 0.271 & 0.638 & 0.122 & 0.217 & 0.574 & 0.005 & 0.244 & 0.985 \\
 & $\beta_{2,1}-\beta_{1,1}$ & -0.377 & 0.277 & 0.174 & -0.059 & 0.246 & 0.810 & -0.191 & 0.266 & 0.472 \\
\midrule
BMPR2 & $\beta_{1,1}$ & 0.238 & 0.122 & 0.050 & 0.157 & 0.029 & <0.001 & 0.172 & 0.062 & 0.006 \\
 & $\beta_{2,1}$ & -0.044 & 0.268 & 0.871 & 0.302 & 0.174 & 0.083 & 0.179 & 0.221 & 0.418 \\
 & $\beta_{2,1}-\beta_{1,1}$ & -0.282 & 0.275 & 0.305 & 0.145 & 0.197 & 0.461 & 0.007 & 0.231 & 0.975 \\
\midrule
FAT1 & $\beta_{1,1}$ & 0.204 & 0.121 & 0.091 & 0.136 & 0.030 & <0.001 & 0.153 & 0.058 & 0.009 \\
 & $\beta_{2,1}$ & 0.210 & 0.298 & 0.480 & 0.533 & 0.229 & 0.020 & 0.344 & 0.264 & 0.193 \\
 & $\beta_{2,1}-\beta_{1,1}$ & 0.006 & 0.303 & 0.984 & 0.397 & 0.253 & 0.117 & 0.191 & 0.281 & 0.497 \\
\midrule
FBXW7 & $\beta_{1,1}$ & 0.240 & 0.124 & 0.052 & 0.188 & 0.041 & <0.001 & 0.198 & 0.068 & 0.004 \\
 & $\beta_{2,1}$ & 0.017 & 0.234 & 0.941 & 0.102 & 0.196 & 0.601 & 0.030 & 0.215 & 0.890 \\
 & $\beta_{2,1}-\beta_{1,1}$ & -0.223 & 0.242 & 0.357 & -0.085 & 0.233 & 0.714 & -0.168 & 0.244 & 0.491 \\
\midrule
GNAS & $\beta_{1,1}$ & 0.229 & 0.121 & 0.058 & 0.176 & 0.032 & <0.001 & 0.182 & 0.064 & 0.004 \\
 & $\beta_{2,1}$ & -0.025 & 0.296 & 0.933 & 0.157 & 0.256 & 0.539 & 0.072 & 0.286 & 0.802 \\
 & $\beta_{2,1}-\beta_{1,1}$ & -0.254 & 0.301 & 0.400 & -0.018 & 0.283 & 0.948 & -0.110 & 0.307 & 0.720 \\
\midrule
KMT2B & $\beta_{1,1}$ & 0.249 & 0.121 & 0.039 & 0.176 & 0.029 & <0.001 & 0.194 & 0.061 & 0.001 \\
 & $\beta_{2,1}$ & -0.199 & 0.294 & 0.499 & 0.153 & 0.215 & 0.477 & -0.032 & 0.263 & 0.904 \\
 & $\beta_{2,1}-\beta_{1,1}$ & -0.448 & 0.299 & 0.135 & -0.024 & 0.237 & 0.921 & -0.226 & 0.283 & 0.424 \\
\midrule
KMT2D & $\beta_{1,1}$ & 0.270 & 0.122 & 0.027 & 0.193 & 0.033 & <0.001 & 0.210 & 0.064 & 0.001 \\
 & $\beta_{2,1}$ & -0.256 & 0.266 & 0.334 & 0.040 & 0.197 & 0.840 & -0.105 & 0.230 & 0.650 \\
 & $\beta_{2,1}-\beta_{1,1}$ & -0.527 & 0.272 & 0.053 & -0.153 & 0.224 & 0.494 & -0.314 & 0.252 & 0.212 \\
\midrule
PCDH10 & $\beta_{1,1}$ & 0.202 & 0.121 & 0.095 & 0.137 & 0.031 & <0.001 & 0.150 & 0.061 & 0.014 \\
 & $\beta_{2,1}$ & 0.230 & 0.293 & 0.433 & 0.516 & 0.233 & 0.027 & 0.372 & 0.291 & 0.201 \\
 & $\beta_{2,1}-\beta_{1,1}$ & 0.028 & 0.298 & 0.925 & 0.379 & 0.258 & 0.142 & 0.222 & 0.309 & 0.473 \\
\midrule
PIK3CA & $\beta_{1,1}$ & 0.182 & 0.123 & 0.137 & 0.136 & 0.039 & <0.001 & 0.143 & 0.068 & 0.035 \\
 & $\beta_{2,1}$ & 0.334 & 0.247 & 0.176 & 0.398 & 0.210 & 0.058 & 0.338 & 0.224 & 0.132 \\
 & $\beta_{2,1}-\beta_{1,1}$ & 0.152 & 0.254 & 0.550 & 0.262 & 0.245 & 0.284 & 0.195 & 0.251 & 0.437 \\
\midrule
RNF43 & $\beta_{1,1}$ & 0.196 & 0.124 & 0.114 & 0.097 & 0.030 & 0.001 & 0.109 & 0.061 & 0.073 \\
 & $\beta_{2,1}$ & 0.245 & 0.226 & 0.278 & 0.537 & 0.117 & <0.001 & 0.467 & 0.172 & 0.007 \\
 & $\beta_{2,1}-\beta_{1,1}$ & 0.049 & 0.235 & 0.836 & 0.440 & 0.141 & 0.002 & 0.359 & 0.185 & 0.053 \\
\midrule
RYR1 & $\beta_{1,1}$ & 0.236 & 0.124 & 0.057 & 0.157 & 0.039 & <0.001 & 0.175 & 0.066 & 0.008 \\
 & $\beta_{2,1}$ & 0.056 & 0.228 & 0.806 & 0.260 & 0.168 & 0.123 & 0.157 & 0.206 & 0.446 \\
 & $\beta_{2,1}-\beta_{1,1}$ & -0.180 & 0.236 & 0.447 & 0.103 & 0.203 & 0.612 & -0.018 & 0.229 & 0.938 \\
\midrule
SOX9 & $\beta_{1,1}$ & 0.171 & 0.123 & 0.163 & 0.125 & 0.039 & 0.002 & 0.130 & 0.062 & 0.037 \\
 & $\beta_{2,1}$ & 0.394 & 0.246 & 0.109 & 0.461 & 0.210 & 0.028 & 0.406 & 0.219 & 0.063 \\
 & $\beta_{2,1}-\beta_{1,1}$ & 0.223 & 0.253 & 0.379 & 0.337 & 0.245 & 0.169 & 0.276 & 0.243 & 0.256 \\
\midrule
SYNE1 & $\beta_{1,1}$ & 0.233 & 0.126 & 0.063 & 0.144 & 0.042 & 0.001 & 0.169 & 0.068 & 0.013 \\
 & $\beta_{2,1}$ & 0.094 & 0.210 & 0.654 & 0.290 & 0.148 & 0.050 & 0.188 & 0.183 & 0.304 \\
 & $\beta_{2,1}-\beta_{1,1}$ & -0.139 & 0.220 & 0.527 & 0.146 & 0.186 & 0.433 & 0.019 & 0.214 & 0.930 \\
\midrule
TP53 & $\beta_{1,1}$ & -0.010 & 0.147 & 0.946 & 0.027 & 0.086 & 0.753 & -0.000 & 0.110 & 0.998 \\
 & $\beta_{2,1}$ & 0.416 & 0.148 & 0.005 & 0.318 & 0.087 & <0.001 & 0.346 & 0.108 & 0.001 \\
 & $\beta_{2,1}-\beta_{1,1}$ & 0.426 & 0.180 & 0.018 & 0.291 & 0.169 & 0.085 & 0.347 & 0.184 & 0.059 \\
\midrule
\bottomrule
\end{tabular}%
}
\end{table}

Results using the other three established markers (BRAF, CIMP, and KRAS) are presented in Tables S6--S8 and Figures S2--S4 of the Supplementary Material. The relative efficiency gains were generally smaller than those obtained using MSI, with several contrasts assessing heterogeneity exhibiting larger standard errors than the corresponding internal-study-only estimates. These findings are consistent with Corollary~\ref{cor:efficiency_firststage}: with fewer external studies, uncertainty in estimating the between-study heterogeneity can offset the efficiency gains from incorporating auxiliary outcome information. Furthermore, $\hat{\beta}_{\text{rGMM}_{\text{W}}}$ generally yielded smaller standard errors than $\hat{\beta}_{\text{mGMM}_{\text{W}}}$ in these analyses.

\section{Discussion} \label{sec:disc}

% Summary of existing methods and advantages supported by simulation and application results

In this article, we proposed a two-stage approach for leveraging external studies to improve the efficiency of internal-study analyses. In the first stage, we developed a random-effects meta-analytic model that combines the internal-study estimator with summary estimates from external studies. The combination of random effects and Huber-type weights enables the meta-analytic estimator to account for between-study heterogeneity and limit the influence of external studies that systematically deviate from the internal study. In the second stage, we used GMM to integrate the meta-analytic estimator into the internal-study model for the target outcome through moment conditions. Unlike most existing methods \citep{Zhai2022}, the proposed approach accommodates external studies reporting an auxiliary outcome that is related to, but distinct from, the internal-study target outcome. Moreover, this framework incorporates multiple external studies without adding estimating equations for each additional study, making it more scalable than existing methods that introduce one set of estimating equations per study \citep{Shao2023, Fang2025}. Finally, unlike constrained empirical likelihood approaches \citep{Zheng2022}, the proposed framework does not require specifying a joint or conditional model relating the target and auxiliary outcomes, thereby improving robustness to model misspecification.

% Gudiance on when to use two proposed estimators

The two proposed GMM methods differ in how the meta-analytic estimates are incorporated into the second-stage GMM estimation. The \texttt{rGMM} method uses the posterior mean as a plug-in estimator, providing a computationally simple and numerically stable implementation. In contrast, \texttt{mGMM} integrates the estimating equations over the posterior distribution of the meta-analytic parameter, thereby accounting for posterior uncertainty in the GMM moments. In our simulation studies, the two methods generally yielded comparable point estimates, standard errors and coverage probabilities. However, in the real-data analysis, \texttt{rGMM} yielded smaller standard errors when fewer studies were available, suggesting that \texttt{rGMM} may be preferred in this setting.

% Limitations and future directions

Our work suggests several directions for future research.
Although we focused on multinomial outcomes, the framework
could be adapted to continuous, survival and ordinal outcomes
by modifying the outcome models and moment conditions.
The framework could also be extended to subtypes jointly
defined by multiple markers by imposing a structure on
the subtype-specific effects, for example through a
two-stage regression model \citep{Chatterjee2004}. The proposed framework incorporates external information on a single auxiliary outcome. An important direction for future research is to develop methods for efficiently integrating information from multiple auxiliary outcomes. Another limitation is that the proposed Huber-type weighting down-weights external studies based on their discrepancies from the internal study, without identifying the sources of those discrepancies. Studies receiving lower weights therefore warrant further investigation to determine whether down-weighting is justified by differences in study populations, study designs, covariate adjustment, or outcome ascertainment.  

\section*{Declaration of the use of generative AI and AI-assisted technologies}
This work was primarily created by authors. After drafting the manuscript, the authors used ChatGPT to check its grammar and math consistency. The authors reviewed suggestions and edited the manuscript as necessary. They take full responsibility for the content of the publication.

\section*{Acknowledgment}
This work was supported by the National Institutes of Health (R01CA297681). The authors thank the participants and study investigators of the Cancer Prevention Study-II (CPS-II), whose data were used in this study, for their invaluable contributions. The CPS-II cohort is funded by the American Cancer Society, with contributions from cancer registries supported by the Centers for Disease Control and Prevention's National Program of Cancer Registries and the National Cancer Institute's Surveillance, Epidemiology, and End Results Program.

\appendix

\appendixone

\bibliographystyle{biometrika}
\bibliography{paper-ref}

\section*{Supplementary material}

\subsection{Assumptions}
\label{sec:additional_assumptions}

The following assumptions are used in the supplementary proofs.

% -------------------------------------------------
\begin{assumption}[Within-study second-order asymptotic representation]
\label{assump:normality}
For $k=0,\ldots,K$, assume that
$$
\hat\theta^{(k)}-\theta^{(k)}
=
n_k^{-1/2}Z_k+n_k^{-1}R_k,
\qquad
\E(Z_k)=0,
\qquad
\Var(Z_k)=V_k,
$$
where $Z_k$ is the leading stochastic term and $R_k$
is the scaled second-order remainder.

Assume that
$$
\sqrt{n_k}\norm{n_kS_k-V_k}
=
O_p(1).
$$
For some constant $C<\infty$,
$$
\E\norm{Z_k}^4+\E\norm{R_k}^2\leq C.
$$
\end{assumption}

% -------------------------------------------------
\begin{assumption}[Asymptotic compatibility of borrowing weights]
\label{assump:weights}
Under the local-heterogeneity regime, assume that
$$
\Prb\left(
\max_{1\leq k\leq K}d_k^2\leq c_{K,n_0}
\right)\to1,
$$
where $d_k^2$ and $c_{K,n_0}$ are defined in the main text.
\end{assumption}

% -------------------------------------------------
\begin{assumption}[Local stability of the iterative update]
\label{assump:stability}
Let $\mathcal T_{n_0,r}(\Omega)$ denote the iterative
method-of-moments update, and let
$\Omega^*=\Psi/(rn_0)$.
For some $0<\delta<\lambda_{\min}(\Psi)$, where
$\lambda_{\min}(\Psi)$ denotes the smallest eigenvalue
of $\Psi$, define
$$
\mathcal N_n
=
\left\{
\Omega\succ0:
\norm{\Omega-\Omega^*}
\leq \frac{\delta}{rn_0}
\right\}.
$$

Assume that there exists a constant $\rho\in[0,1)$
such that, with probability tending to one,
$$
\norm{
\mathcal T_{n_0,r}(\Omega_1)
-
\mathcal T_{n_0,r}(\Omega_2)
}
\leq
\rho\norm{\Omega_1-\Omega_2},
\qquad
\Omega_1,\Omega_2\in\mathcal N_n.
$$

Furthermore, assume that the iterations
$$
\hat\Omega^{(t+1)}
=
\mathcal T_{n_0,r}(\hat\Omega^{(t)}),
\qquad t=0,1,\ldots,
$$
initialized at $\hat\Omega^{(0)}$ as specified
in the main text, are well-defined until entering
$\mathcal N_n$, and satisfy
$$
\Prb\left(
\hat\Omega^{(t)}\in\mathcal N_n
\text{ for some finite }t\geq0
\right)\to1.
$$
\end{assumption}

% -------------------------------------------------
\begin{assumption}[GMM regularity]
\label{assump:gmm}
\leavevmode
\begin{enumerate}[label=(\roman*),leftmargin=2.2em]

\item
Assume that $\beta_0$ lies in the interior of a compact
parameter space $\mathcal P_\beta$, and that
$\E(X^2)<\infty$.

\item
Assume that the covariance matrix of the moment functions,
$$
\Gamma
=
\operatorname{E}
\left[
\mathbf{g}(Y_1,Y_2,X;\beta_0,\theta^{(0)})
\mathbf{g}(Y_1,Y_2,X;\beta_0,\theta^{(0)})^\top
\right]
$$
is positive definite.

\item
Assume that there exists a nonnegative function
$M(Y_2,X)$ with $\E\{M(Y_2,X)\}<\infty$ such that
$$
\sup_{\theta}
\norm{
\nabla_\theta^2\mathbf g_2(Y_2,X;\theta)
}
\leq M(Y_2,X),
$$
where the Hessian is taken componentwise.

\end{enumerate}
\end{assumption}

% -------------------------------------------------
% Supplementary conditions for Theorem 4;

% =================================================
\begin{assumption}[Bootstrap regularity]
\label{assump:boot}

Let $N_i$ denote the number of times observation
$i$ is selected in the internal nonparametric
bootstrap, and write
$$
g_i=
\mathbf g(Y_{i1},Y_{i2},X_i;
\beta_0,\theta^{(0)}).
$$
As $n_0\to\infty$ and $K\to\infty$, assume
the following conditions.

\begin{enumerate}

\item
\textit{Internal influence-function regularity.}

The internal observations are independent and
identically distributed, with
$$
\E(g_i)=0,\qquad
\operatorname{Var}(g_i)=\Gamma.
$$

The internal contribution to the first-stage
meta estimator admits mean-zero influence
functions
$$
\ell_{i,n}
=
\ell_n(Y_{i1},Y_{i2},X_i).
$$

For some $\delta>0$,
$$
\sup_{n_0}
\E\left(
\|g_i\|^{2+\delta}
+
\|\ell_{i,n}\|^{2+\delta}
\right)<\infty.
$$

There exist estimated influence values
$\hat\ell_{i,n}$ satisfying
$$
\frac1{n_0}
\sum_{i=1}^{n_0}
\|\hat\ell_{i,n}-\ell_{i,n}\|^2
=o_p(1),
$$
and
$$
\frac1{n_0}
\sum_{i=1}^{n_0}
\|\hat\ell_{i,n}\|^{2+\delta}
=O_p(1).
$$

\item
\textit{First-stage asymptotic representations.}

The first-stage estimator and its hybrid
bootstrap counterpart admit the representations
$$
\sqrt{n_0}
(\tilde\theta_{RE}-\theta^{(0)})=
\frac1{\sqrt{n_0}}
\sum_{i=1}^{n_0}\ell_{i,n}
+
\sum_{k=1}^K\xi_{k,n}
+
r_n
$$
and
$$
\sqrt{n_0}
\left\{
\tilde\theta_{RE}^*
-
\E^*(\tilde\theta_{RE}^*)
\right\}=
\frac1{\sqrt{n_0}}
\sum_{i=1}^{n_0}
(N_i-1)\hat\ell_{i,n}
+
\sum_{k=1}^K\xi_{k,n}^*
+
r_n^*
$$
where
$$
\E\|r_n\|^2\to0,
\qquad
\E^*\|r_n^*\|^2\xrightarrow{p}0.
$$

The vectors $\xi_{k,n}$ are mean-zero,
independent across external studies, and
independent of the internal observations.
Conditionally on the observed data, the
vectors $\xi_{k,n}^*$ are mean-zero,
independent across external studies, and
independent of the internal bootstrap
resampling.

The aggregate external contributions satisfy
$$
\sup_{n_0}
\E\left\|
\sum_{k=1}^K\xi_{k,n}
\right\|^{2+\delta}<\infty,
$$
and
$$
\E^*
\left\|
\sum_{k=1}^K\xi_{k,n}^*
\right\|^{2+\delta}
=O_p(1).
$$

Furthermore,
$$
\sqrt{n_0}
\left\|
\E^*(\tilde\theta_{RE}^*)
-
\tilde\theta_{RE}
\right\|
=O_p(1).
$$

\item
\textit{External parametric bootstrap
covariance stability.}\\
Let
$$
v_n=
\begin{pmatrix}
\theta^{(0)}\\
\operatorname{vech}(\Omega_n)
\end{pmatrix},
\qquad
\hat v_n=
\begin{pmatrix}
\tilde\theta_{RE}\\
\operatorname{vech}(\hat\Omega)
\end{pmatrix}.
$$

The external first-order contributions
admit the representations
$$
\begin{aligned}
\xi_{k,n}
={}&
h_{k,n}(\hat\theta^{(k)};v_n)
-
\E\{
h_{k,n}(\hat\theta^{(k)};v_n)
\},\\
\xi_{k,n}^*
={}&
h_{k,n}(\hat\theta^{(k)*};\hat v_n)
-
\E^*\{
h_{k,n}(\hat\theta^{(k)*};\hat v_n)
\},
\end{aligned}
$$
where $h_{k,n}$ incorporates the first-order
effects of the borrowing weights and
between-study covariance estimation. Define
$$
\mathcal C_{k,n}(v)
=
\operatorname{Var}_{Z\sim
N\{\mu(v),\Lambda_k(v)\}}
\{h_{k,n}(Z;v)\},
$$
where
$$
\mu(v_n)=\theta^{(0)},
\qquad
\Lambda_k(v_n)=\Omega_n+S_k,
$$
and
$$
\mu(\hat v_n)=\tilde\theta_{RE},
\qquad
\Lambda_k(\hat v_n)=\hat\Omega+S_k.
$$

In a neighborhood containing $v_n$ and
$\hat v_n$ with probability tending to one,
assume that
$$
\|\mathcal C_{k,n}(v)
-\mathcal C_{k,n}(v_n)\|
\leq L_{k,n}\|v-v_n\|,
$$
where $L_{k,n}$ is deterministic and
$$
\|\hat v_n-v_n\|
\sum_{k=1}^K L_{k,n}
=o_p(1).
$$

\item
\textit{Second-stage GMM and moment regularity.}
Conditionally on the observed data,
$$
n_0
\left\|
\hat\beta_{\mathrm{GMM}}^*
-
\hat\beta_{\mathrm{GMM}}
\right\|^2
$$
is asymptotically uniformly integrable
in probability.

\end{enumerate}
\end{assumption}

% =================================================
% =================================================
\newpage
\subsection{Proof of Theorem~1}
\label{sec:proof_thm1}
% =================================================

\noindent\textbf{Restatement of Theorem~1.} Under the regularity conditions stated in \ref{sec:additional_assumptions},
as $n_0\to\infty$, $K\to\infty$, and $K(rn_0)^{-1/2}\to0$,
$$
\norm{\hat\Psi-\Psi}
=
O_p\!\left(K^{-1/2}\right),
\qquad
\norm{\tilde\theta_{RE}-\theta^{(0)}}
=
O_p\!\left((Krn_0)^{-1/2}\right),
\qquad
\norm{\tilde V_{RE}}
=
O_p\!\left((Krn_0)^{-1}\right).
$$

\noindent\textbf{Proof.}
Let $n=rn_0$ and $\Omega^*=\Psi/n$, as defined in
the main text. By Assumption~\ref{assump:weights},
$$
\Prb\left(
\hat\alpha_1=\cdots=\hat\alpha_K=1
\right)\to1.
$$
We establish the required rates on this event, which
suffices because its probability tends to one. From Assumption~\ref{assump:normality},
$$
\hat\theta^{(k)}-\theta^{(k)}
=
n_k^{-1/2}Z_k+n_k^{-1}R_k,
\qquad k=0,\ldots,K.
$$
Under the local-heterogeneity model,
$$
\hat\theta^{(k)}-\theta^{(0)}
=
\gamma^{(k)}
+n_k^{-1/2}Z_k+n_k^{-1}R_k,
\qquad k=1,\ldots,K,
$$
where
$\gamma^{(k)}\stackrel{\mathrm{i.i.d.}}{\sim}N(0,\Psi/n)$
and $\gamma^{(k)}$ is independent of $Z_k$ for each
$k=1,\ldots,K$. The uniform moment bounds and $n_k\asymp n$ imply
$$
\E\norm{
\hat\theta^{(k)}-\theta^{(0)}
}^2
=
O(n^{-1}),
\qquad k=1,\ldots,K.
$$
Similarly,
$$
\norm{\hat\theta^{(0)}-\theta^{(0)}}
=
O_p(n_0^{-1/2}).
$$

\medskip
\noindent
\textbf{Step 1: Posterior quantities evaluated at $\Omega^*$.}

Write $\tilde\theta_{RE}(\Omega)$ and
$\tilde V_{RE}(\Omega)$ for the posterior mean and
covariance evaluated at a given between-study
covariance $\Omega$. By Assumption~\ref{assump:normality} and the uniform
comparability of the study sample sizes,
$$
\max_{1\leq k\leq K}
\norm{
nS_k-\frac{n}{n_k}V_k
}
=
O_p(n^{-1/2}).
$$
Since $\Psi$ is positive definite and the eigenvalues
of $V_k$ are uniformly bounded away from zero and
infinity, the eigenvalues of $nS_k+\Psi$ are uniformly
bounded away from zero and infinity with probability
tending to one.

Since $S_0$ is positive definite with probability
tending to one and the eigenvalues of $nS_k+\Psi$
are uniformly bounded above,
there exists a constant $C<\infty$ such that,
with probability tending to one,
$$
\frac{1}{n}S_0^{-1}
+
\sum_{k=1}^K(nS_k+\Psi)^{-1}
\succeq
\frac{K}{C}I.
$$
Therefore,
$$
\norm{
\left[
\frac{1}{n}S_0^{-1}
+
\sum_{k=1}^K(nS_k+\Psi)^{-1}
\right]^{-1}
}
\leq \frac{C}{K}
=O(K^{-1}).
$$

The posterior covariance evaluated at $\Omega^*$ is
$$
\tilde V_{RE}(\Omega^*)
=
\frac{1}{n}
\left[
\frac{1}{n}S_0^{-1}
+
\sum_{k=1}^K(nS_k+\Psi)^{-1}
\right]^{-1}.
$$
Therefore,
$$
\boxed{
\norm{\tilde V_{RE}(\Omega^*)}
=
O_p((Kn)^{-1}).
}
$$

For the posterior mean,
$$
\begin{aligned}
&\tilde\theta_{RE}(\Omega^*)-\theta^{(0)}
\\
&=
\left[
\frac{1}{n}S_0^{-1}
+
\sum_{k=1}^K(nS_k+\Psi)^{-1}
\right]^{-1}
\\
&\quad\times
\left[
\frac{1}{n}S_0^{-1}
\{\hat\theta^{(0)}-\theta^{(0)}\}
+
\sum_{k=1}^K
(nS_k+\Psi)^{-1}
\{\hat\theta^{(k)}-\theta^{(0)}\}
\right].
\end{aligned}
$$

We first control the external-study contribution.
By Assumption~\ref{assump:normality} and
$n_k\asymp n$ uniformly, since the eigenvalues of $nS_k+\Psi$ and
$(n/n_k)V_k+\Psi$ are uniformly bounded away
from zero with probability tending to one,
the matrix inverse identity yields
$$
\max_{1\leq k\leq K}
\norm{
(nS_k+\Psi)^{-1}
-
\left(\frac{n}{n_k}V_k+\Psi\right)^{-1}
}
=
O_p(n^{-1/2}).
$$
Using the within-study expansion, we have
$$
\begin{aligned}
&\sum_{k=1}^K
(nS_k+\Psi)^{-1}
\{\hat\theta^{(k)}-\theta^{(0)}\}
\\
&=
\sum_{k=1}^K
\left(\frac{n}{n_k}V_k+\Psi\right)^{-1}
\left\{
\gamma^{(k)}+n_k^{-1/2}Z_k
\right\}
\\
&\quad+
\sum_{k=1}^K
\left(\frac{n}{n_k}V_k+\Psi\right)^{-1}
n_k^{-1}R_k
\\
&\quad+
\sum_{k=1}^K
\left[
(nS_k+\Psi)^{-1}
-
\left(\frac{n}{n_k}V_k+\Psi\right)^{-1}
\right]
\{\hat\theta^{(k)}-\theta^{(0)}\}.
\end{aligned}
$$

The first sum consists of independent, mean-zero
random vectors with uniformly bounded covariance
matrices of order $n^{-1}$. Hence,
$$
\norm{
\sum_{k=1}^K
\left(\frac{n}{n_k}V_k+\Psi\right)^{-1}
\left\{
\gamma^{(k)}+n_k^{-1/2}Z_k
\right\}
}
=
O_p\left(\sqrt{\frac{K}{n}}\right).
$$

By the uniform second-moment bound on $R_k$,
$$
\norm{
\sum_{k=1}^K
\left(\frac{n}{n_k}V_k+\Psi\right)^{-1}
n_k^{-1}R_k
}
=
O_p\left(\frac{K}{n}\right).
$$

For the final sum, the Cauchy--Schwarz inequality gives
$$
\begin{aligned}
&\norm{
\sum_{k=1}^K
\left[
(nS_k+\Psi)^{-1}
-
\left(\frac{n}{n_k}V_k+\Psi\right)^{-1}
\right]
\{\hat\theta^{(k)}-\theta^{(0)}\}
}
\\
&\leq
\left[
\sum_{k=1}^K
\norm{
(nS_k+\Psi)^{-1}
-
\left(\frac{n}{n_k}V_k+\Psi\right)^{-1}
}^2
\right]^{1/2}\times
\left[
\sum_{k=1}^K
\norm{
\hat\theta^{(k)}-\theta^{(0)}
}^2
\right]^{1/2}
\\
&=
O_p\left(\frac{K}{n}\right).
\end{aligned}
$$

Since $Kn^{-1/2}\to0$, we have $K/n\to0$, and thus
$$
\frac{K}{n}
=
o\left(\sqrt{\frac{K}{n}}\right).
$$
It follows that
$$
\boxed{
\norm{
\sum_{k=1}^K
(nS_k+\Psi)^{-1}
\{\hat\theta^{(k)}-\theta^{(0)}\}
}
=
O_p\left(\sqrt{\frac{K}{n}}\right).
}
$$

Moreover, Assumption~\ref{assump:normality} gives
$$
\norm{
\frac{1}{n}S_0^{-1}
\{\hat\theta^{(0)}-\theta^{(0)}\}
}
=
O_p(n^{-1/2}).
$$
Combining these bounds with the posterior-mean
representation yields
$$
\boxed{
\norm{
\tilde\theta_{RE}(\Omega^*)-\theta^{(0)}
}
=
O_p((Kn)^{-1/2}).
}
$$

\medskip
\noindent
\textbf{Step 2: One-step iterative update at $\Omega^*$.}

On the event that all borrowing weights equal one,
the iterative method-of-moments update gives
$$
\begin{aligned}
\mathcal T_{n_0,r}(\Omega^*)
={}&
\frac{1}{K}
\sum_{k=1}^K
\left\{
\hat\theta^{(k)}
-
\tilde\theta_{RE}(\Omega^*)
\right\}
\left\{
\hat\theta^{(k)}
-
\tilde\theta_{RE}(\Omega^*)
\right\}^{\top}
\\
&-
\frac{1}{K}\sum_{k=1}^K S_k
+
\tilde V_{RE}(\Omega^*).
\end{aligned}
$$

We first establish the rate of
$$
\frac{n}{K}
\sum_{k=1}^K
\left[
\{\hat\theta^{(k)}-\theta^{(0)}\}
\{\hat\theta^{(k)}-\theta^{(0)}\}^{\top}
-
\left(\frac{\Psi}{n}+S_k\right)
\right].
$$

Using
$$
\hat\theta^{(k)}-\theta^{(0)}
=
\gamma^{(k)}
+n_k^{-1/2}Z_k+n_k^{-1}R_k,
$$
we obtain
$$
\begin{aligned}
&\frac{n}{K}
\sum_{k=1}^K
\left[
\{\hat\theta^{(k)}-\theta^{(0)}\}
\{\hat\theta^{(k)}-\theta^{(0)}\}^{\top}
-
\left(\frac{\Psi}{n}+S_k\right)
\right]
\\
&=
\frac{1}{K}
\sum_{k=1}^K
\left\{
n\gamma^{(k)}\gamma^{(k)\top}-\Psi
\right\}
\\
&\quad+
\frac{n}{K}
\sum_{k=1}^K
\left\{
\gamma^{(k)}n_k^{-1/2}Z_k^\top
+
n_k^{-1/2}Z_k\gamma^{(k)\top}
\right\}
\\
&\quad+
\frac{n}{K}
\sum_{k=1}^K
\left\{
n_k^{-1}Z_kZ_k^\top-\frac{V_k}{n_k}
\right\}
\\
&\quad+
\frac{n}{K}
\sum_{k=1}^K
\left\{
\frac{V_k}{n_k}-S_k
\right\}
+
O_p(n^{-1/2}).
\end{aligned}
$$

The final remainder collects all terms involving
$R_k$. By the Cauchy--Schwarz inequality, the
uniform second-moment bound on $R_k$, and
$n_k\asymp n$, their expected norm after multiplication
by $n$ is of order $n^{-1/2}$. Therefore, their
average is $O_p(n^{-1/2})$.

Since $\sqrt n\,\gamma^{(k)}\sim N(0,\Psi)$,
the first sum consists of independent, centered
random matrices with uniformly bounded second
moments. Consequently,
$$
\norm{
\frac{1}{K}
\sum_{k=1}^K
\left\{
n\gamma^{(k)}\gamma^{(k)\top}-\Psi
\right\}
}
=
O_p(K^{-1/2}).
$$

Similarly, independence of $\gamma^{(k)}$ and $Z_k$
and their uniform second-moment bounds imply
$$
\norm{
\frac{n}{K}
\sum_{k=1}^K
\left\{
\gamma^{(k)}n_k^{-1/2}Z_k^\top
+
n_k^{-1/2}Z_k\gamma^{(k)\top}
\right\}
}
=
O_p(K^{-1/2}).
$$

The uniform fourth-moment bound on $Z_k$ gives
$$
\norm{
\frac{n}{K}
\sum_{k=1}^K
\left\{
n_k^{-1}Z_kZ_k^\top-\frac{V_k}{n_k}
\right\}
}
=
O_p(K^{-1/2}).
$$

Finally, the uniform covariance-estimation condition
in Assumption~\ref{assump:normality} yields
$$
\norm{
\frac{n}{K}
\sum_{k=1}^K
\left\{
\frac{V_k}{n_k}-S_k
\right\}
}
=
O_p(n^{-1/2}).
$$

Because $K/n\to0$,
$$
n^{-1/2}=o(K^{-1/2}).
$$
Combining the preceding bounds,
$$
\boxed{
\norm{
\frac{n}{K}
\sum_{k=1}^K
\left[
\{\hat\theta^{(k)}-\theta^{(0)}\}
\{\hat\theta^{(k)}-\theta^{(0)}\}^{\top}
-
\left(\frac{\Psi}{n}+S_k\right)
\right]
}
=
O_p(K^{-1/2}).
}
$$

Next, independence across external studies and the
within-study representation imply
$$
\begin{aligned}
\norm{
\frac{1}{K}
\sum_{k=1}^K
\{\hat\theta^{(k)}-\theta^{(0)}\}
}
&=
O_p((Kn)^{-1/2})
+
O_p(n^{-1})
\\
&=
O_p((Kn)^{-1/2}),
\end{aligned}
$$
where the last equality follows from $K/n\to0$.

Together with Step~1,
$$
\norm{
\tilde\theta_{RE}(\Omega^*)-\theta^{(0)}
}
=
O_p((Kn)^{-1/2}),
$$
this gives
$$
\begin{aligned}
&n\norm{
\frac{1}{K}
\sum_{k=1}^K
\{\hat\theta^{(k)}-\theta^{(0)}\}
\{\tilde\theta_{RE}(\Omega^*)-\theta^{(0)}\}^{\top}
}
\\
&\qquad=
O_p(K^{-1}).
\end{aligned}
$$

The corresponding transposed term has the same
order. Furthermore,
$$
n\norm{
\tilde\theta_{RE}(\Omega^*)-\theta^{(0)}
}^2
=
O_p(K^{-1}),
$$
and Step~1 gives
$$
n\norm{\tilde V_{RE}(\Omega^*)}
=
O_p(K^{-1}).
$$

Expanding the centered quadratic terms in the
method-of-moments update and combining these bounds,
we obtain
$$
\boxed{
\norm{
n\mathcal T_{n_0,r}(\Omega^*)-\Psi
}
=
O_p(K^{-1/2}).
}
$$

Equivalently,
$$
\boxed{
\norm{
\mathcal T_{n_0,r}(\Omega^*)-\Omega^*
}
=
O_p\left(\frac{1}{n\sqrt K}\right).
}
$$

\medskip
\noindent
\textbf{Step 3: Convergence rate of the local fixed point.}

Recall from Assumption~\ref{assump:stability} that
$$
\mathcal N_n
=
\left\{
\Omega\succ0:
\norm{\Omega-\Omega^*}
\leq\frac{\delta}{n}
\right\},
$$
where $0<\delta<\lambda_{\min}(\Psi)$.

For any $\Omega\in\mathcal N_n$, the contraction
condition gives
$$
\begin{aligned}
\norm{
\mathcal T_{n_0,r}(\Omega)-\Omega^*
}
&\leq
\norm{
\mathcal T_{n_0,r}(\Omega)
-
\mathcal T_{n_0,r}(\Omega^*)
}+
\norm{
\mathcal T_{n_0,r}(\Omega^*)-\Omega^*
}
\\
&\leq
\rho\frac{\delta}{n}
+
O_p\left(\frac{1}{n\sqrt K}\right).
\end{aligned}
$$

Since $K\to\infty$ and $\rho<1$,
$$
\sup_{\Omega\in\mathcal N_n}
\norm{
\mathcal T_{n_0,r}(\Omega)-\Omega^*
}
<
\frac{\delta}{n}
$$
with probability tending to one.

Thus, the update maps $\mathcal N_n$ into itself.
Since $\mathcal N_n$ is a closed neighborhood
contained in the positive-definite cone, the
contraction-mapping theorem gives a unique fixed
point $\hat\Omega\in\mathcal N_n$.
Furthermore, Assumption~\ref{assump:stability}
ensures that the iterations starting from the
initialization specified in the main text enter
$\mathcal N_n$ with probability tending to one.
Consequently, the iterative algorithm converges
to this fixed point with probability tending to one.

At this fixed point,
$$
\hat\Omega=\mathcal T_{n_0,r}(\hat\Omega).
$$
Therefore,
$$
\begin{aligned}
\norm{\hat\Omega-\Omega^*}
&\leq
\norm{
\mathcal T_{n_0,r}(\hat\Omega)
-
\mathcal T_{n_0,r}(\Omega^*)
}
+
\norm{
\mathcal T_{n_0,r}(\Omega^*)-\Omega^*
}
\\
&\leq
\rho\norm{\hat\Omega-\Omega^*}
+
\norm{
\mathcal T_{n_0,r}(\Omega^*)-\Omega^*
}.
\end{aligned}
$$

Hence,
$$
(1-\rho)\norm{\hat\Omega-\Omega^*}
\leq
\norm{
\mathcal T_{n_0,r}(\Omega^*)-\Omega^*
}.
$$

Since $\hat\Psi=n\hat\Omega$ and $\Psi=n\Omega^*$,
Step~2 yields
$$
\boxed{
\norm{\hat\Psi-\Psi}
=
O_p(K^{-1/2}).
}
$$

\medskip
\noindent
\textbf{Step 4: Posterior quantities evaluated at $\hat\Omega$.}

From Step~3,
$$
\norm{\hat\Psi-\Psi}
=
O_p(K^{-1/2}).
$$

By the matrix inverse identity and the uniform
eigenvalue bounds established in Step~1,
$$
\max_{1\leq k\leq K}
\norm{
(nS_k+\hat\Psi)^{-1}
-
(nS_k+\Psi)^{-1}
}
=
O_p(K^{-1/2}).
$$

Consequently,
$$
\norm{
\left[
\frac{1}{n}S_0^{-1}
+
\sum_{k=1}^K(nS_k+\hat\Psi)^{-1}
\right]^{-1}
}
=
O_p(K^{-1}).
$$

The estimated posterior covariance satisfies
$$
\tilde V_{RE}
=
\frac{1}{n}
\left[
\frac{1}{n}S_0^{-1}
+
\sum_{k=1}^K(nS_k+\hat\Psi)^{-1}
\right]^{-1}.
$$

Therefore,
$$
\boxed{
\norm{\tilde V_{RE}}
=
O_p((Kn)^{-1}).
}
$$

For the posterior mean,
$$
\begin{aligned}
&\tilde\theta_{RE}-\theta^{(0)}
\\
&=
\left[
\frac{1}{n}S_0^{-1}
+
\sum_{k=1}^K(nS_k+\hat\Psi)^{-1}
\right]^{-1}
\\
&\quad\times
\left[
\frac{1}{n}S_0^{-1}
\{\hat\theta^{(0)}-\theta^{(0)}\}
+
\sum_{k=1}^K
(nS_k+\hat\Psi)^{-1}
\{\hat\theta^{(k)}-\theta^{(0)}\}
\right].
\end{aligned}
$$

Step~1 established that
$$
\norm{
\sum_{k=1}^K
(nS_k+\Psi)^{-1}
\{\hat\theta^{(k)}-\theta^{(0)}\}
}
=
O_p\left(\sqrt{\frac{K}{n}}\right).
$$

Moreover, the Cauchy--Schwarz inequality gives
$$
\begin{aligned}
&\norm{
\sum_{k=1}^K
\left[
(nS_k+\hat\Psi)^{-1}
-
(nS_k+\Psi)^{-1}
\right]
\{\hat\theta^{(k)}-\theta^{(0)}\}
}
\\
&\leq
\left[
\sum_{k=1}^K
\norm{
(nS_k+\hat\Psi)^{-1}
-
(nS_k+\Psi)^{-1}
}^2
\right]^{1/2}
\left[
\sum_{k=1}^K
\norm{
\hat\theta^{(k)}-\theta^{(0)}
}^2
\right]^{1/2}
\\
&=
O_p\left(\sqrt{\frac{K}{n}}\right).
\end{aligned}
$$

Hence,
$$
\boxed{
\norm{
\sum_{k=1}^K
(nS_k+\hat\Psi)^{-1}
\{\hat\theta^{(k)}-\theta^{(0)}\}
}
=
O_p\left(\sqrt{\frac{K}{n}}\right).
}
$$

The internal-study contribution satisfies
$$
\norm{
\frac{1}{n}S_0^{-1}
\{\hat\theta^{(0)}-\theta^{(0)}\}
}
=
O_p(n^{-1/2}).
$$

Combining these results with the posterior-mean
representation yields
$$
\boxed{
\norm{\tilde\theta_{RE}-\theta^{(0)}}
=
O_p((Kn)^{-1/2}).
}
$$

Finally, recalling that $n=rn_0$, we obtain
$$
\norm{\hat\Psi-\Psi}
=
O_p(K^{-1/2}),
\qquad
\norm{\tilde\theta_{RE}-\theta^{(0)}}
=
O_p\left((Krn_0)^{-1/2}\right),
\qquad
\norm{\tilde V_{RE}}
=
O_p\left((Krn_0)^{-1}\right).
$$
This completes the proof. \hfill$\square$

% =================================================
% =================================================
\newpage

\newpage
\subsection{Proof of Theorem~2}
\label{sec:proof_thm2}
% =================================================

\noindent\textbf{Restatement of Theorem~2.}
Under the regularity conditions stated in the Supplementary Material and  Theorem~1, as $n_0\to\infty$:
\begin{enumerate}
\item[(a)] \textbf{Consistency}
$$
\hat{\beta}_{\mathrm{GMM}}\xrightarrow{P}\beta_0.
$$

\item[(b)] \textbf{Asymptotic normality}
$$
\sqrt{n_0}\,(\hat{\beta}_{\mathrm{GMM}}-\beta_0)
\xrightarrow{d}
N\!\left(
0,
\left(D_\beta^{\top}\Gamma^{-1}D_\beta\right)^{-1}
\right),
$$
where
$$
D_\beta
=
\left.
\frac{\partial}{\partial\beta}
\operatorname{E}
\{\mathbf{g}(Y_1,Y_2,X;\beta,\theta^{(0)})\}
\right|_{\beta=\beta_0}
$$
and
$$
\Gamma
=
\operatorname{E}
\left[
\mathbf{g}(Y_1,Y_2,X;\beta_0,\theta^{(0)})
\mathbf{g}(Y_1,Y_2,X;\beta_0,\theta^{(0)})^\top
\right].
$$
\end{enumerate}

\noindent\textbf{Proof.}
Let
$$
\bar{\mathbf g}_n(\beta,\theta)
=
\frac{1}{n_0}
\sum_{i=1}^{n_0}
\mathbf g(O_i;\beta,\theta),
\qquad
O_i=(Y_{i1},Y_{i2},X_i),
$$
and
$$
m(\beta,\theta)
=
\E\{\mathbf g(O;\beta,\theta)\}.
$$

For notational convenience, let
$\bar{\mathbf g}_n^\dagger(\beta)$ denote the sample
moment used by the proposed estimator:
$$
\bar{\mathbf g}_n^\dagger(\beta)
=
\begin{cases}
\bar{\mathbf g}_n(\beta,\tilde\theta_{RE}),
& \text{for the plug-in formulation},\\[1ex]
\bar{\mathbf g}_n^A
(\beta,\tilde\theta_{RE},\tilde V_{RE}),
& \text{for the uncertainty-propagating formulation}.
\end{cases}
$$

The estimator $\hat\beta_{\mathrm{GMM}}$ minimizes
$$
Q_n(\beta)
=
\bar{\mathbf g}_n^\dagger(\beta)^\top
\hat W_{\mathrm{MLE}}^{-1}
\bar{\mathbf g}_n^\dagger(\beta).
$$

\medskip
\noindent
\textbf{Step 1: Consistency.}

By Theorem~1,
$$
\tilde\theta_{RE}\toP\theta^{(0)},
\qquad
\norm{\tilde V_{RE}}\toP0.
$$

For the plug-in formulation, the smoothness of
the polytomous logistic score functions and
$\E(X^2)<\infty$ imply a uniform law of large
numbers. Together with the consistency of
$\tilde\theta_{RE}$, this gives
$$
\sup_{\beta\in\mathcal P_\beta}
\norm{
\bar{\mathbf g}_n(\beta,\tilde\theta_{RE})
-
m(\beta,\theta^{(0)})
}
=
o_p(1).
$$

For the uncertainty-propagating formulation, let
$U\mid\text{data}\sim N(0,\tilde V_{RE})$.
A second-order Taylor expansion around
$\tilde\theta_{RE}$ gives
$$
\begin{aligned}
\mathbf g_2(Y_{i2},X_i;\tilde\theta_{RE}+U)&=
\mathbf g_2(Y_{i2},X_i;\tilde\theta_{RE})
+
\nabla_\theta\mathbf g_2
(Y_{i2},X_i;\tilde\theta_{RE})U
\\
&\quad+
\int_0^1(1-t)
\nabla_\theta^2\mathbf g_2
(Y_{i2},X_i;\tilde\theta_{RE}+tU)
[U,U]\,dt.
\end{aligned}
$$
Since the posterior distribution is centered at
$\tilde\theta_{RE}$, the conditional expectation
of the first-order term is zero.
By Assumption~\ref{assump:gmm},
$$
\begin{aligned}
&\norm{
\E_{\mathrm{post}}
\left\{
\mathbf g_2(Y_{i2},X_i;\tilde\theta_{RE}+U)
\right\}
-
\mathbf g_2(Y_{i2},X_i;\tilde\theta_{RE})
}
\\
&\qquad\leq
\frac12 M(Y_{i2},X_i)
\E_{\mathrm{post}}(\norm{U}^2)
\\
&\qquad=
\frac12 M(Y_{i2},X_i)
\operatorname{tr}(\tilde V_{RE}).
\end{aligned}
$$
Consequently,
$$
\begin{aligned}
\sup_{\beta\in\mathcal P_\beta}
\norm{
\bar{\mathbf g}_n^A
(\beta,\tilde\theta_{RE},\tilde V_{RE})
-
\bar{\mathbf g}_n(\beta,\tilde\theta_{RE})
}
&\leq
\frac12\operatorname{tr}(\tilde V_{RE})
\frac1{n_0}\sum_{i=1}^{n_0}M(Y_{i2},X_i)
\\
&=
O_p(\norm{\tilde V_{RE}})
=
o_p(1),
\end{aligned}
$$
where the last equality follows from the law of
large numbers, Assumption~\ref{assump:gmm},
and Theorem~1. When the posterior expectation is approximated
using a fixed number $B$ of independent Monte
Carlo draws, let $\bar{\mathbf g}_{n,B}^A$
denote the resulting sample moment.
By the smoothness of $\mathbf g_2$ and
Assumption~\ref{assump:gmm},
$$
\sup_{\beta\in\mathcal P_\beta}
\norm{
\bar{\mathbf g}_{n,B}^A
(\beta,\tilde\theta_{RE},\tilde V_{RE})
-
\bar{\mathbf g}_n^A
(\beta,\tilde\theta_{RE},\tilde V_{RE})
}
=
O_p(\norm{\tilde V_{RE}}^{1/2}).
$$
By Theorem~1,
$$
\sqrt{n_0}\norm{\tilde V_{RE}}^{1/2}
=
O_p((Kr)^{-1/2})
=
o_p(1).
$$
Therefore, the Monte Carlo approximation is
asymptotically negligible at the
$n_0^{1/2}$ scale and does not affect
the consistency or asymptotic distribution
of the proposed estimator.

Hence, for either formulation,
$$
\sup_{\beta\in\mathcal P_\beta}
\norm{
\bar{\mathbf g}_n^\dagger(\beta)
-
m(\beta,\theta^{(0)})
}
=
o_p(1).
$$

Under the correctly specified model, the complete target-outcome score has
expectation zero at $\beta_0$. The positive definiteness of $\Gamma$ implies positive definiteness of the covariance matrix
of the target-outcome score. By the information
identity, the expected target-outcome log-likelihood
is strictly concave and its score has a unique
zero at $\beta_0$. Consequently,
$$
Q(\beta)
=
m(\beta,\theta^{(0)})^\top
\Gamma^{-1}
m(\beta,\theta^{(0)})
$$
is uniquely minimized at $\beta_0$. Furthermore, consistency of the internal-study maximum likelihood estimator, together with
the uniform law of large numbers, gives
$$
\hat W_{\mathrm{MLE}}\toP\Gamma.
$$

Therefore,
$$
\sup_{\beta\in\mathcal P_\beta}
|Q_n(\beta)-Q(\beta)|
=
o_p(1).
$$

Since $\mathcal P_\beta$ is compact and
$\beta_0$ is the unique minimizer of $Q$,
the argmin theorem yields
$$
\boxed{
\hat\beta_{\mathrm{GMM}}\toP\beta_0.
}
$$

\medskip
\noindent
\textbf{Step 2: Asymptotic linear representation.}

By Theorem~1,
$$
\norm{
\tilde\theta_{RE}-\theta^{(0)}
}
=
O_p((Krn_0)^{-1/2}),
$$
and therefore
$$
\sqrt{n_0}
\norm{
\tilde\theta_{RE}-\theta^{(0)}
}
=
O_p((Kr)^{-1/2})
=
o_p(1).
$$

Similarly,
$$
\norm{\tilde V_{RE}}
=
O_p((Krn_0)^{-1}),
$$
so
$$
\sqrt{n_0}
\norm{\tilde V_{RE}}
=
O_p\left(
\frac{1}{Kr\sqrt{n_0}}
\right)
=
o_p(1).
$$

By differentiability of the logistic score
functions and the uniform law of large numbers,
$$
\norm{
\bar{\mathbf g}_n
(\beta_0,\tilde\theta_{RE})
-
\bar{\mathbf g}_n
(\beta_0,\theta^{(0)})
}
=
O_p\left(
\norm{
\tilde\theta_{RE}-\theta^{(0)}
}
\right)
=
o_p(n_0^{-1/2}).
$$

For the uncertainty-propagating formulation,
the Taylor expansion established in Step~1
gives an additional term of order
$O_p(\norm{\tilde V_{RE}})$, which is also
$o_p(n_0^{-1/2})$. Thus, for either formulation,
$$
\boxed{
\bar{\mathbf g}_n^\dagger(\beta_0)
=
\bar{\mathbf g}_n
(\beta_0,\theta^{(0)})
+
o_p(n_0^{-1/2}).
}
$$

By the smoothness of the moment functions,
the uniform law of large numbers for their
derivatives, and the consistency established
in Step~1,
$$
\sup_{0\leq t\leq1}
\norm{
\left.
\frac{\partial}
{\partial\beta^\top}
\bar{\mathbf g}_n^\dagger(\beta)
\right|_{
\beta=\beta_0+
t(\hat\beta_{\mathrm{GMM}}-\beta_0)
}
-
D_\beta
}
=
o_p(1)
$$

Consequently, the integral form of the
mean-value expansion gives
$$
\bar{\mathbf g}_n^\dagger
(\hat\beta_{\mathrm{GMM}})
=
\bar{\mathbf g}_n^\dagger(\beta_0)
+
D_\beta
(\hat\beta_{\mathrm{GMM}}-\beta_0)
+
o_p\left(
\norm{\hat\beta_{\mathrm{GMM}}-\beta_0}
\right)
$$

Since $\beta_0$ lies in the interior of
$\mathcal P_\beta$ and
$\hat\beta_{\mathrm{GMM}}\toP\beta_0$,
the GMM first-order condition holds
with probability tending to one:
$$
0
=
\left[
\left.
\frac{\partial}
{\partial\beta^\top}
\bar{\mathbf g}_n^\dagger(\beta)
\right|_{\beta=\hat\beta_{\mathrm{GMM}}}
\right]^\top
\hat W_{\mathrm{MLE}}^{-1}
\bar{\mathbf g}_n^\dagger
(\hat\beta_{\mathrm{GMM}}).
$$

Combining the preceding expansions with
$\hat W_{\mathrm{MLE}}\toP\Gamma$ gives
$$
0
={}
D_\beta^\top\Gamma^{-1}
\left[
\bar{\mathbf g}_n
(\beta_0,\theta^{(0)})
+
D_\beta
(\hat\beta_{\mathrm{GMM}}-\beta_0)
\right]
+
o_p(n_0^{-1/2})
+
o_p\left(
\norm{\hat\beta_{\mathrm{GMM}}-\beta_0}
\right).
$$

Let
$
J=D_\beta^\top\Gamma^{-1}D_\beta.
$
Under the correctly specified polytomous logistic
model, the information identity gives
$$
\left.\frac{\partial}{\partial\beta^\top}
\E\{\mathbf g_1(Y_1,X;\beta)\}
\right|_{\beta=\beta_0}
=
-\E\left[
\mathbf g_1(Y_1,X;\beta_0)
\mathbf g_1(Y_1,X;\beta_0)^\top
\right]
$$
Since $\Gamma$ is positive definite, the
covariance matrix of the target-outcome score
is positive definite. Thus, $D_\beta$ has
full column rank, and $J$ is positive definite.

Moreover, the multivariate central limit
theorem gives
$$
\bar{\mathbf g}_n
(\beta_0,\theta^{(0)})
=
O_p(n_0^{-1/2}).
$$

Therefore,
$$
J(\hat\beta_{\mathrm{GMM}}-\beta_0)
={}
O_p(n_0^{-1/2})
+
o_p\left(
\norm{\hat\beta_{\mathrm{GMM}}-\beta_0}
\right).
$$

Since $J$ is nonsingular,
$$
\norm{\hat\beta_{\mathrm{GMM}}-\beta_0}
=
O_p(n_0^{-1/2}).
$$

Substituting this rate into the
first-order expansion yields
$$
\hat\beta_{\mathrm{GMM}}-\beta_0
={}
-
J^{-1}D_\beta^\top\Gamma^{-1}
\bar{\mathbf g}_n
(\beta_0,\theta^{(0)})+
o_p(n_0^{-1/2})
$$

Equivalently,
$$
\boxed{
\sqrt{n_0}
(\hat\beta_{\mathrm{GMM}}-\beta_0)=
-
J^{-1}D_\beta^\top\Gamma^{-1}
\sqrt{n_0}
\bar{\mathbf g}_n
(\beta_0,\theta^{(0)})
+
o_p(1).
}
$$

\medskip
\noindent
\textbf{Step 3: Asymptotic normality.}

Since the internal observations are independent
and identically distributed and
$\E(X^2)<\infty$, the multivariate central
limit theorem gives
$$
\sqrt{n_0}
\bar{\mathbf g}_n
(\beta_0,\theta^{(0)})
\toD
N(0,\Gamma).
$$

By Step~2 and Slutsky's theorem,
$$
\sqrt{n_0}
(\hat\beta_{\mathrm{GMM}}-\beta_0)
\toD
N\left(
0,
J^{-1}
D_\beta^\top\Gamma^{-1}
\Gamma
\Gamma^{-1}D_\beta
J^{-1}
\right)
$$

Since
$$
J=D_\beta^\top\Gamma^{-1}D_\beta,
$$
the asymptotic covariance simplifies to
$$
J^{-1}
=
\left(
D_\beta^\top\Gamma^{-1}D_\beta
\right)^{-1}.
$$

Therefore,
$$
\boxed{
\sqrt{n_0}
(\hat\beta_{\mathrm{GMM}}-\beta_0)
\toD
N\left(
0,
\left(
D_\beta^\top\Gamma^{-1}D_\beta
\right)^{-1}
\right).
}
$$

This completes the proof.
\hfill$\square$
% =================================================
\newpage

\newpage
\subsection{Proof of Corollary~1}
\label{sec:proof_cor1}
% =================================================

\noindent\textbf{Restatement of Corollary~1.}
Under Theorem~2, the asymptotic covariance of
the proposed GMM estimator satisfies
$$
\Sigma_{\mathrm{GMM}}
=
\left\{
\Sigma_{\mathrm{MLE}}^{-1}
+
\Gamma_{21}^{\top}
\Gamma_{22\cdot1}^{-1}
\Gamma_{21}
\right\}^{-1}
\preceq
\Sigma_{\mathrm{MLE}}.
$$
Thus, the proposed estimator is asymptotically at least as efficient
as the internal-study-only maximum likelihood estimator. If
$\Gamma_{21}\neq0$, the efficiency gain is strict in at least one direction.
If $\Gamma_{21}$ has full column rank, then
$$
\Sigma_{\mathrm{GMM}}
\prec
\Sigma_{\mathrm{MLE}},
$$
so the efficiency gain is strict in every nonzero direction.

\noindent\textbf{Proof.}
Partition the moment function and its covariance
conformably as
$$
\mathbf g
=
\begin{pmatrix}
\mathbf g_1\\
\mathbf g_2
\end{pmatrix},
\qquad
\Gamma
=
\begin{pmatrix}
\Gamma_{11} & \Gamma_{12}\\
\Gamma_{21} & \Gamma_{22}
\end{pmatrix}.
$$
Since $\mathbf g_2(Y_2,X;\theta^{(0)})$ does not
depend on $\beta$,
$$
D_\beta
=
\begin{pmatrix}
D_1\\
0
\end{pmatrix}\text{ where } D_1
=
\left.
\frac{\partial}
{\partial\beta^\top}
\E\{
\mathbf g_1(Y_1,X;\beta)
\}
\right|_{\beta=\beta_0}.
$$

Because $\mathbf g_1$ is the likelihood score
for $\beta$, the information identity gives
$$
D_1=-\Gamma_{11}.
$$
Consequently, the asymptotic covariance of the
internal-study-only maximum likelihood estimator is
$$
\Sigma_{\mathrm{MLE}}
=
\left(
D_1^\top\Gamma_{11}^{-1}D_1
\right)^{-1}=
\Gamma_{11}^{-1}
$$

Since $\Gamma$ is positive definite, its Schur
complement
$$
\Gamma_{22\cdot1}
=
\Gamma_{22}
-
\Gamma_{21}\Gamma_{11}^{-1}\Gamma_{12}
$$
is also positive definite.
By the block inverse formula,
$$
D_\beta^\top\Gamma^{-1}D_\beta
={}
D_1^\top\Gamma_{11}^{-1}D_1
+
D_1^\top\Gamma_{11}^{-1}
\Gamma_{12}
\Gamma_{22\cdot1}^{-1}
\Gamma_{21}
\Gamma_{11}^{-1}D_1.
$$
Using $D_1=-\Gamma_{11}$ and
$\Gamma_{12}=\Gamma_{21}^\top$, we obtain
$$
D_\beta^\top\Gamma^{-1}D_\beta
=
\Sigma_{\mathrm{MLE}}^{-1}
+
\Gamma_{21}^\top
\Gamma_{22\cdot1}^{-1}
\Gamma_{21}.
$$

Define
$$
M
=
\Gamma_{21}^\top
\Gamma_{22\cdot1}^{-1}
\Gamma_{21}.
$$
Since $\Gamma_{22\cdot1}^{-1}$ is positive
definite,
$$
M\succeq0.
$$
By Theorem~2, the asymptotic covariance of
the proposed GMM estimator is
$$
\begin{aligned}
\Sigma_{\mathrm{GMM}}
=
\left(
D_\beta^\top\Gamma^{-1}D_\beta
\right)^{-1}
=
\left(
\Sigma_{\mathrm{MLE}}^{-1}+M
\right)^{-1}.
\end{aligned}
$$
Since
$$
\Sigma_{\mathrm{MLE}}^{-1}+M
\succeq
\Sigma_{\mathrm{MLE}}^{-1},
$$
the order-reversing property of matrix
inversion gives
$$
\boxed{
\Sigma_{\mathrm{GMM}}
\preceq
\Sigma_{\mathrm{MLE}}.
}
$$
Thus, the proposed estimator is asymptotically
at least as efficient as the internal-study-only
maximum likelihood estimator. Since
$\Gamma_{22\cdot1}^{-1}\succ0$,
$$
M=0
\quad\Longleftrightarrow\quad
\Gamma_{21}=0.
$$
Therefore,
$$
\Gamma_{21}=0
\quad\Longleftrightarrow\quad
\Sigma_{\mathrm{GMM}}
=
\Sigma_{\mathrm{MLE}},
$$
which establishes the no-gain condition.

If $\Gamma_{21}\neq0$, then $M$ is nonzero
and positive semidefinite. Consequently,
$$
\Sigma_{\mathrm{MLE}}-\Sigma_{\mathrm{GMM}}
\succeq0,
\qquad
\Sigma_{\mathrm{MLE}}-\Sigma_{\mathrm{GMM}}
\neq0.
$$
Hence, there exists a nonzero vector $a$ such that
$$
a^\top\Sigma_{\mathrm{GMM}}a
<
a^\top\Sigma_{\mathrm{MLE}}a,
$$
so the efficiency gain is strict in at least
one direction.

Finally, if $\Gamma_{21}$ has full column rank,
then for every nonzero vector $a$,
$$
\begin{aligned}
a^\top M a
&=
(\Gamma_{21}a)^\top
\Gamma_{22\cdot1}^{-1}
(\Gamma_{21}a)>0.
\end{aligned}
$$
Thus $M\succ0$, and the order-reversing
property of matrix inversion gives
$$
\boxed{
\Sigma_{\mathrm{GMM}}
\prec
\Sigma_{\mathrm{MLE}}.
}
$$
Hence, the efficiency gain is strict in
every nonzero direction.
\hfill$\square$

\newpage

\newpage
\subsection{Proof of Theorem~3}
\label{sec:proof_thm3}
% =================================================
\noindent\textbf{Restatement of Theorem~3.}
Under the fixed-heterogeneity regime $\Omega=\Psi$ and the regularity conditions stated in the Supplementary Material, as $n_0\to\infty$, $n_k \asymp n$, $K\to\infty$, and $K(rn_0)^{-1/2}\to0$,
$$
\sqrt{n_0}
\left(\tilde{\theta}_{RE}-\hat{\theta}^{(0)}\right)
\xrightarrow{P}0,
\qquad
n_0\left\|\tilde V_{RE}-S^{(0)}\right\|
\xrightarrow{P}0.
$$
Moreover,
$$
\sqrt{n_0}
\left(
\hat{\beta}_{\mathrm{GMM}}
-
\hat{\beta}_{\mathrm{MLE}}
\right)
\xrightarrow{P}0.
$$

\noindent\textbf{Proof.}
Define the internal and external precision matrices
$$
P_0=S_0^{-1},
\qquad
P_k
=
\hat\alpha_k(\hat\Omega+S_k)^{-1},
\quad k=1,\ldots,K,
$$
and let $P_E=\sum_{k=1}^K P_k$.  The first-stage posterior mean and covariance are
$$
\tilde\theta_{RE}
=
\tilde V_{RE}
\left\{
P_0\hat\theta^{(0)}
+
\sum_{k=1}^K P_k\hat\theta^{(k)}
\right\}\text{ with }
\tilde V_{RE}=(P_0+P_E)^{-1}
$$

\medskip
\noindent
\textbf{Step 1: Convergence of the MM algorithm and
equivalence of the posterior covariance and the
internal-study covariance.}

We first establish that the smallest eigenvalue
of the MM covariance estimator is bounded away
from zero with probability tending to one. Recall that the Huber-type weights are
$$
\hat\alpha_k
=
\min\left\{
1,\sqrt{\frac{c_{K,n_0}}{d_k^2}}
\right\},
$$
where
$$
d_k^2
=
(\hat\theta^{(k)}-\hat\theta^{(0)})^\top
(S_k+S_0)^{-1}
(\hat\theta^{(k)}-\hat\theta^{(0)}).
$$
Let $A=\sum_{k=1}^K\hat\alpha_k$ and $h_n=\sqrt{\frac{c_{K,n_0}}{rn_0}}$. Since $c_{K,n_0}=O\{\log(Kn_0)\}$, the stated growth conditions imply
$h_n\to0$. By Assumption~\ref{assump:normality}
and $n_k\asymp rn_0$ uniformly, there exist
constants $0<c<C<\infty$ such that,
with probability tending to one,
$$
cI
\preceq
rn_0(S_k+S_0)
\preceq
CI,
\qquad k=1,\ldots,K.
$$
Consequently,
$$
\min\left\{
1,
\frac{\sqrt c\,h_n}
{\norm{\hat\theta^{(k)}-\hat\theta^{(0)}}}
\right\}
\leq
\hat\alpha_k
\leq
\min\left\{
1,
\frac{\sqrt C\,h_n}
{\norm{\hat\theta^{(k)}-\hat\theta^{(0)}}}
\right\}
$$

Under the fixed-heterogeneity regime,
$$
\hat\theta^{(k)}-\hat\theta^{(0)}
=
\gamma^{(k)}
+
\{\hat\theta^{(k)}-\theta^{(k)}\}
-
\{\hat\theta^{(0)}-\theta^{(0)}\},
$$
where $\gamma^{(k)}
\stackrel{\mathrm{i.i.d.}}{\sim}N(0,\Psi),
\text{ with }\Psi\succ0$.

We next establish an upper bound on $A$. Let $\varepsilon_n=h_n^{3/4}$. Recall from Assumption~\ref{assump:normality}
that
$$
\hat\theta^{(k)}-\theta^{(k)}
=
n_k^{-1/2}Z_k+n_k^{-1}R_k,
\qquad k=0,\ldots,K,
$$
where the fourth moments of $Z_k$ and
the second moments of $R_k$ are
uniformly bounded. Let $M_Z,M_R<\infty$ denote fixed
constants satisfying
$$
\sup_{0\leq k\leq K}\E\norm{Z_k}^4
\leq M_Z,
\qquad
\sup_{0\leq k\leq K}\E\norm{R_k}^2
\leq M_R.
$$

By the triangle inequality,
Markov's inequality, and the union bound,
$$
\begin{aligned}
\Prb\left\{
\max_{0\leq k\leq K}
\norm{\hat\theta^{(k)}-\theta^{(k)}}
>\delta\varepsilon_n
\right\}
&\leq
\sum_{k=0}^K
\Prb\left\{
\norm{\hat\theta^{(k)}-\theta^{(k)}}
>\delta\varepsilon_n
\right\}
\\
&\leq
\sum_{k=0}^K
\left\{
\frac{16\E\norm{Z_k}^4}
{\delta^4n_k^2\varepsilon_n^4}
+
\frac{4\E\norm{R_k}^2}
{\delta^2n_k^2\varepsilon_n^2}
\right\}
\\
&\leq
C_\delta
\left\{
\frac{K+1}{(rn_0)^2\varepsilon_n^4}
+
\frac{K+1}{(rn_0)^2\varepsilon_n^2}
\right\}.
\end{aligned}
$$
where $C_\delta
=
\frac1{c_s^2}
\max\left\{
\frac{16M_Z}{\delta^4},
\frac{4M_R}{\delta^2}
\right\}$ with a constant $c_s>0$ such that, for all
sufficiently large $n_0$,
$ n_k\geq c_srn_0,k=0,\ldots,K$. 
Since $\varepsilon_n^4=
h_n^3
=
\left(
\frac{c_{K,n_0}}{rn_0}
\right)^{3/2},
$
we have
$$
\frac{K+1}{(rn_0)^2\varepsilon_n^4}
=
\frac{K+1}
{\sqrt{rn_0}\,c_{K,n_0}^{3/2}}
=\frac{K+1}
{\sqrt{rn_0}}\frac{1}
{c_{K,n_0}^{3/2}}=o(1),
$$
where the last equality follows from
$K/\sqrt{rn_0}\to0$. Similarly, given $\varepsilon_n^2=\left(\frac{c_{K,n_0}}{rn_0}\right)^{3/4}$,
$$
\frac{K+1}{(rn_0)^2\varepsilon_n^2}
=
\frac{K+1}
{(rn_0)^{5/4}c_{K,n_0}^{3/4}} = 
\frac{K+1}{(rn_0)^{1/2}} \frac{1}{(rn_0)^{3/4}} \frac{1}{c_{K,n_0}^{3/4}} 
=o(1)
$$

Therefore, for every fixed $\delta>0$,
$$
\Prb\left\{
\max_{0\leq k\leq K}
\norm{\hat\theta^{(k)}-\theta^{(k)}}
>\delta\varepsilon_n
\right\}
\to0.
$$
Consequently,
$$
\boxed{
\max_{0\leq k\leq K}
\norm{\hat\theta^{(k)}-\theta^{(k)}}
=o_p(\varepsilon_n).
}
$$

With probability tending to one,
the estimation errors are therefore
bounded by $\varepsilon_n$ uniformly
over all studies. For every external study satisfying
$\norm{\gamma^{(k)}}>4\varepsilon_n$,
the reverse triangle inequality gives
$$
\begin{aligned}
\norm{\hat\theta^{(k)}-\hat\theta^{(0)}}&\geq
\norm{\gamma^{(k)}}
-
\norm{\hat\theta^{(k)}-\theta^{(k)}}
-
\norm{\hat\theta^{(0)}-\theta^{(0)}}
\\
&\geq
\norm{\gamma^{(k)}}-2\varepsilon_n
\\
&\geq
\frac12\norm{\gamma^{(k)}}.
\end{aligned}
$$

Using the upper bound on the Huber weights,
we obtain, with probability tending to one,
$$
\begin{aligned}
\frac{A}{Kh_n}
&=
\frac1{Kh_n}\sum_{k=1}^K\hat\alpha_k
\\
&\leq
\frac1{Kh_n}
\sum_{k=1}^K
\left[
\mathbf 1\left\{
\norm{\gamma^{(k)}}\leq4\varepsilon_n
\right\}
+
\frac{2\sqrt C\,h_n}{\norm{\gamma^{(k)}}}
\mathbf 1\left\{
\norm{\gamma^{(k)}}>4\varepsilon_n
\right\}
\right]
\\
&=
\frac1{Kh_n}
\sum_{k=1}^K
\mathbf 1\left\{
\norm{\gamma^{(k)}}\leq4\varepsilon_n
\right\}
+
\frac{2\sqrt C}{K}
\sum_{k=1}^K
\frac{
\mathbf 1\left\{
\norm{\gamma^{(k)}}>4\varepsilon_n
\right\}
}{
\norm{\gamma^{(k)}}
}
\\
&\leq
\frac1{Kh_n}
\sum_{k=1}^K
\mathbf 1\left\{
\norm{\gamma^{(k)}}\leq4\varepsilon_n
\right\}
+
\frac{2\sqrt C}{K}
\sum_{k=1}^K
\norm{\gamma^{(k)}}^{-1}.
\end{aligned}
$$

Since $\Psi\succ0$, the Gaussian density
is bounded. Moreover, both the intercept and slope parameters are included in $\theta^{(0)}$, we have $p=\dim(\theta^{(0)})=2L\geq2$ and 
$$
\Prb\left\{
\norm{\gamma^{(k)}}\leq4\varepsilon_n
\right\}
=
O(\varepsilon_n^p).
$$

By Markov's inequality,
$$
\frac1{Kh_n}
\sum_{k=1}^K
\mathbf 1\left\{
\norm{\gamma^{(k)}}\leq4\varepsilon_n
\right\}
=
O_p\left(
\frac{\varepsilon_n^p}{h_n}
\right)
=
O_p\left(
h_n^{3p/4-1}
\right)
=o_p(1)
$$
where the last equality follows because
$p\geq2$ and $h_n\to0$. Moreover, since $p\geq2$,
$$
\E\norm{\gamma^{(k)}}^{-1}<\infty.
$$
By the law of large numbers,
$$
\frac1K\sum_{k=1}^K
\norm{\gamma^{(k)}}^{-1}
\toP
\E\norm{\gamma^{(k)}}^{-1}
<\infty.
$$

Combining these results, for a sufficiently
large constant $C_1<\infty$,
$$
\boxed{
\Prb\left(A\leq C_1Kh_n\right)\to1.
}
$$

Next, for any fixed constants
$0<a<b<\infty$, consider the external
studies satisfying
$$
a\leq\norm{\gamma^{(k)}}\leq b.
$$

Since the Gaussian distribution has
a strictly positive density,
$$
\pi_{a,b}
=
\Prb\left(
a\leq\norm{\gamma^{(k)}}\leq b
\right)>0.
$$
By the law of large numbers,
$$
\frac1K\sum_{k=1}^K
\mathbf 1\left\{
a\leq\norm{\gamma^{(k)}}\leq b
\right\}
\toP\pi_{a,b}.
$$
Thus, a positive proportion of external
studies satisfy this condition with
probability tending to one.

Since the within-study estimation errors
converge uniformly to zero, for these
studies,
$$
\norm{\hat\theta^{(k)}-\hat\theta^{(0)}}
\leq
\norm{\gamma^{(k)}}+o_p(1)\leq b+o_p(1)
$$
uniformly over the selected studies. Consequently, with probability tending
to one,
$$
\begin{aligned}
\hat\alpha_k
&\geq
\min\left\{
1,
\frac{\sqrt c\,h_n}
{\norm{\hat\theta^{(k)}-\hat\theta^{(0)}}}
\right\}
\\
&\geq
\min\left\{
1,\frac{\sqrt c\,h_n}{b+1}
\right\}=
\frac{\sqrt c}{b+1}h_n,
\end{aligned}
$$
where the last equality holds for
sufficiently large $n_0$ because
$h_n\to0$.

Thus,
$$
\hat\alpha_k\geq c_1h_n
$$
uniformly over the selected studies,
where $c_1=\frac{\sqrt c}{b+1}>0$. Because the proportion of selected
studies converges to $\pi_{a,b}>0$,
we also obtain, for some constant
$c_0>0$,
$$
\boxed{
\Prb\left(A\geq c_0Kh_n\right)\to1.
}
$$

Combining the upper and lower bounds,
$$
\boxed{
c_0Kh_n\leq A\leq C_1Kh_n
}
$$
with probability tending to one. For any $\mu\in\mathbb R^p$,
the matrix $(\hat\theta^{(k)}-\mu)(\hat\theta^{(k)}-\mu)^\top$ is positive semidefinite.

Therefore, uniformly over
$\mu\in\mathbb R^p$,
$$
\begin{aligned}
\frac1A\sum_{k=1}^K
\hat\alpha_k
(\hat\theta^{(k)}-\mu)
(\hat\theta^{(k)}-\mu)^\top
&\succeq
\sum_{\substack{1\leq k\leq K:\\
a\leq\norm{\gamma^{(k)}}\leq b}}
\frac{\hat\alpha_k}{A}
(\hat\theta^{(k)}-\mu)
(\hat\theta^{(k)}-\mu)^\top
\\
&\succeq
\frac{c_1h_n}{C_1Kh_n}
\sum_{\substack{1\leq k\leq K:\\
a\leq\norm{\gamma^{(k)}}\leq b}}
(\hat\theta^{(k)}-\mu)
(\hat\theta^{(k)}-\mu)^\top
\\
&=
\frac{c_1}{C_1K}
\sum_{\substack{1\leq k\leq K:\\
a\leq\norm{\gamma^{(k)}}\leq b}}
(\hat\theta^{(k)}-\mu)
(\hat\theta^{(k)}-\mu)^\top.
\end{aligned}
$$

Let $\bar\theta_{a,b}$ denote the sample
mean of the external estimates satisfying
$a\leq\norm{\gamma^{(k)}}\leq b$.

By the sample-mean identity,
$$
\sum_{\substack{1\leq k\leq K:\\
a\leq\norm{\gamma^{(k)}}\leq b}}
(\hat\theta^{(k)}-\mu)
(\hat\theta^{(k)}-\mu)^\top
\succeq
\sum_{\substack{1\leq k\leq K:\\
a\leq\norm{\gamma^{(k)}}\leq b}}
(\hat\theta^{(k)}-\bar\theta_{a,b})
(\hat\theta^{(k)}-\bar\theta_{a,b})^\top.
$$

By the law of large numbers, the symmetry
of the centered Gaussian distribution,
and the uniform convergence of the
within-study estimation errors,
$$
\frac1K
\sum_{\substack{1\leq k\leq K:\\
a\leq\norm{\gamma^{(k)}}\leq b}}
(\hat\theta^{(k)}-\bar\theta_{a,b})
(\hat\theta^{(k)}-\bar\theta_{a,b})^\top
\toP
\E\left[
\gamma^{(k)}\gamma^{(k)\top}
\mathbf 1\left\{
a\leq\norm{\gamma^{(k)}}\leq b
\right\}
\right]
$$

The limiting matrix is positive definite.
Indeed, for every nonzero vector $v$,
$$
\E\left[
(v^\top\gamma^{(k)})^2
\mathbf 1\left\{
a\leq\norm{\gamma^{(k)}}\leq b
\right\}
\right]>0,
$$
because $\Psi\succ0$ and the Gaussian
distribution has a strictly positive
density throughout $\mathbb R^p$.

It follows that, for some constant
$c_2>0$,
$$
\boxed{
\begin{aligned}
\Prb\Bigg\{
\inf_{\mu\in\mathbb R^p}
\lambda_{\min}\Bigg(
&\frac1A\sum_{k=1}^K
\hat\alpha_k
(\hat\theta^{(k)}-\mu)
(\hat\theta^{(k)}-\mu)^\top
\Bigg)
\geq c_2
\Bigg\}
\to1.
\end{aligned}
}
$$

Furthermore, Assumption~\ref{assump:normality}
implies
$$
\norm{
\frac1A\sum_{k=1}^K\hat\alpha_kS_k
}
\leq
\max_{1\leq k\leq K}\norm{S_k}
=
O_p(n_0^{-1})
=o_p(1),
$$
and
$\norm{S_0}=O_p(n_0^{-1})=o_p(1)$. Recall that the original MM initialization is
$$
\begin{aligned}
\hat\Omega^{(0)}
={}&
\frac1A\sum_{k=1}^K
\hat\alpha_k
(\hat\theta^{(k)}-\hat\theta^{(0)})
(\hat\theta^{(k)}-\hat\theta^{(0)})^\top-
\frac1A\sum_{k=1}^K\hat\alpha_kS_k
-S_0.
\end{aligned}
$$

The preceding bounds imply that, for
some constant $\underline\omega>0$,
$$
\Prb\left\{
\lambda_{\min}(\hat\Omega^{(0)})
\geq\underline\omega
\right\}\to1.
$$

We next establish convergence of the original
MM iterations. For any covariance matrix
$\Omega\succeq\underline\omega I$,
let $\tilde V_{RE}(\Omega)$ and
$\tilde\theta_{RE}(\Omega)$ denote the
previously defined posterior covariance
and mean, respectively, evaluated with
$\hat\Omega$ replaced by $\Omega$.

The original MM covariance update can
then be written as
$$
\begin{aligned}
\mathcal T_n(\Omega)
={}&
\frac1A\sum_{k=1}^K
\hat\alpha_k
\{\hat\theta^{(k)}-\tilde\theta_{RE}(\Omega)\}
\{\hat\theta^{(k)}-\tilde\theta_{RE}(\Omega)\}^\top
-
\frac1A\sum_{k=1}^K\hat\alpha_kS_k
+
\tilde V_{RE}(\Omega).
\end{aligned}
$$

Since $\tilde V_{RE}(\Omega)\succeq0$,
the uniform weighted-covariance lower
bound established above gives
$$
\mathcal T_n(\Omega)
\succeq
\underline\omega I
$$
with probability tending to one,
uniformly over
$\Omega\succeq\underline\omega I$,
after choosing
$\underline\omega<c_2/2$.

Thus, the MM update maps the closed set
$$
\mathcal D
=
\left\{
\Omega:\Omega\succeq\underline\omega I
\right\}
$$
into itself, and the original
initialization belongs to $\mathcal D$
with probability tending to one.

We now show that $\mathcal T_n$ is
contractive on $\mathcal D$. For any $\Omega\in\mathcal D$,
$$
\norm{(\Omega+S_k)^{-1}}
\leq\underline\omega^{-1}.
$$

By the matrix inverse identity, for
any $\Omega_1,\Omega_2\in\mathcal D$,
$$
\begin{aligned}
\norm{
(\Omega_1+S_k)^{-1}
-
(\Omega_2+S_k)^{-1}
}&\leq
\norm{(\Omega_1+S_k)^{-1}}
\norm{\Omega_2-\Omega_1}
\norm{(\Omega_2+S_k)^{-1}} \\
&\leq
\underline\omega^{-1}
\norm{\Omega_2-\Omega_1}
\underline\omega^{-1}\\
&=
\underline\omega^{-2}
\norm{\Omega_1-\Omega_2}.
\end{aligned}
$$

Moreover, the upper bound on the
Huber weights gives
$$
\sum_{k=1}^K
\hat\alpha_k
\norm{\hat\theta^{(k)}-\hat\theta^{(0)}}
\leq
\sqrt C\,Kh_n.
$$

Since
$\tilde V_{RE}(\Omega)\preceq S_0$,
$$
\sup_{\Omega\in\mathcal D}
\norm{\tilde V_{RE}(\Omega)}
=
O_p(n_0^{-1}).
$$

By the matrix inverse identity, submultiplicativity and the
triangle inequality give
$$
\begin{aligned}
\norm{
\tilde V_{RE}(\Omega_1)
-
\tilde V_{RE}(\Omega_2)
}
&\leq
\norm{\tilde V_{RE}(\Omega_1)}
\norm{\tilde V_{RE}(\Omega_2)}
\sum_{k=1}^K\hat\alpha_k
\norm{
(\Omega_2+S_k)^{-1}
-
(\Omega_1+S_k)^{-1}
}
\\
&\leq
\norm{S_0}^2
\sum_{k=1}^K\hat\alpha_k
\underline\omega^{-2}
\norm{\Omega_1-\Omega_2} \\
&\leq
\frac{A\norm{S_0}^2}{\underline\omega^2}
\norm{\Omega_1-\Omega_2} \\
&\leq O_p\left(\frac{Kh_n}{n_0^2}\right)\norm{\Omega_1-\Omega_2} 
\end{aligned}
$$

Furthermore, from the posterior mean formula,
$$
\tilde\theta_{RE}(\Omega)-\hat\theta^{(0)}
=
\tilde V_{RE}(\Omega)
\sum_{k=1}^K
\hat\alpha_k
(\Omega+S_k)^{-1}
(\hat\theta^{(k)}-\hat\theta^{(0)}).
$$
By submultiplicativity of the matrix norm and
the triangle inequality,
$$
\begin{aligned}
\norm{\tilde\theta_{RE}(\Omega)-\hat\theta^{(0)}}
&\leq
\norm{\tilde V_{RE}(\Omega)}
\sum_{k=1}^K
\hat\alpha_k
\norm{(\Omega+S_k)^{-1}}
\norm{\hat\theta^{(k)}-\hat\theta^{(0)}}\\
&\leq
\norm{S_0}\,
\underline\omega^{-1}
\sum_{k=1}^K
\hat\alpha_k
\norm{\hat\theta^{(k)}-\hat\theta^{(0)}}\\
&\leq
\norm{S_0}\,
\underline\omega^{-1}
\sqrt C\,Kh_n\\
&=
O_p\left(\frac{Kh_n}{n_0}\right).
\end{aligned}
$$
Next, since $A\geq c_0Kh_n$
with probability tending to one,
$$
\frac1A\sum_{k=1}^K
\hat\alpha_k
\norm{\hat\theta^{(k)}-\hat\theta^{(0)}}
\leq
\frac{\sqrt C\,Kh_n}{A}
=
O_p(1).
$$

Using the difference-of-outer-products identity
$$
\begin{aligned}
&\norm{
(\hat\theta^{(k)}-\tilde\theta_{RE}(\Omega_1))
(\hat\theta^{(k)}-\tilde\theta_{RE}(\Omega_1))^\top
-
(\hat\theta^{(k)}-\tilde\theta_{RE}(\Omega_2))
(\hat\theta^{(k)}-\tilde\theta_{RE}(\Omega_2))^\top
}
\\
\leq&
\left\{
\norm{\hat\theta^{(k)}-\tilde\theta_{RE}(\Omega_1)}
+
\norm{\hat\theta^{(k)}-\tilde\theta_{RE}(\Omega_2)}
\right\}
\norm{\tilde\theta_{RE}(\Omega_1)-\tilde\theta_{RE}(\Omega_2)}
\\
\leq&
\left\{
2\norm{\hat\theta^{(k)}-\hat\theta^{(0)}}
+
\sum_{i=1}^2
\norm{\tilde\theta_{RE}(\Omega_i)-\hat\theta^{(0)}}
\right\}
\norm{\tilde\theta_{RE}(\Omega_1)-\tilde\theta_{RE}(\Omega_2)},
\end{aligned}
$$
where the second inequality follows from the triangle inequality.

Since the term
$A^{-1}\sum_{k=1}^K\hat\alpha_kS_k$
does not depend on $\Omega$, it cancels when
subtracting the two MM updates. Therefore,
$$
\begin{aligned}
&\norm{\mathcal T_n(\Omega_1)-\mathcal T_n(\Omega_2)}
\\
\leq&
\frac1A\sum_{k=1}^K\hat\alpha_k
\Bigg\{
2\norm{\hat\theta^{(k)}-\hat\theta^{(0)}}
+
\sum_{i=1}^2
\norm{\tilde\theta_{RE}(\Omega_i)-\hat\theta^{(0)}}
\Bigg\}
\norm{\tilde\theta_{RE}(\Omega_1)-\tilde\theta_{RE}(\Omega_2)}
\\
&\quad+
\norm{\tilde V_{RE}(\Omega_1)-\tilde V_{RE}(\Omega_2)}
\\
=&
\Bigg\{
\frac2A\sum_{k=1}^K
\hat\alpha_k
\norm{\hat\theta^{(k)}-\hat\theta^{(0)}}
+
\sum_{i=1}^2
\norm{\tilde\theta_{RE}(\Omega_i)-\hat\theta^{(0)}}
\Bigg\}
\norm{\tilde\theta_{RE}(\Omega_1)-\tilde\theta_{RE}(\Omega_2)}
\\
&\quad+
\norm{\tilde V_{RE}(\Omega_1)-\tilde V_{RE}(\Omega_2)},
\end{aligned}
$$
From the previously established bounds,
$$
\sum_{k=1}^K
\hat\alpha_k
\norm{\hat\theta^{(k)}-\hat\theta^{(0)}}
\leq \sqrt C\,Kh_n
$$
and, with probability tending to one,
$$
A\geq c_0Kh_n.
$$
Consequently,
$$
\begin{aligned}
\frac2A\sum_{k=1}^K
\hat\alpha_k
\norm{\hat\theta^{(k)}-\hat\theta^{(0)}}
&\leq
\frac{2\sqrt C\,Kh_n}{A}\\
&\leq
\frac{2\sqrt C}{c_0}
=O_p(1).
\end{aligned}
$$

Furthermore, we have established that
$$
\sup_{\Omega\in\mathcal D}
\norm{\tilde\theta_{RE}(\Omega)-\hat\theta^{(0)}}
=
O_p\left(\frac{Kh_n}{n_0}\right).
$$
Therefore,
$$
\sum_{i=1}^2
\norm{\tilde\theta_{RE}(\Omega_i)-\hat\theta^{(0)}}
=
O_p\left(\frac{Kh_n}{n_0}\right).
$$

Combining these results gives
$$
\begin{aligned}
\frac2A\sum_{k=1}^K
\hat\alpha_k
\norm{\hat\theta^{(k)}-\hat\theta^{(0)}}
+
\sum_{i=1}^2
\norm{\tilde\theta_{RE}(\Omega_i)-\hat\theta^{(0)}}
=
O_p(1)+O_p\left(\frac{Kh_n}{n_0}\right)
=O_p(1),
\end{aligned}
$$
where the last equality follows because
$Kh_n/n_0\to0$. Since
$$
\begin{aligned}
\tilde\theta_{RE}(\Omega_1)
-\tilde\theta_{RE}(\Omega_2)
&=
\{\tilde V_{RE}(\Omega_1)
-\tilde V_{RE}(\Omega_2)\}
\sum_{k=1}^K
\hat\alpha_k(\Omega_1+S_k)^{-1}
(\hat\theta^{(k)}-\hat\theta^{(0)})
\\
&\quad+
\tilde V_{RE}(\Omega_2)
\sum_{k=1}^K
\hat\alpha_k
\{(\Omega_1+S_k)^{-1}
-(\Omega_2+S_k)^{-1}\}
(\hat\theta^{(k)}-\hat\theta^{(0)}).
\end{aligned}
$$
Then
$$
\begin{aligned}
\norm{
\tilde\theta_{RE}(\Omega_1)
-\tilde\theta_{RE}(\Omega_2)
}
&\leq
\norm{
\tilde V_{RE}(\Omega_1)
-\tilde V_{RE}(\Omega_2)
}
\sum_{k=1}^K
\hat\alpha_k
\norm{(\Omega_1+S_k)^{-1}}
\norm{\hat\theta^{(k)}-\hat\theta^{(0)}}
\\
&\quad+
\norm{\tilde V_{RE}(\Omega_2)}
\sum_{k=1}^K
\hat\alpha_k
\norm{
(\Omega_1+S_k)^{-1}
-(\Omega_2+S_k)^{-1}
}
\norm{\hat\theta^{(k)}-\hat\theta^{(0)}}.
\end{aligned}
$$

Using the previously established bounds
$$
\begin{aligned}
    \norm{
\tilde V_{RE}(\Omega_1)
-\tilde V_{RE}(\Omega_2)
}
&\leq
O_p\left(\frac{Kh_n}{n_0^2}\right)
\norm{\Omega_1-\Omega_2} \\
\norm{\tilde V_{RE}(\Omega_2)} &=
O_p(n_0^{-1}) \\
\norm{(\Omega_i+S_k)^{-1}}
&\leq\underline\omega^{-1},
\qquad i=1,2,\\
\norm{
(\Omega_1+S_k)^{-1}
-(\Omega_2+S_k)^{-1}
}
&\leq
\underline\omega^{-2}
\norm{\Omega_1-\Omega_2} \\
\sum_{k=1}^K
\hat\alpha_k
\norm{\hat\theta^{(k)}-\hat\theta^{(0)}}
&\leq
\sqrt C\,Kh_n,
\end{aligned}
$$
we obtain
$$
\begin{aligned}
\norm{
\tilde\theta_{RE}(\Omega_1)
-\tilde\theta_{RE}(\Omega_2)
}
&\leq
O_p\left(\frac{Kh_n}{n_0^2}\right)
O(Kh_n)
\norm{\Omega_1-\Omega_2}
+
O_p(n_0^{-1})
O(Kh_n)
\norm{\Omega_1-\Omega_2}
\\
&=
O_p\left(
\frac{K^2h_n^2}{n_0^2}
+
\frac{Kh_n}{n_0}
\right)
\norm{\Omega_1-\Omega_2}.
\end{aligned}
$$

Since $Kh_n/n_0\to0$,
$$
\frac{K^2h_n^2}{n_0^2}
=
o\left(\frac{Kh_n}{n_0}\right).
$$
Consequently,
$$
\norm{
\tilde\theta_{RE}(\Omega_1)
-\tilde\theta_{RE}(\Omega_2)
}
\leq
O_p\left(\frac{Kh_n}{n_0}\right)
\norm{\Omega_1-\Omega_2},
$$
uniformly over
$\Omega_1,\Omega_2\in\mathcal D$.

Therefore
$$
\begin{aligned}
\norm{\mathcal T_n(\Omega_1)
-\mathcal T_n(\Omega_2)} &\leq
\left\{
O_p\left(\frac{Kh_n}{n_0}\right)
+
O_p\left(\frac{Kh_n}{n_0^2}\right)
\right\}
\norm{\Omega_1-\Omega_2}\\
&=
O_p\left(\frac{Kh_n}{n_0}\right)
\norm{\Omega_1-\Omega_2}.
\end{aligned}
$$

Hence, there exists a nonnegative random
coefficient $L_n$, independent of
$\Omega_1$ and $\Omega_2$, such that
$$
\norm{\mathcal T_n(\Omega_1)
-\mathcal T_n(\Omega_2)}
\leq
L_n\norm{\Omega_1-\Omega_2},
$$
uniformly over
$\Omega_1,\Omega_2\in\mathcal D$, where
$L_n=O_p\left(\frac{Kh_n}{n_0}\right)$ and $L_n=o_p(1)$.  
With probability tending to one,
$\mathcal T_n$ is thus a contraction
mapping from $\mathcal D$ into itself.
Since $\mathcal D$ is complete under
the matrix norm and
$\hat\Omega^{(0)}\in\mathcal D$,
the Banach fixed-point theorem
establishes that the original MM
iterations converge to a unique
fixed point $\hat\Omega$ in
$\mathcal D$. In particular,
$$
\boxed{
\Prb\left\{
\lambda_{\min}(\hat\Omega)
\geq\underline\omega
\right\}\to1.
}
$$

We now establish the equivalence
of the posterior covariance and
the internal-study covariance.

Since $S_k\succeq0$ and
$0\leq\hat\alpha_k\leq1$,
the preceding eigenvalue bound gives
$$
\norm{P_E}
=
\norm{
\sum_{k=1}^K
\hat\alpha_k
(\hat\Omega+S_k)^{-1}
}
\leq
\underline\omega^{-1}A
=
O_p(Kh_n)
=
O_p(K)
$$

Moreover, because $P_E\succeq0$,
$$
0
\prec
\tilde V_{RE}
=
(P_0+P_E)^{-1}
\preceq
P_0^{-1}
=
S_0.
$$

Thus,
$$
\norm{\tilde V_{RE}}
=
O_p(n_0^{-1}).
$$

Using the matrix inverse identity,
$$
\tilde V_{RE}-S_0
=
-\tilde V_{RE}P_ES_0.
$$

It follows that
$$
\norm{\tilde V_{RE}-S_0}=
O_p\left(\frac{Kh_n}{n_0^2}\right)
=
O_p\left(\frac{K}{n_0^2}\right).
$$

Since $r\to r_0\in(0,\infty)$
and $K(rn_0)^{-1/2}\to0$,
we have $K/\sqrt{n_0}\to0$.
Consequently,
$$
n_0\norm{\tilde V_{RE}-S_0}
=
O_p\left(\frac{Kh_n}{n_0}\right)
=
o_p(1).
$$

Therefore,
$$
\boxed{
n_0\norm{\tilde V_{RE}-S_0}
\toP0.
}
$$

\medskip
\noindent
\textbf{Step 2: Equivalence of the posterior mean
and the internal-study estimator.}

From the definition of $\tilde\theta_{RE}$,
$$
\tilde\theta_{RE}-\hat\theta^{(0)}
=
\tilde V_{RE}
\sum_{k=1}^K
P_k
(\hat\theta^{(k)}-\hat\theta^{(0)}).
$$

Since
$$
\sup_{1\leq k\leq K}
\norm{(\hat\Omega+S_k)^{-1}}
=
O_p(1),
$$
and
$$
\sum_{k=1}^K
\hat\alpha_k
\norm{\hat\theta^{(k)}-\hat\theta^{(0)}}
\leq
\sqrt C\,Kh_n,
$$
we obtain
$$
\norm{
\sum_{k=1}^K
P_k
(\hat\theta^{(k)}-\hat\theta^{(0)})
}
=
O_p(Kh_n).
$$

Together with
$\norm{\tilde V_{RE}}=O_p(n_0^{-1})$,
this yields
$$
\begin{aligned}
\norm{\tilde\theta_{RE}-\hat\theta^{(0)}}
&=
O_p\left(\frac{Kh_n}{n_0}\right)
\\
&=
O_p\left(\frac{K}{n_0}\right).
\end{aligned}
$$

Therefore,
$$
\begin{aligned}
&\sqrt{n_0}
\norm{\tilde\theta_{RE}-\hat\theta^{(0)}}
\\
&\qquad=
O_p\left(\frac{Kh_n}{\sqrt{n_0}}\right)
=o_p(1).
\end{aligned}
$$

Consequently,
$$
\boxed{
\sqrt{n_0}
(\tilde\theta_{RE}-\hat\theta^{(0)})
\toP0.
}
$$

\medskip
\noindent
\textbf{Step 3: Equivalence of the GMM and
internal-study-only estimators.}

Since $\hat\beta_{\mathrm{MLE}}$ and
$\hat\theta^{(0)}$ solve their respective
internal-study score equations,
$$
\bar{\mathbf g}_n
(\hat\beta_{\mathrm{MLE}},\hat\theta^{(0)})
=0.
$$

For the plug-in formulation, the smoothness
of the auxiliary score functions gives
$$
\norm{
\bar{\mathbf g}_n
(\hat\beta_{\mathrm{MLE}},\tilde\theta_{RE})
-
\bar{\mathbf g}_n
(\hat\beta_{\mathrm{MLE}},\hat\theta^{(0)})
}
=
O_p\left(
\norm{\tilde\theta_{RE}-\hat\theta^{(0)}}
\right).
$$

For the uncertainty-propagating formulation, we consider the exact
posterior expectation. The second-order Taylor expansion
established in the proof of Theorem~2 gives
$$
\sup_{\beta\in\mathcal P_\beta}
\norm{
\bar{\mathbf g}_n^A
(\beta,\tilde\theta_{RE},\tilde V_{RE})
-
\bar{\mathbf g}_n
(\beta,\tilde\theta_{RE})
}
=
O_p(\norm{\tilde V_{RE}}).
$$

Consequently, for either formulation,
$$
\norm{
\bar{\mathbf g}_n^\dagger
(\hat\beta_{\mathrm{MLE}})
}
=
O_p\left(
\norm{\tilde\theta_{RE}-\hat\theta^{(0)}}
+
\norm{\tilde V_{RE}}
\right)
=
O_p\left(
\frac{Kh_n}{n_0}+\frac1{n_0}
\right).
$$

Furthermore, Assumption~\ref{assump:normality}
and Step~2 imply
$$
\tilde\theta_{RE}\toP\theta^{(0)},
\qquad
\norm{\tilde V_{RE}}\toP0.
$$

Under the uniform convergence and
identification conditions used in
Theorem~2, these results give
$$
\hat\beta_{\mathrm{GMM}}\toP\beta_0,
\qquad
\hat W_{\mathrm{MLE}}\toP\Gamma.
$$

By the smoothness of the moment functions,
the uniform law of large numbers for
their derivatives, and the consistency
of both estimators,
$$
\bar{\mathbf g}_n^\dagger
(\hat\beta_{\mathrm{GMM}})
=
\bar{\mathbf g}_n^\dagger
(\hat\beta_{\mathrm{MLE}})
+
\{D_\beta+o_p(1)\}
(\hat\beta_{\mathrm{GMM}}
-\hat\beta_{\mathrm{MLE}}).
$$

Since $\hat\beta_{\mathrm{GMM}}$ minimizes
the GMM criterion, its first-order
condition gives
$$
0
=
\{D_\beta^\top\Gamma^{-1}+o_p(1)\}
\bar{\mathbf g}_n^\dagger
(\hat\beta_{\mathrm{MLE}})
+
\{D_\beta^\top\Gamma^{-1}D_\beta
+o_p(1)\}
(\hat\beta_{\mathrm{GMM}}
-\hat\beta_{\mathrm{MLE}})
$$

As established in Theorem~2,
$D_\beta$ has full column rank and
$\Gamma$ is positive definite.
Hence,
$$
D_\beta^\top\Gamma^{-1}D_\beta
\succ0.
$$

Since this matrix is nonsingular,
the preceding expansion yields
$$
\norm{
\hat\beta_{\mathrm{GMM}}
-\hat\beta_{\mathrm{MLE}}
}
=
O_p\left(
\norm{
\bar{\mathbf g}_n^\dagger
(\hat\beta_{\mathrm{MLE}})
}
\right)
=
O_p\left(
\frac{Kh_n}{n_0}+\frac1{n_0}
\right)
$$

Because $K/\sqrt{n_0}\to0$
and $h_n\to0$,
$$
\frac{Kh_n}{n_0}+\frac1{n_0}
=
o(n_0^{-1/2}).
$$

Therefore,
$$
\norm{
\hat\beta_{\mathrm{GMM}}
-\hat\beta_{\mathrm{MLE}}
}
=
o_p(n_0^{-1/2}).
$$

Consequently,
$$
\boxed{
\sqrt{n_0}
(\hat\beta_{\mathrm{GMM}}
-\hat\beta_{\mathrm{MLE}})
\toP0.
}
$$

This completes the proof.
\hfill$\square$
% =================================================
% =================================================
\newpage
% =================================================

\newpage
% =================================================
\subsection{Proof of Theorem~4}
\label{sec:proof_boot}
% =================================================

\noindent\textbf{Restatement of Theorem~4.}

Consider an asymptotic regime in which,
as $n_0\to\infty$ and $K\to\infty$,
$$
\left(
\sqrt{n_0}
\bar{\mathbf g}_n(\beta_0,\theta^{(0)}),
\sqrt{n_0}
(\tilde\theta_{RE}-\theta^{(0)})
\right)
\xrightarrow{d}
(\mathbb G,\mathbb T),
$$
where $\mathbb G$ is a mean-zero Gaussian random
vector with covariance matrix $\Gamma$, and
$\mathbb T$ is a mean-zero, nondegenerate random
vector with finite covariance matrix. Under the bootstrap regularity conditions stated in the Supplementary Material,
$$
\sqrt{n_0}
(\hat\beta_{\mathrm{GMM}}-\beta_0)
\xrightarrow{d}\mathbb B,
$$
where
$$
\mathbb B
=
-J^{-1}D_\beta^\top\Gamma^{-1}
(\mathbb G+D_\theta\mathbb T).
$$

Moreover, the hybrid bootstrap consistently
estimates the asymptotic covariance of the
proposed estimator:
$$
n_0\operatorname{Var}^*
(\hat\beta_{\mathrm{GMM}}^*)
\xrightarrow{p}
\operatorname{Var}(\mathbb B).
$$

\noindent\textbf{Proof.}

Throughout the proof, $n_0\to\infty$
and $K\to\infty$. Write
$$
g_i
=
\mathbf g(Y_{i1},Y_{i2},X_i;
\beta_0,\theta^{(0)})
$$

The original and bootstrap first-stage
estimators use the same meta-analytic
procedure, with the borrowing weights
and between-study covariance re-estimated
in each bootstrap replicate.

\medskip
\noindent
\textbf{Step 1: Asymptotic distribution
of the original GMM estimator.}

The posterior covariance is
$$
\tilde V_{RE}
=
\left[
S_0^{-1}
+
\sum_{k=1}^K
\hat\alpha_k(\hat\Omega+S_k)^{-1}
\right]^{-1}.
$$

Since the borrowing weights are
nonnegative and the precision matrices
are positive definite,
$$
0\prec\tilde V_{RE}\preceq S_0.
$$

The same argument applies to the
bootstrap posterior covariance:
$$
0\prec\tilde V_{RE}^*\preceq S_0^*.
$$

Consequently, by
Assumption~\ref{assump:boot}(iv),
$$
\|\tilde V_{RE}\|
=
O_p(n_0^{-1}),
$$
and
$$
\|\tilde V_{RE}^*\|
=
O_{p^*}(n_0^{-1})
$$
conditionally in probability.

The second-derivative bound on the
moment functions in
Assumption~\ref{assump:gmm}
implies that posterior-integrated
and plug-in moments differ by
$O_p(n_0^{-1})$ in the original sample
and $O_{p^*}(n_0^{-1})$ in the
bootstrap sample. Thus, posterior integration does not
affect the first-order GMM expansions.

Under the joint convergence regime considered in Theorem~4,
$$
\tilde\theta_{RE}-\theta^{(0)}
=
O_p(n_0^{-1/2}).
$$

The standard GMM expansion gives
$$
\sqrt{n_0}
(\hat\beta_{\mathrm{GMM}}-\beta_0)=
-J^{-1}D_\beta^\top\Gamma^{-1}
\left[
\sqrt{n_0}
\bar{\mathbf g}_n(\beta_0,\theta^{(0)})
+
D_\theta\sqrt{n_0}
(\tilde\theta_{RE}-\theta^{(0)})
\right]
+o_p(1)
$$

Therefore, by the continuous mapping
theorem and Slutsky's theorem,
$$
\boxed{
\sqrt{n_0}
(\hat\beta_{\mathrm{GMM}}-\beta_0)
\xrightarrow{d}\mathbb B,
}
$$
where
$$
\mathbb B
=
-J^{-1}D_\beta^\top\Gamma^{-1}(\mathbb G+D_\theta\mathbb T).
$$

This establishes the first conclusion.

\medskip
\noindent
\textbf{Step 2: First-order expansion
of the bootstrap GMM estimator.}

Since the internal observations are
resampled jointly,
$$
\sqrt{n_0}
\left\{
\bar{\mathbf g}_n^*(\beta_0,\theta^{(0)})
-
\bar{\mathbf g}_n(\beta_0,\theta^{(0)})
\right\}=
\frac1{\sqrt{n_0}}
\sum_{i=1}^{n_0}(N_i-1)g_i.
$$

By Assumption~\ref{assump:boot}(ii),
$$
\sqrt{n_0}
\left\{
\tilde\theta_{RE}^*
-
\E^*(\tilde\theta_{RE}^*)
\right\}=
\frac1{\sqrt{n_0}}
\sum_{i=1}^{n_0}
(N_i-1)\hat\ell_{i,n}
+
\sum_{k=1}^K\xi_{k,n}^*
+
r_n^*
$$
where
$$
\E^*\|r_n^*\|^2\xrightarrow{p}0.
$$

Furthermore,
$$
\sqrt{n_0}
\left\|
\E^*(\tilde\theta_{RE}^*)
-
\tilde\theta_{RE}
\right\|
=O_p(1).
$$

By Assumption~\ref{assump:boot}(ii),
$$
\sqrt{n_0}
\{\tilde\theta_{RE}^*-\E^*(\tilde\theta_{RE}^*)\}
=
\frac1{\sqrt{n_0}}
\sum_{i=1}^{n_0}(N_i-1)\hat\ell_{i,n}
+
\sum_{k=1}^K\xi_{k,n}^*
+
r_n^*
$$

For the internal bootstrap contribution,
the multinomial covariance identity gives
$$
\begin{aligned}
\E^*
\left\|
\frac1{\sqrt{n_0}}
\sum_{i=1}^{n_0}
(N_i-1)\hat\ell_{i,n}
\right\|^2&=
\frac1{n_0}
\sum_{i=1}^{n_0}
\|\hat\ell_{i,n}-\bar{\hat\ell}_n\|^2\\
&\leq
\frac1{n_0}
\sum_{i=1}^{n_0}\|\hat\ell_{i,n}\|^2
=O_p(1),
\end{aligned}
$$
where the last equality follows from
Assumption~\ref{assump:boot}(i).

For the external contribution, the
conditional moment bound in
Assumption~\ref{assump:boot}(ii) implies
$$
\begin{aligned}
\E^*
\left\|
\sum_{k=1}^K\xi_{k,n}^*
\right\|^2
&\leq
\left\{
\E^*
\left\|
\sum_{k=1}^K\xi_{k,n}^*
\right\|^{2+\delta}
\right\}^{2/(2+\delta)}\\
&=O_p(1).
\end{aligned}
$$

Together with
$\E^*\|r_n^*\|^2\xrightarrow{p}0$,
these bounds yield
$$
\begin{aligned}
&\E^*
\left\|
\sqrt{n_0}
\{\tilde\theta_{RE}^*-\E^*(\tilde\theta_{RE}^*)\}
\right\|^2=O_p(1).
\end{aligned}
$$

Thus, by conditional Markov's inequality,
$$
\sqrt{n_0}
\{\tilde\theta_{RE}^*
-\E^*(\tilde\theta_{RE}^*)\}
=O_{p^*}(1)
$$
conditionally in probability.

Furthermore, Assumption~\ref{assump:boot}(ii)
gives
$$
\sqrt{n_0}
\left\|
\E^*(\tilde\theta_{RE}^*)
-\tilde\theta_{RE}
\right\|
=O_p(1).
$$

Consequently,
$$
\boxed{
\sqrt{n_0}
(\tilde\theta_{RE}^*-\tilde\theta_{RE})
=O_{p^*}(1)
}
$$
conditionally in probability.

The bootstrap GMM expansion around
the original estimator gives
$$
\sqrt{n_0}
(\hat\beta_{\mathrm{GMM}}^*
-\hat\beta_{\mathrm{GMM}})=
-J^{-1}D_\beta^\top\Gamma^{-1}
\left[
\frac1{\sqrt{n_0}}
\sum_{i=1}^{n_0}
(N_i-1)g_i
+
D_\theta\sqrt{n_0}
(\tilde\theta_{RE}^*-\tilde\theta_{RE})
\right]
+
R_n^*
$$
where
$$
R_n^*=o_{p^*}(1).
$$

We next establish that this remainder
is negligible in conditional $L^2$.

By Assumption~\ref{assump:boot}(iv),
$$
n_0
\|\hat\beta_{\mathrm{GMM}}^*
-\hat\beta_{\mathrm{GMM}}\|^2
$$
is conditionally asymptotically uniformly
integrable in probability.

The conditional $(2+\delta)$-moment
bounds for the internal influence values
and aggregate external contributions,
together with the multinomial bootstrap
moment inequality, imply that the
leading term in the bootstrap GMM
expansion is also conditionally
asymptotically uniformly integrable
in squared norm.

It follows that $\|R_n^*\|^2$ is
conditionally asymptotically uniformly
integrable. Combining this property with
$R_n^*=o_{p^*}(1)$ yields
$$
\boxed{
\E^*\|R_n^*\|^2\xrightarrow{p}0.
}
$$

Taking conditional expectations of
the bootstrap GMM expansion and using
$$
\E^*(N_i-1)=0
$$
gives
$$
\sqrt{n_0}
\left\{
\E^*(\hat\beta_{\mathrm{GMM}}^*)
-
\hat\beta_{\mathrm{GMM}}
\right\}=
-J^{-1}D_\beta^\top\Gamma^{-1}D_\theta\sqrt{n_0}
\left\{
\E^*(\tilde\theta_{RE}^*)
-
\tilde\theta_{RE}
\right\}
+
\E^*(R_n^*)
$$

Subtracting this expression from
the bootstrap GMM expansion yields
$$
\begin{aligned}
\sqrt{n_0}
\left\{
\hat\beta_{\mathrm{GMM}}^*
-
\E^*(\hat\beta_{\mathrm{GMM}}^*)
\right\}&=
-J^{-1}D_\beta^\top\Gamma^{-1}
\left[
\frac1{\sqrt{n_0}}
\sum_{i=1}^{n_0}(N_i-1)g_i
+
D_\theta\sqrt{n_0}
\left\{
\tilde\theta_{RE}^*
-
\E^*(\tilde\theta_{RE}^*)
\right\}
\right]\\
&\quad+
R_n^*-\E^*(R_n^*).
\end{aligned}
$$

Substituting the first-stage
bootstrap representation gives
$$
\begin{aligned}
&\sqrt{n_0}
\left\{
\hat\beta_{\mathrm{GMM}}^*
-
\E^*(\hat\beta_{\mathrm{GMM}}^*)
\right\}=
-J^{-1}D_\beta^\top\Gamma^{-1}
\left[
\frac1{\sqrt{n_0}}
\sum_{i=1}^{n_0}
(N_i-1)
\{g_i+D_\theta\hat\ell_{i,n}\}
+
D_\theta\sum_{k=1}^K\xi_{k,n}^*
\right]
+
\widetilde R_n^*,
\end{aligned}
$$
where
$$
\widetilde R_n^*
=
R_n^*-\E^*(R_n^*)
-
J^{-1}D_\beta^\top\Gamma^{-1}D_\theta r_n^*.
$$

By the conditional variance identity,
$$
\begin{aligned}
\E^*
\|R_n^*-\E^*(R_n^*)\|^2
&=
\E^*\|R_n^*\|^2
-
\|\E^*(R_n^*)\|^2\\
&\leq
\E^*\|R_n^*\|^2.
\end{aligned}
$$

Consequently,
$$
\begin{aligned}
\E^*\|\widetilde R_n^*\|^2
\leq{}&
2\E^*\|R_n^*\|^2+
2\|J^{-1}D_\beta^\top\Gamma^{-1}D_\theta\|^2
\E^*\|r_n^*\|^2\\
\E^*\|\widetilde R_n^*\|^2\xrightarrow{p}&0
\end{aligned}
$$

Thus,
$$
\boxed{
\widetilde R_n^*
=
o_{L^2(P^*)}(1).
}
$$

\medskip
\noindent
\textbf{Step 3: Conditional covariance
decomposition.}

Define
$$
u_{i,n}
=
g_i+D_\theta\ell_{i,n},\qquad
\hat u_{i,n}
=
g_i+D_\theta\hat\ell_{i,n},
\qquad
\bar{\hat u}_n
=
\frac1{n_0}
\sum_{i=1}^{n_0}\hat u_{i,n}.
$$

The centered bootstrap expansion from
Step~2 becomes
$$
\begin{aligned}
\sqrt{n_0}
\left\{
\hat\beta_{\mathrm{GMM}}^*
-
\E^*(\hat\beta_{\mathrm{GMM}}^*)
\right\}=
-J^{-1}D_\beta^\top\Gamma^{-1}
\left[
\frac1{\sqrt{n_0}}
\sum_{i=1}^{n_0}
(N_i-1)\hat u_{i,n}
+
D_\theta\sum_{k=1}^K\xi_{k,n}^*
\right]
+
\widetilde R_n^*.
\end{aligned}
$$

For the multinomial nonparametric
bootstrap,
$$
\begin{aligned}
&\operatorname{Var}^*
\left\{
\frac1{\sqrt{n_0}}
\sum_{i=1}^{n_0}
(N_i-1)\hat u_{i,n}
\right\}=
\frac1{n_0}
\sum_{i=1}^{n_0}
(\hat u_{i,n}-\bar{\hat u}_n)
(\hat u_{i,n}-\bar{\hat u}_n)^\top.
\end{aligned}
$$

Because the external bootstrap
contributions are conditionally
independent across studies and
independent of the internal resampling,
$$
\begin{aligned}
\operatorname{Var}^*
\left[
\frac1{\sqrt{n_0}}
\sum_{i=1}^{n_0}
(N_i-1)\hat u_{i,n}
+
D_\theta\sum_{k=1}^K\xi_{k,n}^*
\right]&=
\frac1{n_0}
\sum_{i=1}^{n_0}
(\hat u_{i,n}-\bar{\hat u}_n)
(\hat u_{i,n}-\bar{\hat u}_n)^\top\\
&\quad+
D_\theta
\left\{
\sum_{k=1}^K
\operatorname{Var}^*(\xi_{k,n}^*)
\right\}
D_\theta^\top.
\end{aligned}
$$

The leading term has bounded
conditional second moments in
probability. Since
$$
\E^*\|\widetilde R_n^*\|^2
\xrightarrow{p}0,
$$
the Cauchy--Schwarz inequality implies
that the covariance contributions
involving $\widetilde R_n^*$ are
$o_p(1)$.

Therefore,
\begin{equation}
\begin{aligned}
n_0\operatorname{Var}^*
(\hat\beta_{\mathrm{GMM}}^*)&=
J^{-1}D_\beta^\top\Gamma^{-1}
\Bigg[
\frac1{n_0}
\sum_{i=1}^{n_0}
(\hat u_{i,n}-\bar{\hat u}_n)
(\hat u_{i,n}-\bar{\hat u}_n)^\top\\
&\qquad+
D_\theta
\left\{
\sum_{k=1}^K
\operatorname{Var}^*(\xi_{k,n}^*)
\right\}
D_\theta^\top
\Bigg]
\Gamma^{-1}D_\beta J^{-1}+o_p(1).
\end{aligned}
\label{eq:boot_cov_decomp}
\end{equation}

\medskip
\noindent
\textbf{Step 4: Consistency of the internal
covariance contribution.}

Define
$$
V_{I,n}
=
\operatorname{Var}(u_{i,n}).
$$

Recall that
$$
u_{i,n}
=
g_i+D_\theta\ell_{i,n},
\qquad
\hat u_{i,n}
=
g_i+D_\theta\hat\ell_{i,n}.
$$

By Assumption~\ref{assump:boot}(i),
$$
\begin{aligned}
\frac1{n_0}
\sum_{i=1}^{n_0}
\|\hat u_{i,n}-u_{i,n}\|^2
&\leq
\|D_\theta\|^2
\frac1{n_0}
\sum_{i=1}^{n_0}
\|\hat\ell_{i,n}-\ell_{i,n}\|^2\\
&=o_p(1).
\end{aligned}
$$

The Cauchy--Schwarz inequality gives
$$
\begin{aligned}
\left\|
\frac1{n_0}
\sum_{i=1}^{n_0}
\hat u_{i,n}\hat u_{i,n}^\top
-
\frac1{n_0}
\sum_{i=1}^{n_0}
u_{i,n}u_{i,n}^\top
\right\|&\leq
\left\{
\frac1{n_0}
\sum_{i=1}^{n_0}
\|\hat u_{i,n}-u_{i,n}\|^2
\right\}^{1/2}
\left\{
\frac1{n_0}
\sum_{i=1}^{n_0}
\|\hat u_{i,n}+u_{i,n}\|^2
\right\}^{1/2}\\
&=o_p(1),
\end{aligned}
$$
where the second factor is $O_p(1)$
by the uniform moment bound and
empirical $L^2$ consistency.

Similarly,
$$
\begin{aligned}
\|\bar{\hat u}_n-\bar u_n\|^2
&\leq
\frac1{n_0}
\sum_{i=1}^{n_0}
\|\hat u_{i,n}-u_{i,n}\|^2\\
&=o_p(1),
\end{aligned}
$$
where
$$
\bar u_n
=
\frac1{n_0}
\sum_{i=1}^{n_0}u_{i,n},
\qquad
\bar{\hat u}_n
=
\frac1{n_0}
\sum_{i=1}^{n_0}\hat u_{i,n}.
$$

Since $\bar u_n=O_p(1)$, it follows that
$$
\bar{\hat u}_n\bar{\hat u}_n^\top
-
\bar u_n\bar u_n^\top
=o_p(1).
$$

Consequently,
$$
\frac1{n_0}
\sum_{i=1}^{n_0}
(\hat u_{i,n}-\bar{\hat u}_n)
(\hat u_{i,n}-\bar{\hat u}_n)^\top=
\frac1{n_0}
\sum_{i=1}^{n_0}
(u_{i,n}-\bar u_n)
(u_{i,n}-\bar u_n)^\top
+o_p(1)
$$

Since the internal observations are
independent and identically distributed
and $u_{i,n}$ has uniformly bounded
$(2+\delta)$-moments, the triangular-array
law of large numbers gives
$$
\begin{aligned}
&\frac1{n_0}
\sum_{i=1}^{n_0}
(u_{i,n}-\bar u_n)
(u_{i,n}-\bar u_n)^\top=
V_{I,n}+o_p(1).
\end{aligned}
$$

Combining these results yields
\begin{equation}
\boxed{
\frac1{n_0}
\sum_{i=1}^{n_0}
(\hat u_{i,n}-\bar{\hat u}_n)
(\hat u_{i,n}-\bar{\hat u}_n)^\top=
V_{I,n}+o_p(1)
}
\label{eq:boot_internal_cov}
\end{equation}

\medskip
\noindent
\textbf{Step 5: Consistency of the
external covariance contribution.}

By Assumption~\ref{assump:boot}(iii),
the original and bootstrap external
influence contributions have the
representations
$$
\begin{aligned}
\xi_{k,n}
={}&
h_{k,n}(\hat\theta^{(k)};v_n)
-
\E\{
h_{k,n}(\hat\theta^{(k)};v_n)
\},\\
\xi_{k,n}^*
={}&
h_{k,n}(\hat\theta^{(k)*};\hat v_n)
-
\E^*\{
h_{k,n}(\hat\theta^{(k)*};\hat v_n)
\}.
\end{aligned}
$$

Under the original Gaussian
random-effects model,
$$
\operatorname{Var}(\xi_{k,n})
=
\mathcal C_{k,n}(v_n).
$$

Under the fitted external parametric
bootstrap law,
$$
\operatorname{Var}^*(\xi_{k,n}^*)
=
\mathcal C_{k,n}(\hat v_n).
$$

The covariance-continuity condition
therefore gives
$$
\begin{aligned}
\left\|
\sum_{k=1}^K
\operatorname{Var}^*(\xi_{k,n}^*)
-
\sum_{k=1}^K
\operatorname{Var}(\xi_{k,n})
\right\|&=
\left\|
\sum_{k=1}^K
\{
\mathcal C_{k,n}(\hat v_n)
-
\mathcal C_{k,n}(v_n)
\}
\right\|\\
&\leq
\sum_{k=1}^K
\|
\mathcal C_{k,n}(\hat v_n)
-
\mathcal C_{k,n}(v_n)
\|\\
&\leq
\|\hat v_n-v_n\|
\sum_{k=1}^K L_{k,n}\\
&=o_p(1).
\end{aligned}
$$

Hence,
\begin{equation}
\boxed{
\sum_{k=1}^K
\operatorname{Var}^*(\xi_{k,n}^*)
=
\sum_{k=1}^K
\operatorname{Var}(\xi_{k,n})
+o_p(1).
}
\label{eq:boot_external_cov}
\end{equation}

\medskip
\noindent
\textbf{Step 6: Identification of the
limiting covariance.}

By the original first-stage
representation in
Assumption~\ref{assump:boot}(ii),
$$
\begin{aligned}
\sqrt{n_0}
\bar{\mathbf g}_n(\beta_0,\theta^{(0)})
+
D_\theta\sqrt{n_0}
(\tilde\theta_{RE}-\theta^{(0)})=
\frac1{\sqrt{n_0}}
\sum_{i=1}^{n_0}u_{i,n}
+
D_\theta\sum_{k=1}^K\xi_{k,n}
+
D_\theta r_n.
\end{aligned}
$$

Define
$$
V_{E,n}
=
D_\theta
\left\{
\sum_{k=1}^K
\operatorname{Var}(\xi_{k,n})
\right\}
D_\theta^\top.
$$

The internal and external leading
contributions are mean-zero and
independent.

Moreover,
$$
\E\|r_n\|^2\to0,
$$
and their second moments are
uniformly bounded by
Assumptions~\ref{assump:boot}(i)--(ii).

Consequently,
$$
\operatorname{Var}
\left\{
\sqrt{n_0}
\bar{\mathbf g}_n(\beta_0,\theta^{(0)})
+
D_\theta\sqrt{n_0}
(\tilde\theta_{RE}-\theta^{(0)})
\right\}=
V_{I,n}+V_{E,n}+o(1)
$$

The uniformly bounded
$(2+\delta)$-moments of the leading
contributions, together with the
$L^2$-negligibility of $r_n$, imply
uniform integrability of the squared
norm of
$$
\sqrt{n_0}
\bar{\mathbf g}_n(\beta_0,\theta^{(0)})
+
D_\theta\sqrt{n_0}
(\tilde\theta_{RE}-\theta^{(0)}).
$$

Under the joint convergence regime considered in Theorem~4,
$$
\sqrt{n_0}
\bar{\mathbf g}_n(\beta_0,\theta^{(0)})
+
D_\theta\sqrt{n_0}
(\tilde\theta_{RE}-\theta^{(0)})\xrightarrow{d}
\mathbb G+D_\theta\mathbb T.
$$

Weak convergence and uniform
integrability therefore yield
\begin{equation}
\boxed{
V_{I,n}+V_{E,n}
\longrightarrow
\operatorname{Var}
(\mathbb G+D_\theta\mathbb T).
}
\label{eq:boot_lim_cov}
\end{equation}

Importantly, the internal covariance
contribution $V_{I,n}$ incorporates
the dependence between the internal
estimating equations and the internal
component of the meta estimator.

Thus, the limiting covariance includes
the cross-covariance between
$\mathbb G$ and $\mathbb T$.

Finally, substituting
\eqref{eq:boot_internal_cov}
and \eqref{eq:boot_external_cov}
into \eqref{eq:boot_cov_decomp}
gives
$$
\begin{aligned}
n_0\operatorname{Var}^*
(\hat\beta_{\mathrm{GMM}}^*)=
J^{-1}D_\beta^\top\Gamma^{-1}(V_{I,n}+V_{E,n})\Gamma^{-1}D_\beta J^{-1}
+o_p(1).
\end{aligned}
$$

Applying \eqref{eq:boot_lim_cov},
we obtain
$$
\begin{aligned}
n_0\operatorname{Var}^*
(\hat\beta_{\mathrm{GMM}}^*)&\xrightarrow{p}
J^{-1}D_\beta^\top\Gamma^{-1}\operatorname{Var}
(\mathbb G+D_\theta\mathbb T)\Gamma^{-1}D_\beta J^{-1}\\
&=
\operatorname{Var}
\{-J^{-1}D_\beta^\top\Gamma^{-1}(\mathbb G+D_\theta\mathbb T)\}\\
&=
\operatorname{Var}(\mathbb B).
\end{aligned}
$$

Therefore,
$$
\boxed{
n_0\operatorname{Var}^*
(\hat\beta_{\mathrm{GMM}}^*)
\xrightarrow{p}
\operatorname{Var}(\mathbb B).
}
$$

This completes the proof.
\hfill$\square$

\newpage
\subsection{Proof of Corollary~2}
\label{sec:proof_cor2}
% =================================================

\noindent\textbf{Restatement of Corollary~2.}
Under the assumptions of Theorem~4, suppose that
$(\mathbb G,\mathbb T)$ has finite second moments.
The proposed estimator is asymptotically at least as
efficient as the internal-study-only maximum
likelihood estimator if and only if
$$
\mathcal H_T
\preceq
\mathcal G.
$$

\noindent\textbf{Proof.}
By Theorem~4,
$$
\sqrt{n_0}
(\hat\beta_{\mathrm{GMM}}-\beta_0)
\toD
-J^{-1}D_\beta^\top\Gamma^{-1}
\left(
\mathbb G+D_\theta\mathbb T
\right).
$$

Since $\operatorname{Var}(\mathbb G)=\Gamma$,
the asymptotic covariance matrix of the
proposed estimator is
$$
\begin{aligned}
\Sigma_{\mathrm{GMM},T}
={}&
J^{-1}D_\beta^\top\Gamma^{-1}
\operatorname{Var}
\left(
\mathbb G+D_\theta\mathbb T
\right)
\Gamma^{-1}D_\beta J^{-1}\\
={}&
J^{-1}D_\beta^\top\Gamma^{-1}
(\Gamma+\Delta_T)
\Gamma^{-1}D_\beta J^{-1},
\end{aligned}
$$
where, as defined in the manuscript,
$$
\Delta_T
=
D_\theta V_TD_\theta^\top
+
C_{GT}D_\theta^\top
+
D_\theta C_{GT}^\top,
$$
with
$$
V_T
=
\operatorname{Var}(\mathbb T),
\qquad
C_{GT}
=
\operatorname{Cov}(\mathbb G,\mathbb T).
$$

Expanding the covariance expression gives
$$
\Sigma_{\mathrm{GMM},T}
=
J^{-1}D_\beta^\top\Gamma^{-1}
D_\beta J^{-1}+
J^{-1}D_\beta^\top\Gamma^{-1}
\Delta_T
\Gamma^{-1}D_\beta J^{-1}.
$$

Since $J=D_\beta^\top\Gamma^{-1}D_\beta$,
the first term satisfies
$$
J^{-1}D_\beta^\top\Gamma^{-1}
D_\beta J^{-1}
=
J^{-1}
=
\Sigma_0.
$$

Moreover, by the definition of
$\mathcal H_T$ in the manuscript,
$$
\mathcal H_T
=
J^{-1}D_\beta^\top\Gamma^{-1}
\Delta_T
\Gamma^{-1}D_\beta J^{-1}.
$$

Therefore,
$$
\boxed{
\Sigma_{\mathrm{GMM},T}
=
\Sigma_0+\mathcal H_T.
}
$$

Recall that
$$
\mathcal G
=
\Sigma_{\mathrm{MLE}}-\Sigma_0.
$$

It follows that
$$
\begin{aligned}
\Sigma_{\mathrm{MLE}}
-
\Sigma_{\mathrm{GMM},T}
&=
\Sigma_{\mathrm{MLE}}
-
\Sigma_0
-
\mathcal H_T\\
&=
\mathcal G-\mathcal H_T.
\end{aligned}
$$

Consequently,
$$
\Sigma_{\mathrm{GMM},T}
\preceq
\Sigma_{\mathrm{MLE}}
$$
if and only if
$$
\boxed{
\mathcal H_T
\preceq
\mathcal G.
}
$$
This completes the proof.
\hfill$\square$

% =================================================
% =================================================

% =================================================
\newpage
\subsection{Additional empirical results}
\label{sec:additional_empirical_results}
% =================================================

This section reports additional simulation results and supplementary results from the data application.

\begin{table}[htbp]
\centering
\caption{Simulation results for $\theta$ with $\tau^2 = 0.03$ and varying numbers of external studies.}
\label{tab:compare_meta_models_varying_K}
\scriptsize
\setlength{\tabcolsep}{1.5pt}
\renewcommand{\arraystretch}{0.95}
\begin{adjustbox}{width=\textwidth}
\begin{tabular}{llcccccccccccccccccc}
\toprule
 &  & \multicolumn{2}{c}{MLE(Int)} & \multicolumn{4}{c}{FE} & \multicolumn{4}{c}{$\text{wFE}$} & \multicolumn{4}{c}{RE} & \multicolumn{4}{c}{$\text{wRE}$} \\
\cmidrule(lr){3-4} \cmidrule(lr){5-8} \cmidrule(lr){9-12} \cmidrule(lr){13-16} \cmidrule(lr){17-20}
$K$ & Parameter & Bias & ESD & Bias & ESD & SE & CP & Bias & ESD & SE & CP & Bias & ESD & SE & CP & Bias & ESD & SE & CP \\
\midrule
\multirow{6}{*}{3} & $\theta_{11}$ & 0.08 & 14.99 & 1.13 & 9.82 & 6.36 & 78.85 & 0.89 & 9.38 & 6.60 & 82.90 & 0.01 & 10.34 & 10.65 & 95.46 & 0.03 & 11.27 & 11.48 & 95.66 \\
 & $\theta_{21}$ & 0.35 & 18.06 & 0.73 & 10.91 & 7.74 & 83.40 & 0.59 & 10.46 & 8.01 & 86.15 & 0.38 & 11.67 & 12.59 & 96.33 & 0.43 & 13.06 & 13.72 & 96.01 \\
 & $\theta_{21} - \theta_{11}$ & 0.43 & 19.39 & 0.40 & 11.53 & 8.30 & 84.65 & 0.30 & 11.36 & 8.60 & 86.30 & 0.37 & 13.34 & 14.20 & 96.84 & 0.39 & 14.64 & 15.25 & 96.32 \\
 & $\theta_{12}$ & 0.25 & 15.64 & 0.90 & 9.87 & 6.70 & 81.70 & 0.90 & 9.28 & 6.95 & 85.70 & 0.38 & 10.21 & 11.02 & 96.84 & 0.29 & 11.27 & 11.96 & 96.42 \\
 & $\theta_{22}$ & 0.55 & 17.96 & 0.66 & 11.30 & 7.73 & 81.75 & 0.72 & 10.87 & 8.01 & 85.00 & 0.14 & 11.88 & 12.61 & 95.92 & 0.07 & 13.07 & 13.74 & 95.71 \\
 & $\theta_{22} - \theta_{12}$ & 0.31 & 19.68 & 1.56 & 11.83 & 8.47 & 82.60 & 1.61 & 11.63 & 8.77 & 84.60 & 0.52 & 13.85 & 14.44 & 95.92 & 0.37 & 14.91 & 15.53 & 96.32 \\
\midrule
\multirow{6}{*}{5} & $\theta_{11}$ & 0.08 & 14.99 & 1.46 & 8.06 & 5.11 & 77.80 & 1.15 & 7.77 & 5.33 & 81.55 & 0.09 & 8.01 & 8.29 & 95.30 & 0.05 & 8.66 & 9.39 & 96.13 \\
 & $\theta_{21}$ & 0.35 & 18.06 & 0.91 & 9.07 & 6.21 & 82.25 & 0.77 & 8.75 & 6.45 & 85.40 & 0.24 & 9.12 & 9.67 & 95.30 & 0.29 & 9.90 & 11.08 & 96.13 \\
 & $\theta_{21} - \theta_{11}$ & 0.43 & 19.39 & 0.56 & 9.54 & 6.68 & 83.60 & 0.39 & 9.44 & 6.93 & 84.70 & 0.15 & 10.22 & 10.99 & 95.95 & 0.24 & 11.07 & 12.45 & 96.98 \\
 & $\theta_{12}$ & 0.25 & 15.64 & 0.70 & 8.05 & 5.39 & 81.00 & 0.71 & 7.64 & 5.61 & 85.60 & 0.20 & 7.82 & 8.47 & 96.00 & 0.21 & 8.48 & 9.70 & 97.13 \\
 & $\theta_{22}$ & 0.55 & 17.96 & 1.03 & 8.96 & 6.20 & 81.80 & 1.04 & 8.70 & 6.44 & 84.35 & 0.19 & 8.81 & 9.67 & 95.60 & 0.12 & 9.55 & 11.10 & 96.93 \\
 & $\theta_{22} - \theta_{12}$ & 0.31 & 19.68 & 1.73 & 9.60 & 6.79 & 83.10 & 1.75 & 9.47 & 7.05 & 84.75 & 0.40 & 10.34 & 11.16 & 94.84 & 0.33 & 11.23 & 12.68 & 96.63 \\
\midrule
\multirow{6}{*}{7} & $\theta_{11}$ & 0.08 & 14.99 & 1.57 & 6.88 & 4.40 & 78.90 & 1.25 & 6.65 & 4.60 & 81.80 & 0.12 & 6.76 & 6.68 & 93.60 & 0.10 & 7.20 & 7.70 & 95.30 \\
 & $\theta_{21}$ & 0.35 & 18.06 & 0.90 & 7.80 & 5.33 & 82.00 & 0.70 & 7.55 & 5.54 & 84.20 & 0.11 & 7.72 & 7.68 & 94.45 & 0.11 & 8.12 & 8.93 & 96.15 \\
 & $\theta_{21} - \theta_{11}$ & 0.43 & 19.39 & 0.67 & 8.26 & 5.73 & 83.50 & 0.55 & 8.20 & 5.96 & 85.80 & 0.01 & 8.66 & 8.80 & 94.30 & 0.01 & 9.11 & 10.18 & 96.30 \\
 & $\theta_{12}$ & 0.25 & 15.64 & 0.67 & 7.14 & 4.63 & 78.70 & 0.67 & 6.86 & 4.83 & 83.30 & 0.14 & 6.77 & 6.76 & 93.95 & 0.18 & 7.12 & 7.85 & 95.90 \\
 & $\theta_{22}$ & 0.55 & 17.96 & 1.03 & 7.74 & 5.33 & 81.15 & 1.01 & 7.51 & 5.55 & 84.15 & 0.10 & 7.37 & 7.71 & 95.00 & 0.02 & 7.83 & 8.95 & 96.50 \\
 & $\theta_{22} - \theta_{12}$ & 0.31 & 19.68 & 1.70 & 8.42 & 5.84 & 81.65 & 1.68 & 8.34 & 6.08 & 82.90 & 0.24 & 8.68 & 8.94 & 94.25 & 0.20 & 9.18 & 10.33 & 96.15 \\
\midrule
\multirow{6}{*}{10} & $\theta_{11}$ & 0.08 & 14.99 & 1.65 & 5.92 & 3.73 & 77.35 & 1.31 & 5.71 & 3.91 & 81.55 & 0.04 & 5.63 & 5.39 & 93.60 & 0.02 & 5.88 & 6.02 & 94.95 \\
 & $\theta_{21}$ & 0.35 & 18.06 & 0.97 & 6.58 & 4.53 & 82.75 & 0.73 & 6.35 & 4.71 & 85.15 & 0.05 & 6.45 & 6.20 & 93.25 & 0.08 & 6.71 & 6.94 & 95.00 \\
 & $\theta_{21} - \theta_{11}$ & 0.43 & 19.39 & 0.68 & 6.95 & 4.86 & 83.60 & 0.58 & 6.92 & 5.07 & 85.05 & 0.00 & 7.18 & 7.08 & 94.25 & 0.06 & 7.49 & 7.96 & 95.80 \\
 & $\theta_{12}$ & 0.25 & 15.64 & 0.77 & 6.20 & 3.93 & 78.75 & 0.77 & 5.93 & 4.11 & 82.65 & 0.13 & 5.68 & 5.44 & 92.85 & 0.18 & 5.84 & 6.09 & 94.90 \\
 & $\theta_{22}$ & 0.55 & 17.96 & 0.96 & 6.65 & 4.52 & 81.35 & 0.97 & 6.47 & 4.71 & 83.85 & 0.02 & 6.26 & 6.22 & 94.10 & 0.08 & 6.52 & 6.95 & 95.00 \\
 & $\theta_{22} - \theta_{12}$ & 0.31 & 19.68 & 1.73 & 7.26 & 4.95 & 79.55 & 1.74 & 7.23 & 5.16 & 82.15 & 0.11 & 7.28 & 7.17 & 92.70 & 0.10 & 7.54 & 8.04 & 94.90 \\
\bottomrule
\end{tabular}
\end{adjustbox}
\vspace{2pt}
\begin{minipage}{\textwidth}\footnotesize
Note: $K$ denotes the number of external studies. The between-study heterogeneity is fixed at $\tau^2 = 0.03$. Bias (x100) is the mean estimation error multiplied by 100; ESD (x100) is the empirical standard deviation multiplied by 100; SE (x100) is the mean estimated standard error multiplied by 100; CP is the coverage probability of the 95\% Wald confidence interval (\%). MLE(Int) is based only on the internal sample.
\end{minipage}
\end{table}

\newpage

\begin{table}[htbp]
\centering
\caption{Simulation results for $\theta$ with $K=15$, $\tau^2=0.03$, and varying external study sample size $n_k$.}
\label{tab:compare_meta_models_varying_nk}
\scriptsize
\setlength{\tabcolsep}{1.5pt}
\renewcommand{\arraystretch}{0.95}
\begin{adjustbox}{width=\textwidth}
\begin{tabular}{llcccccccccccccccccc}
\toprule
 &  & \multicolumn{2}{c}{MLE(Int)} & \multicolumn{4}{c}{FE} & \multicolumn{4}{c}{$\text{wFE}$} & \multicolumn{4}{c}{RE} & \multicolumn{4}{c}{$\text{wRE}$} \\
\cmidrule(lr){3-4} \cmidrule(lr){5-8} \cmidrule(lr){9-12} \cmidrule(lr){13-16} \cmidrule(lr){17-20}
$n_k$ & Parameter & Bias & ESD & Bias & ESD & SE & CP & Bias & ESD & SE & CP & Bias & ESD & SE & CP & Bias & ESD & SE & CP \\
\midrule
\multirow{6}{*}{100} & $\theta_{11}$ & 0.08 & 14.99 & 2.37 & 9.51 & 9.49 & 94.70 & 2.34 & 9.52 & 9.55 & 94.95 & 1.08 & 9.98 & 10.59 & 96.70 & 1.05 & 10.02 & 10.68 & 96.80 \\
 & $\theta_{21}$ & 0.35 & 18.06 & 2.14 & 10.96 & 11.47 & 95.50 & 2.13 & 10.96 & 11.54 & 95.55 & 1.40 & 11.63 & 12.79 & 96.55 & 1.38 & 11.67 & 12.87 & 96.80 \\
 & $\theta_{21} - \theta_{11}$ & 0.43 & 19.39 & 0.23 & 12.38 & 12.32 & 94.60 & 0.21 & 12.37 & 12.40 & 94.80 & 0.32 & 13.06 & 13.94 & 96.20 & 0.33 & 13.09 & 14.04 & 96.40 \\
 & $\theta_{12}$ & 0.25 & 15.64 & 2.50 & 9.62 & 9.98 & 94.80 & 2.50 & 9.64 & 10.04 & 95.05 & 1.67 & 10.13 & 11.13 & 96.40 & 1.63 & 10.17 & 11.21 & 96.75 \\
 & $\theta_{22}$ & 0.55 & 17.96 & 1.56 & 11.25 & 11.47 & 95.25 & 1.53 & 11.24 & 11.54 & 95.55 & 0.37 & 11.88 & 12.84 & 96.55 & 0.34 & 11.93 & 12.95 & 96.45 \\
 & $\theta_{22} - \theta_{12}$ & 0.31 & 19.68 & 4.06 & 12.31 & 12.54 & 94.50 & 4.03 & 12.32 & 12.62 & 94.70 & 2.04 & 13.07 & 14.23 & 96.70 & 1.97 & 13.12 & 14.34 & 96.60 \\
\midrule
\multirow{6}{*}{300} & $\theta_{11}$ & 0.08 & 14.99 & 2.01 & 6.71 & 6.37 & 92.65 & 1.87 & 6.74 & 6.48 & 93.05 & 0.48 & 7.02 & 7.52 & 96.40 & 0.46 & 7.10 & 7.70 & 96.35 \\
 & $\theta_{21}$ & 0.35 & 18.06 & 1.67 & 8.24 & 7.72 & 92.25 & 1.61 & 8.23 & 7.85 & 92.30 & 0.88 & 8.35 & 9.00 & 96.15 & 0.89 & 8.42 & 9.21 & 96.15 \\
 & $\theta_{21} - \theta_{11}$ & 0.43 & 19.39 & 0.33 & 8.74 & 8.30 & 93.65 & 0.26 & 8.76 & 8.43 & 94.20 & 0.40 & 9.10 & 9.86 & 96.60 & 0.43 & 9.19 & 10.12 & 96.85 \\
 & $\theta_{12}$ & 0.25 & 15.64 & 1.19 & 7.25 & 6.71 & 92.55 & 1.17 & 7.22 & 6.83 & 92.70 & 0.52 & 7.40 & 7.87 & 95.75 & 0.50 & 7.50 & 8.05 & 95.90 \\
 & $\theta_{22}$ & 0.55 & 17.96 & 1.63 & 8.41 & 7.72 & 91.70 & 1.61 & 8.40 & 7.85 & 92.15 & 0.64 & 8.55 & 9.00 & 95.50 & 0.60 & 8.65 & 9.22 & 96.30 \\
 & $\theta_{22} - \theta_{12}$ & 0.31 & 19.68 & 2.83 & 9.06 & 8.45 & 92.15 & 2.79 & 9.03 & 8.59 & 92.70 & 1.17 & 9.36 & 10.07 & 96.20 & 1.10 & 9.47 & 10.31 & 96.50 \\
\midrule
\multirow{6}{*}{500} & $\theta_{11}$ & 0.08 & 14.99 & 1.82 & 6.15 & 5.12 & 88.55 & 1.59 & 6.11 & 5.26 & 90.10 & 0.11 & 6.10 & 6.31 & 95.55 & 0.08 & 6.21 & 6.53 & 95.60 \\
 & $\theta_{21}$ & 0.35 & 18.06 & 1.44 & 7.25 & 6.21 & 90.95 & 1.27 & 7.22 & 6.37 & 92.00 & 0.46 & 7.31 & 7.45 & 95.25 & 0.45 & 7.44 & 7.71 & 95.90 \\
 & $\theta_{21} - \theta_{11}$ & 0.43 & 19.39 & 0.38 & 7.55 & 6.67 & 91.90 & 0.32 & 7.52 & 6.84 & 92.90 & 0.35 & 7.68 & 8.22 & 96.75 & 0.37 & 7.81 & 8.52 & 97.00 \\
 & $\theta_{12}$ & 0.25 & 15.64 & 1.22 & 6.33 & 5.39 & 89.80 & 1.19 & 6.23 & 5.53 & 91.45 & 0.56 & 6.27 & 6.52 & 95.85 & 0.55 & 6.37 & 6.74 & 95.90 \\
 & $\theta_{22}$ & 0.55 & 17.96 & 0.89 & 7.15 & 6.20 & 91.40 & 0.86 & 7.08 & 6.35 & 92.40 & 0.05 & 7.20 & 7.43 & 95.05 & 0.11 & 7.32 & 7.69 & 95.50 \\
 & $\theta_{22} - \theta_{12}$ & 0.31 & 19.68 & 2.12 & 7.58 & 6.79 & 91.05 & 2.05 & 7.52 & 6.95 & 92.40 & 0.52 & 7.77 & 8.36 & 96.80 & 0.44 & 7.88 & 8.65 & 97.05 \\
\midrule
\multirow{6}{*}{1000} & $\theta_{11}$ & 0.08 & 14.99 & 1.93 & 5.44 & 3.73 & 79.45 & 1.61 & 5.30 & 3.88 & 83.25 & 0.27 & 5.16 & 4.96 & 93.60 & 0.25 & 5.26 & 5.21 & 94.60 \\
 & $\theta_{21}$ & 0.35 & 18.06 & 1.40 & 6.06 & 4.53 & 85.05 & 1.18 & 5.93 & 4.69 & 87.45 & 0.47 & 5.77 & 5.78 & 94.25 & 0.51 & 5.92 & 6.05 & 94.65 \\
 & $\theta_{21} - \theta_{11}$ & 0.43 & 19.39 & 0.53 & 6.28 & 4.86 & 86.70 & 0.44 & 6.22 & 5.03 & 88.85 & 0.19 & 6.25 & 6.45 & 95.20 & 0.26 & 6.40 & 6.80 & 95.30 \\
 & $\theta_{12}$ & 0.25 & 15.64 & 0.88 & 5.34 & 3.93 & 83.95 & 0.83 & 5.23 & 4.09 & 86.30 & 0.28 & 5.18 & 5.13 & 94.55 & 0.27 & 5.32 & 5.36 & 95.15 \\
 & $\theta_{22}$ & 0.55 & 17.96 & 1.15 & 6.07 & 4.51 & 84.95 & 1.15 & 5.94 & 4.68 & 87.65 & 0.18 & 5.87 & 5.78 & 94.65 & 0.15 & 6.03 & 6.05 & 94.60 \\
 & $\theta_{22} - \theta_{12}$ & 0.31 & 19.68 & 2.03 & 6.66 & 4.94 & 82.50 & 1.98 & 6.66 & 5.12 & 84.45 & 0.47 & 6.76 & 6.57 & 94.25 & 0.42 & 6.92 & 6.91 & 94.60 \\
\bottomrule
\end{tabular}
\end{adjustbox}
\vspace{2pt}
\begin{minipage}{\textwidth}\footnotesize
Note: $n_k$ denotes the sample size of each external study, with $K=15$ external studies. The between-study heterogeneity is fixed at $\tau^2=0.03$. Bias ($\times 100$) is the mean estimation error multiplied by 100; ESD ($\times 100$) is the empirical standard deviation multiplied by 100; SE ($\times 100$) is the mean estimated standard error multiplied by 100; CP is the coverage probability of the 95\% Wald confidence interval (\%). MLE(Int) is based only on the internal sample.
\end{minipage}
\end{table}

\newpage

\begin{table}[htbp]
\centering
\scriptsize
\setlength{\tabcolsep}{1.8pt}
\renewcommand{\arraystretch}{0.90}
\caption{Simulation results for $\beta$ under no location shift with $\tau = 0.03$ and varying numbers of external studies.}
\label{tab:sim_results_all}
\begin{adjustbox}{max width=\textwidth}
\begin{tabular}{llcccccccccccccccccccccc}
\toprule
 &  & \multicolumn{2}{c}{MLE} & \multicolumn{4}{c}{EL} & \multicolumn{4}{c}{mGMM} & \multicolumn{4}{c}{$\text{mGMM}_{\text{W}}$} & \multicolumn{4}{c}{rGMM} & \multicolumn{4}{c}{$\text{rGMM}_{\text{W}}$} \\
\cmidrule(lr){3-4} \cmidrule(lr){5-8} \cmidrule(lr){9-12} \cmidrule(lr){13-16} \cmidrule(lr){17-20} \cmidrule(lr){21-24}
$K$ & Parameter & Bias & ESD & Bias & ESD & SE & CP & Bias & ESD & SE & CP & Bias & ESD & SE & CP & Bias & ESD & SE & CP & Bias & ESD & SE & CP \\
\midrule
\multirow{6}{*}{3} & $\beta_{11}$ & 1.37 & 15.35 & 3.02 & 11.84 & 9.24 & 86.85 & 0.61 & 13.74 & 13.77 & 95.16 & 0.70 & 12.90 & 12.98 & 95.05 & 0.26 & 13.77 & 13.79 & 94.99 & 0.60 & 12.94 & 13.01 & 94.91 \\
 & $\beta_{21}$ & 0.47 & 16.40 & 0.80 & 11.97 & 9.83 & 89.40 & 0.03 & 14.28 & 14.32 & 95.31 & 0.02 & 13.12 & 13.39 & 95.40 & 0.15 & 14.21 & 14.37 & 95.76 & 0.03 & 13.13 & 13.41 & 95.57 \\
 & $\beta_{21} - \beta_{11}$ & 0.49 & 18.71 & 1.58 & 16.31 & 15.66 & 94.20 & 0.58 & 17.76 & 17.86 & 94.75 & 0.71 & 17.39 & 17.50 & 95.25 & 0.59 & 17.63 & 17.90 & 94.99 & 0.59 & 17.45 & 17.52 & 94.86 \\
 & $\beta_{12}$ & 0.04 & 17.05 & 0.79 & 12.86 & 10.15 & 87.50 & 0.19 & 14.65 & 14.89 & 95.57 & 0.11 & 13.56 & 13.98 & 95.60 & 0.19 & 14.62 & 14.87 & 95.86 & 0.14 & 13.65 & 13.99 & 95.11 \\
 & $\beta_{22}$ & 0.88 & 16.25 & 0.26 & 11.99 & 9.62 & 87.65 & 1.45 & 14.23 & 14.24 & 94.80 & 0.72 & 12.96 & 13.27 & 95.85 & 1.40 & 14.23 & 14.31 & 95.35 & 0.51 & 12.96 & 13.32 & 95.83 \\
 & $\beta_{22} - \beta_{12}$ & 0.34 & 19.06 & 0.41 & 17.31 & 16.28 & 93.25 & 0.41 & 18.46 & 18.57 & 95.21 & 0.19 & 18.06 & 18.19 & 95.55 & 0.39 & 18.46 & 18.64 & 95.61 & 0.10 & 18.08 & 18.25 & 95.52 \\
\midrule
\multirow{6}{*}{5} & $\beta_{11}$ & 1.37 & 15.35 & 4.10 & 10.60 & 8.60 & 88.30 & 0.59 & 12.50 & 12.75 & 94.94 & 0.66 & 10.97 & 11.48 & 95.67 & 0.58 & 12.53 & 12.75 & 95.24 & 0.46 & 11.01 & 11.45 & 96.07 \\
 & $\beta_{21}$ & 0.47 & 16.40 & 1.25 & 10.80 & 9.23 & 90.15 & 0.11 & 12.98 & 13.27 & 95.55 & 0.01 & 11.15 & 11.83 & 95.41 & 0.02 & 13.07 & 13.28 & 96.15 & 0.03 & 11.20 & 11.86 & 95.56 \\
 & $\beta_{21} - \beta_{11}$ & 0.49 & 18.71 & 1.82 & 15.94 & 15.45 & 94.10 & 0.39 & 17.07 & 17.43 & 95.30 & 0.68 & 16.32 & 16.87 & 95.31 & 0.67 & 17.18 & 17.42 & 95.39 & 0.57 & 16.47 & 16.88 & 95.41 \\
 & $\beta_{12}$ & 0.04 & 17.05 & 0.46 & 11.39 & 9.61 & 90.35 & 0.10 & 13.19 & 13.83 & 96.16 & 0.11 & 11.67 & 12.39 & 96.12 & 0.06 & 13.24 & 13.82 & 95.79 & 0.03 & 11.68 & 12.41 & 96.17 \\
 & $\beta_{22}$ & 0.88 & 16.25 & 0.74 & 10.67 & 9.04 & 89.85 & 1.15 & 12.87 & 13.14 & 95.70 & 0.47 & 10.74 & 11.68 & 96.52 & 1.08 & 12.91 & 13.18 & 95.54 & 0.58 & 10.80 & 11.71 & 96.12 \\
 & $\beta_{22} - \beta_{12}$ & 0.34 & 19.06 & 0.56 & 16.70 & 16.09 & 93.85 & 0.36 & 17.77 & 18.10 & 94.84 & 0.10 & 17.12 & 17.52 & 95.21 & 0.43 & 17.84 & 18.11 & 94.93 & 0.23 & 17.18 & 17.53 & 95.56 \\
\midrule
\multirow{6}{*}{7} & $\beta_{11}$ & 1.37 & 15.35 & 4.44 & 9.70 & 8.26 & 89.90 & 0.84 & 11.59 & 11.96 & 96.11 & 0.65 & 9.88 & 10.28 & 95.51 & 0.69 & 11.61 & 11.95 & 95.90 & 0.38 & 10.03 & 10.26 & 95.11 \\
 & $\beta_{21}$ & 0.47 & 16.40 & 1.34 & 10.15 & 8.92 & 91.40 & 0.03 & 12.14 & 12.53 & 95.85 & 0.09 & 10.31 & 10.70 & 95.46 & 0.13 & 12.19 & 12.53 & 96.35 & 0.09 & 13.75 & 10.71 & 95.57 \\
 & $\beta_{21} - \beta_{11}$ & 0.49 & 18.71 & 1.98 & 15.69 & 15.34 & 94.40 & 0.96 & 16.69 & 17.07 & 95.60 & 0.88 & 16.00 & 16.42 & 95.51 & 1.00 & 16.77 & 17.08 & 95.49 & 0.22 & 19.25 & 16.42 & 95.21 \\
 & $\beta_{12}$ & 0.04 & 17.05 & 0.43 & 10.87 & 9.33 & 90.75 & 0.01 & 12.48 & 13.02 & 96.21 & 0.13 & 10.89 & 11.21 & 95.81 & 0.07 & 12.48 & 13.05 & 96.25 & 0.09 & 10.88 & 11.26 & 96.22 \\
 & $\beta_{22}$ & 0.88 & 16.25 & 0.78 & 10.03 & 8.73 & 90.70 & 1.08 & 12.05 & 12.34 & 95.45 & 0.70 & 9.97 & 10.48 & 95.91 & 1.07 & 11.94 & 12.39 & 95.80 & 0.82 & 9.99 & 10.49 & 95.62 \\
 & $\beta_{22} - \beta_{12}$ & 0.34 & 19.06 & 0.56 & 16.57 & 15.99 & 94.50 & 0.39 & 17.46 & 17.71 & 95.40 & 0.34 & 16.86 & 17.01 & 95.15 & 0.44 & 17.40 & 17.76 & 95.44 & 0.36 & 16.83 & 17.05 & 95.47 \\
\midrule
\multirow{6}{*}{10} & $\beta_{11}$ & 1.37 & 15.35 & 4.72 & 8.96 & 7.98 & 91.20 & 0.71 & 10.60 & 11.11 & 96.17 & 0.43 & 9.00 & 9.27 & 95.27 & 0.63 & 10.63 & 11.13 & 96.37 & 0.21 & 9.02 & 9.29 & 95.36 \\
 & $\beta_{21}$ & 0.47 & 16.40 & 1.47 & 9.57 & 8.65 & 92.55 & 0.12 & 11.22 & 11.69 & 96.07 & 0.23 & 9.70 & 9.75 & 94.51 & 0.10 & 11.42 & 11.72 & 96.12 & 0.20 & 9.69 & 9.78 & 94.65 \\
 & $\beta_{21} - \beta_{11}$ & 0.49 & 18.71 & 2.00 & 15.32 & 15.26 & 94.70 & 1.00 & 16.26 & 16.67 & 95.56 & 0.93 & 15.68 & 15.98 & 95.17 & 0.88 & 16.51 & 16.70 & 95.06 & 0.64 & 15.68 & 16.04 & 95.26 \\
 & $\beta_{12}$ & 0.04 & 17.05 & 0.60 & 10.21 & 9.08 & 91.40 & 0.02 & 11.62 & 12.22 & 96.67 & 0.07 & 10.15 & 10.29 & 95.32 & 0.16 & 11.61 & 12.24 & 96.37 & 0.19 & 10.14 & 10.30 & 94.86 \\
 & $\beta_{22}$ & 0.88 & 16.25 & 0.57 & 9.60 & 8.48 & 91.30 & 1.10 & 11.13 & 11.56 & 95.61 & 0.94 & 9.37 & 9.55 & 95.02 & 1.72 & 12.74 & 11.54 & 95.51 & 0.94 & 9.41 & 9.53 & 94.96 \\
 & $\beta_{22} - \beta_{12}$ & 0.34 & 19.06 & 0.59 & 16.33 & 15.91 & 94.50 & 0.39 & 17.10 & 17.38 & 95.06 & 0.39 & 16.52 & 16.68 & 94.86 & 0.73 & 18.49 & 17.35 & 95.26 & 0.47 & 16.55 & 16.66 & 95.36 \\
\bottomrule
\end{tabular}
\end{adjustbox}
\vspace{2pt}
\begin{minipage}{\textwidth}
\footnotesize
Note: Bias, relative bias ($\%$); ESD (x100), empirical standard deviation displayed after multiplying by 100; SE (x100), mean of bootstrap standard errors displayed after multiplying by 100; CP, coverage probability of 95\% confidence intervals ($\%$). MLE is based only on the internal sample, so its results are the same across values of $K$.
\end{minipage}
\end{table}

\newpage 

\begin{table}[htbp]
\centering
\scriptsize
\setlength{\tabcolsep}{1.8pt}
\renewcommand{\arraystretch}{0.90}
\caption{Simulation results for $\beta$ under no location shift with $\tau = 0.03$, $K = 15$, and varying external-study sample size $n_k$.}
\label{tab:sim_results_all}
\begin{adjustbox}{max width=\textwidth}
\begin{tabular}{llcccccccccccccccccccccc}
\toprule
 &  & \multicolumn{2}{c}{MLE} & \multicolumn{4}{c}{EL} & \multicolumn{4}{c}{mGMM} & \multicolumn{4}{c}{$\text{mGMM}_{\text{W}}$} & \multicolumn{4}{c}{rGMM} & \multicolumn{4}{c}{$\text{rGMM}_{\text{W}}$} \\
\cmidrule(lr){3-4} \cmidrule(lr){5-8} \cmidrule(lr){9-12} \cmidrule(lr){13-16} \cmidrule(lr){17-20} \cmidrule(lr){21-24}
$n_k$ & Parameter & Bias & ESD & Bias & ESD & SE & CP & Bias & ESD & SE & CP & Bias & ESD & SE & CP & Bias & ESD & SE & CP & Bias & ESD & SE & CP \\
\midrule
\multirow{6}{*}{100} & $\beta_{11}$ & 1.37 & 15.35 & 6.79 & 11.54 & 11.09 & 92.90 & 3.57 & 12.32 & 12.81 & 95.76 & 3.84 & 12.00 & 12.45 & 95.36 & 3.32 & 12.35 & 12.81 & 95.40 & 3.42 & 12.00 & 12.45 & 94.86 \\
 & $\beta_{21}$ & 0.47 & 16.40 & 2.83 & 11.72 & 11.55 & 94.85 & 1.81 & 12.54 & 13.33 & 96.07 & 1.86 & 12.12 & 12.99 & 96.27 & 1.64 & 12.57 & 13.36 & 95.96 & 1.68 & 12.13 & 12.98 & 96.12 \\
 & $\beta_{21} - \beta_{11}$ & 0.49 & 18.71 & 1.42 & 16.51 & 16.31 & 94.60 & 0.07 & 16.95 & 17.39 & 95.26 & 0.27 & 16.84 & 17.22 & 94.85 & 0.17 & 16.99 & 17.39 & 94.90 & 0.18 & 16.84 & 17.24 & 94.51 \\
 & $\beta_{12}$ & 0.04 & 17.05 & 3.16 & 12.07 & 11.55 & 93.90 & 2.26 & 12.89 & 13.87 & 96.07 & 2.41 & 12.48 & 13.50 & 96.01 & 2.08 & 12.89 & 13.91 & 96.11 & 2.25 & 12.52 & 13.55 & 96.27 \\
 & $\beta_{22}$ & 0.88 & 16.25 & 0.00 & 12.06 & 11.24 & 92.55 & 1.97 & 12.74 & 13.30 & 95.71 & 1.93 & 12.46 & 12.92 & 95.61 & 2.04 & 12.81 & 13.28 & 95.66 & 2.02 & 12.49 & 12.92 & 95.42 \\
 & $\beta_{22} - \beta_{12}$ & 0.34 & 19.06 & 2.00 & 17.11 & 16.86 & 94.80 & 0.71 & 17.52 & 18.14 & 95.36 & 0.82 & 17.40 & 17.96 & 95.66 & 0.58 & 17.56 & 18.13 & 95.81 & 0.69 & 17.44 & 17.96 & 95.47 \\
\midrule
\multirow{6}{*}{300} & $\beta_{11}$ & 1.37 & 15.35 & 5.78 & 9.53 & 9.25 & 94.25 & 1.85 & 10.67 & 11.37 & 96.21 & 1.82 & 9.87 & 10.37 & 95.76 & 1.69 & 10.69 & 11.32 & 96.32 & 1.74 & 9.84 & 10.35 & 95.76 \\
 & $\beta_{21}$ & 0.47 & 16.40 & 2.28 & 10.39 & 9.83 & 93.45 & 0.78 & 11.30 & 11.84 & 95.61 & 0.88 & 10.52 & 10.89 & 95.21 & 0.64 & 11.36 & 11.89 & 95.87 & 0.77 & 10.47 & 10.90 & 95.70 \\
 & $\beta_{21} - \beta_{11}$ & 0.49 & 18.71 & 1.48 & 15.84 & 15.66 & 94.55 & 0.36 & 16.45 & 16.77 & 95.51 & 0.12 & 16.17 & 16.40 & 95.01 & 0.50 & 16.49 & 16.78 & 95.02 & 0.26 & 16.06 & 16.41 & 95.20 \\
 & $\beta_{12}$ & 0.04 & 17.05 & 1.12 & 10.79 & 10.14 & 93.45 & 0.47 & 11.70 & 12.38 & 96.97 & 0.30 & 11.61 & 11.41 & 96.12 & 0.46 & 11.73 & 12.43 & 96.63 & 0.32 & 10.96 & 11.43 & 95.91 \\
 & $\beta_{22}$ & 0.88 & 16.25 & 1.51 & 10.39 & 9.62 & 92.70 & 0.87 & 11.28 & 11.73 & 95.46 & 0.26 & 10.56 & 10.73 & 95.16 & 0.78 & 11.27 & 11.75 & 95.57 & 0.32 & 10.47 & 10.76 & 95.20 \\
 & $\beta_{22} - \beta_{12}$ & 0.34 & 19.06 & 1.26 & 16.46 & 16.27 & 94.85 & 0.02 & 16.95 & 17.43 & 95.76 & 0.10 & 17.34 & 17.05 & 95.71 & 0.01 & 16.97 & 17.49 & 96.07 & 0.08 & 16.55 & 17.08 & 95.96 \\
\midrule
\multirow{6}{*}{500} & $\beta_{11}$ & 1.37 & 15.35 & 5.25 & 9.21 & 8.60 & 92.60 & 0.79 & 10.29 & 10.83 & 95.52 & 0.94 & 9.31 & 9.61 & 95.21 & 0.69 & 10.28 & 10.85 & 95.45 & 0.63 & 9.34 & 9.61 & 95.01 \\
 & $\beta_{21}$ & 0.47 & 16.40 & 1.99 & 10.00 & 9.23 & 93.10 & 0.25 & 10.95 & 11.41 & 96.28 & 0.20 & 10.14 & 10.15 & 95.26 & 0.12 & 10.99 & 11.45 & 96.36 & 0.14 & 10.15 & 10.17 & 95.16 \\
 & $\beta_{21} - \beta_{11}$ & 0.49 & 18.71 & 1.50 & 15.68 & 15.45 & 94.85 & 0.33 & 16.22 & 16.56 & 95.52 & 0.60 & 15.93 & 16.10 & 95.41 & 0.49 & 16.21 & 16.61 & 95.40 & 0.38 & 15.99 & 16.12 & 94.70 \\
 & $\beta_{12}$ & 0.04 & 17.05 & 1.32 & 10.22 & 9.61 & 93.10 & 0.61 & 11.24 & 11.94 & 96.58 & 0.71 & 10.55 & 10.67 & 95.36 & 0.59 & 11.25 & 11.94 & 96.26 & 0.52 & 10.41 & 10.68 & 95.86 \\
 & $\beta_{22}$ & 0.88 & 16.25 & 0.00 & 9.74 & 9.04 & 93.00 & 1.74 & 10.84 & 11.27 & 95.98 & 1.33 & 10.07 & 9.95 & 95.16 & 1.66 & 10.81 & 11.28 & 95.61 & 1.55 & 9.80 & 9.98 & 95.01 \\
 & $\beta_{22} - \beta_{12}$ & 0.34 & 19.06 & 0.83 & 16.20 & 16.08 & 95.25 & 0.25 & 16.94 & 17.24 & 95.37 & 0.03 & 17.08 & 16.78 & 95.16 & 0.23 & 16.91 & 17.28 & 95.71 & 0.23 & 16.57 & 16.81 & 95.01 \\
\midrule
\multirow{6}{*}{1000} & $\beta_{11}$ & 1.37 & 15.35 & 5.57 & 8.84 & 7.98 & 91.55 & 1.15 & 9.88 & 10.41 & 95.96 & 1.14 & 8.97 & 8.86 & 94.56 & 0.96 & 9.89 & 10.42 & 95.90 & 1.10 & 8.90 & 8.85 & 94.53 \\
 & $\beta_{21}$ & 0.47 & 16.40 & 2.05 & 9.41 & 8.65 & 92.85 & 0.23 & 10.47 & 11.01 & 96.26 & 0.41 & 9.50 & 9.41 & 95.21 & 0.20 & 10.48 & 11.01 & 96.00 & 0.25 & 9.32 & 9.41 & 95.29 \\
 & $\beta_{21} - \beta_{11}$ & 0.49 & 18.71 & 1.73 & 15.43 & 15.26 & 94.65 & 0.76 & 16.02 & 16.35 & 95.30 & 0.38 & 15.92 & 15.84 & 95.31 & 0.61 & 16.00 & 16.42 & 95.24 & 0.67 & 15.64 & 15.85 & 94.84 \\
 & $\beta_{12}$ & 0.04 & 17.05 & 0.75 & 9.66 & 9.07 & 93.25 & 0.15 & 10.75 & 11.53 & 96.52 & 0.01 & 11.04 & 9.96 & 94.91 & 0.11 & 10.76 & 11.54 & 96.71 & 0.14 & 9.79 & 9.93 & 95.14 \\
 & $\beta_{22}$ & 0.88 & 16.25 & 0.90 & 9.32 & 8.47 & 91.80 & 1.14 & 10.34 & 10.90 & 96.31 & 0.88 & 10.23 & 9.27 & 94.30 & 1.04 & 10.42 & 10.88 & 95.95 & 0.65 & 9.19 & 9.22 & 94.38 \\
 & $\beta_{22} - \beta_{12}$ & 0.34 & 19.06 & 0.80 & 16.21 & 15.91 & 94.70 & 0.32 & 16.78 & 17.08 & 95.30 & 0.32 & 18.97 & 16.55 & 94.96 & 0.31 & 16.85 & 17.07 & 95.09 & 0.15 & 16.37 & 16.49 & 94.69 \\
\bottomrule
\end{tabular}
\end{adjustbox}
\vspace{2pt}
\begin{minipage}{\textwidth}
\footnotesize
Note: Bias, relative bias ($\%$); ESD (x100), empirical standard deviation displayed after multiplying by 100; SE (x100), mean of bootstrap standard errors displayed after multiplying by 100; CP, coverage probability of 95\% confidence intervals ($\%$). MLE is based only on the internal sample, so its results are the same the same across values of $n_k$.
\end{minipage}
\end{table}

% Requires \usepackage{booktabs}. SE highlighting is disabled; all SE values are printed as plain numbers.
% Requires \usepackage{booktabs}. SE highlighting is disabled; all SE values are printed as plain numbers.
\begin{table}[!htbp]
\centering
\caption{Log(OR) estimates, bootstrap SEs, and p-values for the association of smoking with the novel tumor markers using the established tumor marker MSI.}
\label{tab:smoking_novel_markers_mle_el_ppmgmm_pprgmm}
\resizebox{\textwidth}{!}{%
\begin{tabular}{llrrrrrrrrrrrr}
\toprule
 & & \multicolumn{3}{c}{MLE} & \multicolumn{3}{c}{EL} & \multicolumn{3}{c}{$\text{mGMM}_{\text{W}}$} & \multicolumn{3}{c}{$\text{rGMM}_{\text{W}}$} \\
\cmidrule(lr){3-5} \cmidrule(lr){6-8} \cmidrule(lr){9-11} \cmidrule(lr){12-14}
Tumor marker & Parameter & log(OR) & SE & p-value & log(OR) & SE & p-value & log(OR) & SE & p-value & log(OR) & SE & p-value \\
\midrule
APC & $\beta_{1,1}$ & 0.370 & 0.154 & 0.016 & 0.455 & 0.089 & <0.001 & 0.391 & 0.119 & 0.001 & 0.389 & 0.119 & 0.001 \\
 & $\beta_{2,1}$ & 0.069 & 0.142 & 0.626 & -0.051 & 0.072 & 0.480 & -0.007 & 0.096 & 0.945 & -0.007 & 0.096 & 0.939 \\
 & $\beta_{2,1}-\beta_{1,1}$ & -0.301 & 0.180 & 0.095 & -0.506 & 0.157 & 0.001 & -0.398 & 0.181 & 0.028 & -0.396 & 0.181 & 0.028 \\
\midrule
ARID1A & $\beta_{1,1}$ & 0.209 & 0.122 & 0.085 & 0.142 & 0.033 & <0.001 & 0.156 & 0.062 & 0.013 & 0.155 & 0.062 & 0.013 \\
 & $\beta_{2,1}$ & 0.167 & 0.271 & 0.538 & 0.414 & 0.214 & 0.053 & 0.298 & 0.245 & 0.223 & 0.299 & 0.245 & 0.222 \\
 & $\beta_{2,1}-\beta_{1,1}$ & -0.043 & 0.277 & 0.877 & 0.272 & 0.243 & 0.263 & 0.143 & 0.262 & 0.586 & 0.144 & 0.262 & 0.582 \\
\midrule
BCL9L & $\beta_{1,1}$ & 0.250 & 0.122 & 0.040 & 0.181 & 0.034 & <0.001 & 0.196 & 0.065 & 0.003 & 0.195 & 0.065 & 0.003 \\
 & $\beta_{2,1}$ & -0.128 & 0.271 & 0.638 & 0.122 & 0.217 & 0.574 & 0.005 & 0.244 & 0.985 & -0.001 & 0.244 & 0.996 \\
 & $\beta_{2,1}-\beta_{1,1}$ & -0.377 & 0.277 & 0.174 & -0.059 & 0.246 & 0.810 & -0.191 & 0.266 & 0.472 & -0.196 & 0.266 & 0.461 \\
\midrule
BMPR2 & $\beta_{1,1}$ & 0.238 & 0.122 & 0.050 & 0.157 & 0.029 & <0.001 & 0.172 & 0.062 & 0.006 & 0.170 & 0.062 & 0.006 \\
 & $\beta_{2,1}$ & -0.044 & 0.268 & 0.871 & 0.302 & 0.174 & 0.083 & 0.179 & 0.221 & 0.418 & 0.182 & 0.220 & 0.407 \\
 & $\beta_{2,1}-\beta_{1,1}$ & -0.282 & 0.275 & 0.305 & 0.145 & 0.197 & 0.461 & 0.007 & 0.231 & 0.975 & 0.012 & 0.230 & 0.958 \\
\midrule
FAT1 & $\beta_{1,1}$ & 0.204 & 0.121 & 0.091 & 0.136 & 0.030 & <0.001 & 0.153 & 0.058 & 0.009 & 0.148 & 0.058 & 0.011 \\
 & $\beta_{2,1}$ & 0.210 & 0.298 & 0.480 & 0.533 & 0.229 & 0.020 & 0.344 & 0.264 & 0.193 & 0.390 & 0.265 & 0.141 \\
 & $\beta_{2,1}-\beta_{1,1}$ & 0.006 & 0.303 & 0.984 & 0.397 & 0.253 & 0.117 & 0.191 & 0.281 & 0.497 & 0.241 & 0.281 & 0.391 \\
\midrule
FBXW7 & $\beta_{1,1}$ & 0.240 & 0.124 & 0.052 & 0.188 & 0.041 & <0.001 & 0.198 & 0.068 & 0.004 & 0.199 & 0.068 & 0.004 \\
 & $\beta_{2,1}$ & 0.017 & 0.234 & 0.941 & 0.102 & 0.196 & 0.601 & 0.030 & 0.215 & 0.890 & 0.028 & 0.216 & 0.898 \\
 & $\beta_{2,1}-\beta_{1,1}$ & -0.223 & 0.242 & 0.357 & -0.085 & 0.233 & 0.714 & -0.168 & 0.244 & 0.491 & -0.171 & 0.244 & 0.483 \\
\midrule
GNAS & $\beta_{1,1}$ & 0.229 & 0.121 & 0.058 & 0.176 & 0.032 & <0.001 & 0.182 & 0.064 & 0.004 & 0.182 & 0.063 & 0.004 \\
 & $\beta_{2,1}$ & -0.025 & 0.296 & 0.933 & 0.157 & 0.256 & 0.539 & 0.072 & 0.286 & 0.802 & 0.074 & 0.286 & 0.796 \\
 & $\beta_{2,1}-\beta_{1,1}$ & -0.254 & 0.301 & 0.400 & -0.018 & 0.283 & 0.948 & -0.110 & 0.307 & 0.720 & -0.108 & 0.307 & 0.725 \\
\midrule
KMT2B & $\beta_{1,1}$ & 0.249 & 0.121 & 0.039 & 0.176 & 0.029 & <0.001 & 0.194 & 0.061 & 0.001 & 0.196 & 0.061 & 0.001 \\
 & $\beta_{2,1}$ & -0.199 & 0.294 & 0.499 & 0.153 & 0.215 & 0.477 & -0.032 & 0.263 & 0.904 & -0.051 & 0.263 & 0.846 \\
 & $\beta_{2,1}-\beta_{1,1}$ & -0.448 & 0.299 & 0.135 & -0.024 & 0.237 & 0.921 & -0.226 & 0.283 & 0.424 & -0.247 & 0.283 & 0.382 \\
\midrule
KMT2D & $\beta_{1,1}$ & 0.270 & 0.122 & 0.027 & 0.193 & 0.033 & <0.001 & 0.210 & 0.064 & 0.001 & 0.210 & 0.064 & 0.001 \\
 & $\beta_{2,1}$ & -0.256 & 0.266 & 0.334 & 0.040 & 0.197 & 0.840 & -0.105 & 0.230 & 0.650 & -0.098 & 0.230 & 0.669 \\
 & $\beta_{2,1}-\beta_{1,1}$ & -0.527 & 0.272 & 0.053 & -0.153 & 0.224 & 0.494 & -0.314 & 0.252 & 0.212 & -0.308 & 0.252 & 0.221 \\
\midrule
PCDH10 & $\beta_{1,1}$ & 0.202 & 0.121 & 0.095 & 0.137 & 0.031 & <0.001 & 0.150 & 0.061 & 0.014 & 0.149 & 0.061 & 0.015 \\
 & $\beta_{2,1}$ & 0.230 & 0.293 & 0.433 & 0.516 & 0.233 & 0.027 & 0.372 & 0.291 & 0.201 & 0.377 & 0.291 & 0.195 \\
 & $\beta_{2,1}-\beta_{1,1}$ & 0.028 & 0.298 & 0.925 & 0.379 & 0.258 & 0.142 & 0.222 & 0.309 & 0.473 & 0.228 & 0.309 & 0.461 \\
\midrule
PIK3CA & $\beta_{1,1}$ & 0.182 & 0.123 & 0.137 & 0.136 & 0.039 & <0.001 & 0.143 & 0.068 & 0.035 & 0.143 & 0.068 & 0.035 \\
 & $\beta_{2,1}$ & 0.334 & 0.247 & 0.176 & 0.398 & 0.210 & 0.058 & 0.338 & 0.224 & 0.132 & 0.337 & 0.224 & 0.132 \\
 & $\beta_{2,1}-\beta_{1,1}$ & 0.152 & 0.254 & 0.550 & 0.262 & 0.245 & 0.284 & 0.195 & 0.251 & 0.437 & 0.194 & 0.251 & 0.439 \\
\midrule
RNF43 & $\beta_{1,1}$ & 0.196 & 0.124 & 0.114 & 0.097 & 0.030 & 0.001 & 0.109 & 0.061 & 0.073 & 0.110 & 0.061 & 0.069 \\
 & $\beta_{2,1}$ & 0.245 & 0.226 & 0.278 & 0.537 & 0.117 & <0.001 & 0.467 & 0.172 & 0.007 & 0.454 & 0.172 & 0.008 \\
 & $\beta_{2,1}-\beta_{1,1}$ & 0.049 & 0.235 & 0.836 & 0.440 & 0.141 & 0.002 & 0.359 & 0.185 & 0.053 & 0.344 & 0.185 & 0.063 \\
\midrule
RYR1 & $\beta_{1,1}$ & 0.236 & 0.124 & 0.057 & 0.157 & 0.039 & <0.001 & 0.175 & 0.066 & 0.008 & 0.175 & 0.066 & 0.008 \\
 & $\beta_{2,1}$ & 0.056 & 0.228 & 0.806 & 0.260 & 0.168 & 0.123 & 0.157 & 0.206 & 0.446 & 0.155 & 0.206 & 0.451 \\
 & $\beta_{2,1}-\beta_{1,1}$ & -0.180 & 0.236 & 0.447 & 0.103 & 0.203 & 0.612 & -0.018 & 0.229 & 0.938 & -0.020 & 0.229 & 0.930 \\
\midrule
SOX9 & $\beta_{1,1}$ & 0.171 & 0.123 & 0.163 & 0.125 & 0.039 & 0.002 & 0.130 & 0.062 & 0.037 & 0.128 & 0.062 & 0.039 \\
 & $\beta_{2,1}$ & 0.394 & 0.246 & 0.109 & 0.461 & 0.210 & 0.028 & 0.406 & 0.219 & 0.063 & 0.419 & 0.219 & 0.056 \\
 & $\beta_{2,1}-\beta_{1,1}$ & 0.223 & 0.253 & 0.379 & 0.337 & 0.245 & 0.169 & 0.276 & 0.243 & 0.256 & 0.291 & 0.243 & 0.231 \\
\midrule
SYNE1 & $\beta_{1,1}$ & 0.233 & 0.126 & 0.063 & 0.144 & 0.042 & 0.001 & 0.169 & 0.068 & 0.013 & 0.168 & 0.068 & 0.013 \\
 & $\beta_{2,1}$ & 0.094 & 0.210 & 0.654 & 0.290 & 0.148 & 0.050 & 0.188 & 0.183 & 0.304 & 0.184 & 0.183 & 0.314 \\
 & $\beta_{2,1}-\beta_{1,1}$ & -0.139 & 0.220 & 0.527 & 0.146 & 0.186 & 0.433 & 0.019 & 0.214 & 0.930 & 0.015 & 0.213 & 0.942 \\
\midrule
TP53 & $\beta_{1,1}$ & -0.010 & 0.147 & 0.946 & 0.027 & 0.086 & 0.753 & -0.000 & 0.110 & 0.998 & -0.004 & 0.110 & 0.973 \\
 & $\beta_{2,1}$ & 0.416 & 0.148 & 0.005 & 0.318 & 0.087 & <0.001 & 0.346 & 0.108 & 0.001 & 0.345 & 0.107 & 0.001 \\
 & $\beta_{2,1}-\beta_{1,1}$ & 0.426 & 0.180 & 0.018 & 0.291 & 0.169 & 0.085 & 0.347 & 0.184 & 0.059 & 0.349 & 0.184 & 0.058 \\
\midrule
\bottomrule
\end{tabular}%
}
\end{table}

\begin{figure}[h]
    \centering
    \hspace{-2cm}
    \includegraphics[scale=0.7]{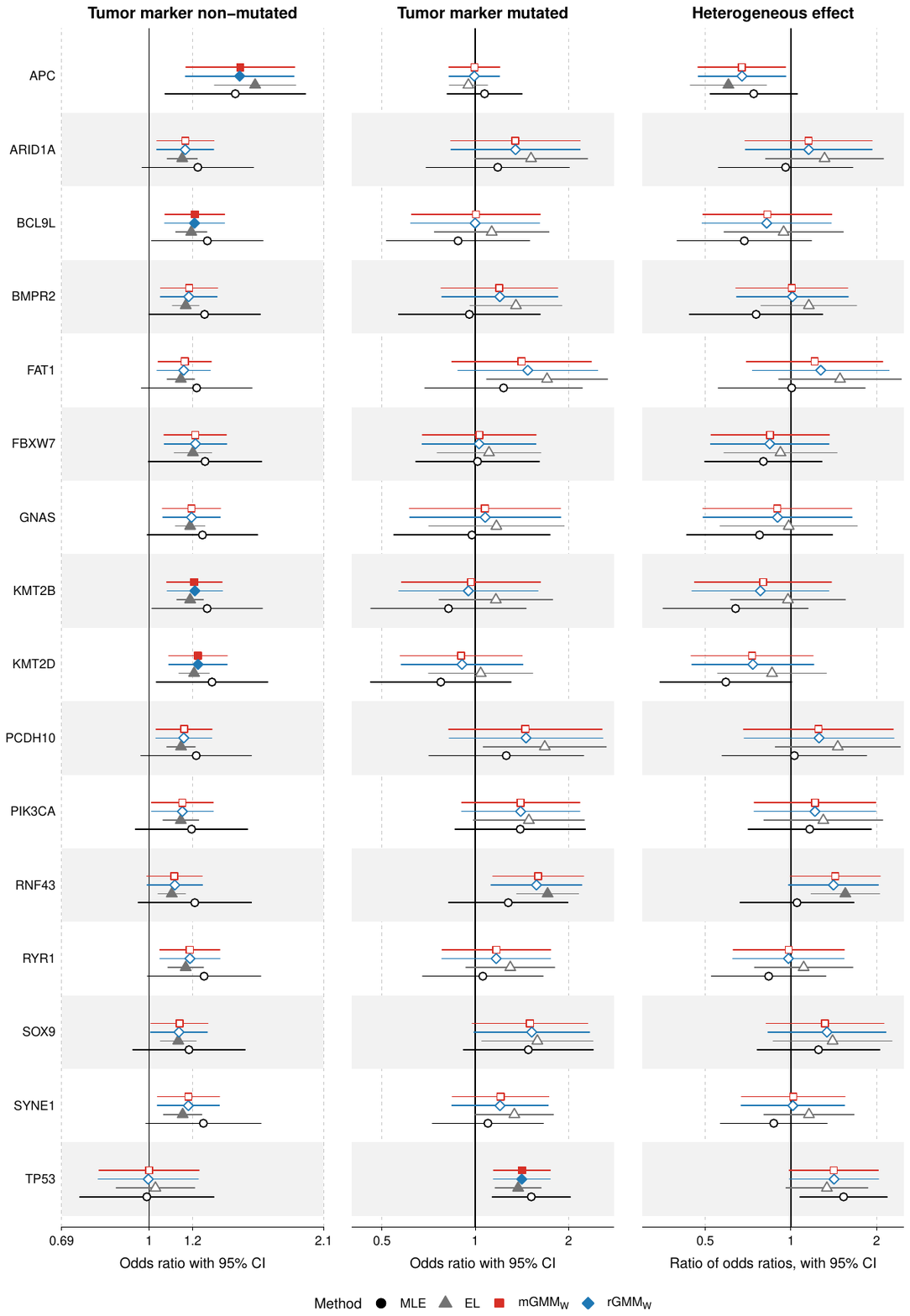}
    \caption{Forest plot of ORs of smoking on novel tumor markers using external established marker MSI. $\text{mGMM}_{\text{W}}$ and $\text{rGMM}_{\text{W}}$ are proposed approach using the huber-type weight. The nonsignificant results are displayed as hollow points using Bonferroni correction to account for multiple testing, with significance threshold 0.05/16 $\approx$ 0.003. The ratio of ORs is the ratio of the OR of smoking on the risk of developing CRC with mutated marker and non-mutated marker.}
    \label{fig:forest_plot_MSI}
\end{figure}

% Requires \usepackage{booktabs}. SE highlighting is disabled; all SE values are printed as plain numbers.
% Requires \usepackage{booktabs}. SE highlighting is disabled; all SE values are printed as plain numbers.
% Requires \usepackage{booktabs}. SE highlighting is disabled; all SE values are printed as plain numbers.
\begin{table}[!htbp]
\centering
\caption{Log(OR) estimates, bootstrap SEs, and p-values for the association of smoking with the novel tumor markers using the established marker BRAF}
\label{tab:smoking_novel_markers_mle_el_ppmgmm_pprgmm}
\resizebox{\textwidth}{!}{%
\begin{tabular}{llrrrrrrrrrrrr}
\toprule
 & & \multicolumn{3}{c}{MLE} & \multicolumn{3}{c}{EL} & \multicolumn{3}{c}{$\text{mGMM}_{\text{W}}$} & \multicolumn{3}{c}{$\text{rGMM}_{\text{W}}$} \\
\cmidrule(lr){3-5} \cmidrule(lr){6-8} \cmidrule(lr){9-11} \cmidrule(lr){12-14}
Tumor marker & Parameter & log(OR) & SE & p-value & log(OR) & SE & p-value & log(OR) & SE & p-value & log(OR) & SE & p-value \\
\midrule
APC & $\beta_{1,1}$ & 0.370 & 0.154 & 0.016 & 0.455 & 0.089 & <0.001 & 0.381 & 0.124 & 0.002 & 0.388 & 0.114 & <0.001 \\
 & $\beta_{2,1}$ & 0.069 & 0.142 & 0.626 & -0.051 & 0.072 & 0.480 & 0.050 & 0.113 & 0.662 & 0.043 & 0.098 & 0.661 \\
 & $\beta_{2,1}-\beta_{1,1}$ & -0.301 & 0.180 & 0.095 & -0.506 & 0.157 & 0.001 & -0.331 & 0.189 & 0.080 & -0.345 & 0.187 & 0.065 \\
\midrule
ARID1A & $\beta_{1,1}$ & 0.209 & 0.122 & 0.085 & 0.142 & 0.033 & <0.001 & 0.202 & 0.071 & 0.004 & 0.202 & 0.063 & 0.001 \\
 & $\beta_{2,1}$ & 0.167 & 0.271 & 0.538 & 0.414 & 0.214 & 0.053 & 0.173 & 0.270 & 0.522 & 0.169 & 0.252 & 0.503 \\
 & $\beta_{2,1}-\beta_{1,1}$ & -0.043 & 0.277 & 0.877 & 0.272 & 0.243 & 0.263 & -0.029 & 0.294 & 0.921 & -0.033 & 0.280 & 0.906 \\
\midrule
BCL9L & $\beta_{1,1}$ & 0.250 & 0.122 & 0.040 & 0.181 & 0.034 & <0.001 & 0.234 & 0.073 & 0.001 & 0.235 & 0.060 & <0.001 \\
 & $\beta_{2,1}$ & -0.128 & 0.271 & 0.638 & 0.122 & 0.217 & 0.574 & -0.076 & 0.268 & 0.775 & -0.076 & 0.259 & 0.769 \\
 & $\beta_{2,1}-\beta_{1,1}$ & -0.377 & 0.277 & 0.174 & -0.059 & 0.246 & 0.810 & -0.310 & 0.300 & 0.301 & -0.311 & 0.287 & 0.277 \\
\midrule
BMPR2 & $\beta_{1,1}$ & 0.238 & 0.122 & 0.050 & 0.157 & 0.029 & <0.001 & 0.220 & 0.077 & 0.005 & 0.228 & 0.061 & <0.001 \\
 & $\beta_{2,1}$ & -0.044 & 0.268 & 0.871 & 0.302 & 0.174 & 0.083 & 0.031 & 0.263 & 0.907 & -0.032 & 0.248 & 0.897 \\
 & $\beta_{2,1}-\beta_{1,1}$ & -0.282 & 0.275 & 0.305 & 0.145 & 0.197 & 0.461 & -0.189 & 0.288 & 0.512 & -0.261 & 0.277 & 0.346 \\
\midrule
FAT1 & $\beta_{1,1}$ & 0.204 & 0.121 & 0.091 & 0.136 & 0.030 & <0.001 & 0.193 & 0.070 & 0.006 & 0.193 & 0.057 & <0.001 \\
 & $\beta_{2,1}$ & 0.210 & 0.298 & 0.480 & 0.533 & 0.229 & 0.020 & 0.242 & 0.277 & 0.381 & 0.244 & 0.270 & 0.366 \\
 & $\beta_{2,1}-\beta_{1,1}$ & 0.006 & 0.303 & 0.984 & 0.397 & 0.253 & 0.117 & 0.049 & 0.301 & 0.870 & 0.051 & 0.296 & 0.863 \\
\midrule
FBXW7 & $\beta_{1,1}$ & 0.240 & 0.124 & 0.052 & 0.188 & 0.041 & <0.001 & 0.232 & 0.069 & <0.001 & 0.233 & 0.066 & <0.001 \\
 & $\beta_{2,1}$ & 0.017 & 0.234 & 0.941 & 0.102 & 0.196 & 0.601 & 0.016 & 0.214 & 0.942 & 0.016 & 0.211 & 0.941 \\
 & $\beta_{2,1}-\beta_{1,1}$ & -0.223 & 0.242 & 0.357 & -0.085 & 0.233 & 0.714 & -0.217 & 0.246 & 0.378 & -0.218 & 0.244 & 0.374 \\
\midrule
GNAS & $\beta_{1,1}$ & 0.229 & 0.121 & 0.058 & 0.176 & 0.032 & <0.001 & 0.222 & 0.061 & <0.001 & 0.222 & 0.059 & <0.001 \\
 & $\beta_{2,1}$ & -0.025 & 0.296 & 0.933 & 0.157 & 0.256 & 0.539 & -0.024 & 0.295 & 0.935 & -0.027 & 0.289 & 0.927 \\
 & $\beta_{2,1}-\beta_{1,1}$ & -0.254 & 0.301 & 0.400 & -0.018 & 0.283 & 0.948 & -0.246 & 0.320 & 0.442 & -0.249 & 0.315 & 0.430 \\
\midrule
KMT2B & $\beta_{1,1}$ & 0.249 & 0.121 & 0.039 & 0.176 & 0.029 & <0.001 & 0.229 & 0.066 & <0.001 & 0.229 & 0.060 & <0.001 \\
 & $\beta_{2,1}$ & -0.199 & 0.294 & 0.499 & 0.153 & 0.215 & 0.477 & -0.088 & 0.273 & 0.746 & -0.087 & 0.261 & 0.739 \\
 & $\beta_{2,1}-\beta_{1,1}$ & -0.448 & 0.299 & 0.135 & -0.024 & 0.237 & 0.921 & -0.317 & 0.295 & 0.283 & -0.317 & 0.288 & 0.272 \\
\midrule
KMT2D & $\beta_{1,1}$ & 0.270 & 0.122 & 0.027 & 0.193 & 0.033 & <0.001 & 0.256 & 0.067 & <0.001 & 0.255 & 0.060 & <0.001 \\
 & $\beta_{2,1}$ & -0.256 & 0.266 & 0.334 & 0.040 & 0.197 & 0.840 & -0.199 & 0.249 & 0.423 & -0.197 & 0.241 & 0.415 \\
 & $\beta_{2,1}-\beta_{1,1}$ & -0.527 & 0.272 & 0.053 & -0.153 & 0.224 & 0.494 & -0.455 & 0.274 & 0.097 & -0.452 & 0.270 & 0.095 \\
\midrule
PCDH10 & $\beta_{1,1}$ & 0.202 & 0.121 & 0.095 & 0.137 & 0.031 & <0.001 & 0.194 & 0.071 & 0.006 & 0.193 & 0.064 & 0.002 \\
 & $\beta_{2,1}$ & 0.230 & 0.293 & 0.433 & 0.516 & 0.233 & 0.027 & 0.242 & 0.299 & 0.419 & 0.241 & 0.293 & 0.412 \\
 & $\beta_{2,1}-\beta_{1,1}$ & 0.028 & 0.298 & 0.925 & 0.379 & 0.258 & 0.142 & 0.047 & 0.327 & 0.885 & 0.048 & 0.324 & 0.883 \\
\midrule
PIK3CA & $\beta_{1,1}$ & 0.182 & 0.123 & 0.137 & 0.136 & 0.039 & <0.001 & 0.173 & 0.073 & 0.017 & 0.175 & 0.060 & 0.003 \\
 & $\beta_{2,1}$ & 0.334 & 0.247 & 0.176 & 0.398 & 0.210 & 0.058 & 0.343 & 0.232 & 0.139 & 0.331 & 0.222 & 0.136 \\
 & $\beta_{2,1}-\beta_{1,1}$ & 0.152 & 0.254 & 0.550 & 0.262 & 0.245 & 0.284 & 0.170 & 0.252 & 0.500 & 0.156 & 0.248 & 0.529 \\
\midrule
RNF43 & $\beta_{1,1}$ & 0.196 & 0.124 & 0.114 & 0.097 & 0.030 & 0.001 & 0.184 & 0.095 & 0.054 & 0.172 & 0.067 & 0.010 \\
 & $\beta_{2,1}$ & 0.245 & 0.226 & 0.278 & 0.537 & 0.117 & <0.001 & 0.268 & 0.218 & 0.218 & 0.322 & 0.199 & 0.106 \\
 & $\beta_{2,1}-\beta_{1,1}$ & 0.049 & 0.235 & 0.836 & 0.440 & 0.141 & 0.002 & 0.084 & 0.244 & 0.730 & 0.149 & 0.238 & 0.530 \\
\midrule
RYR1 & $\beta_{1,1}$ & 0.236 & 0.124 & 0.057 & 0.157 & 0.039 & <0.001 & 0.224 & 0.630 & 0.722 & 0.226 & 0.065 & <0.001 \\
 & $\beta_{2,1}$ & 0.056 & 0.228 & 0.806 & 0.260 & 0.168 & 0.123 & 0.070 & 0.212 & 0.740 & 0.071 & 0.204 & 0.727 \\
 & $\beta_{2,1}-\beta_{1,1}$ & -0.180 & 0.236 & 0.447 & 0.103 & 0.203 & 0.612 & -0.154 & 0.707 & 0.828 & -0.154 & 0.236 & 0.514 \\
\midrule
SOX9 & $\beta_{1,1}$ & 0.171 & 0.123 & 0.163 & 0.125 & 0.039 & 0.002 & 0.165 & 0.065 & 0.011 & 0.164 & 0.062 & 0.008 \\
 & $\beta_{2,1}$ & 0.394 & 0.246 & 0.109 & 0.461 & 0.210 & 0.028 & 0.393 & 0.221 & 0.075 & 0.393 & 0.218 & 0.072 \\
 & $\beta_{2,1}-\beta_{1,1}$ & 0.223 & 0.253 & 0.379 & 0.337 & 0.245 & 0.169 & 0.228 & 0.249 & 0.359 & 0.229 & 0.247 & 0.354 \\
\midrule
SYNE1 & $\beta_{1,1}$ & 0.233 & 0.126 & 0.063 & 0.144 & 0.042 & 0.001 & 0.224 & 0.074 & 0.002 & 0.215 & 0.071 & 0.002 \\
 & $\beta_{2,1}$ & 0.094 & 0.210 & 0.654 & 0.290 & 0.148 & 0.050 & 0.113 & 0.188 & 0.549 & 0.133 & 0.182 & 0.465 \\
 & $\beta_{2,1}-\beta_{1,1}$ & -0.139 & 0.220 & 0.527 & 0.146 & 0.186 & 0.433 & -0.112 & 0.225 & 0.621 & -0.082 & 0.223 & 0.712 \\
\midrule
TP53 & $\beta_{1,1}$ & -0.010 & 0.147 & 0.946 & 0.027 & 0.086 & 0.753 & -0.007 & 0.115 & 0.954 & -0.008 & 0.104 & 0.941 \\
 & $\beta_{2,1}$ & 0.416 & 0.148 & 0.005 & 0.318 & 0.087 & <0.001 & 0.402 & 0.125 & 0.001 & 0.403 & 0.108 & <0.001 \\
 & $\beta_{2,1}-\beta_{1,1}$ & 0.426 & 0.180 & 0.018 & 0.291 & 0.169 & 0.085 & 0.409 & 0.191 & 0.032 & 0.410 & 0.188 & 0.029 \\
\midrule
\bottomrule
\end{tabular}%
}
\end{table}

\begin{figure}[!htbp]
    \centering
    \hspace{-2cm}
    \includegraphics[scale=0.7]{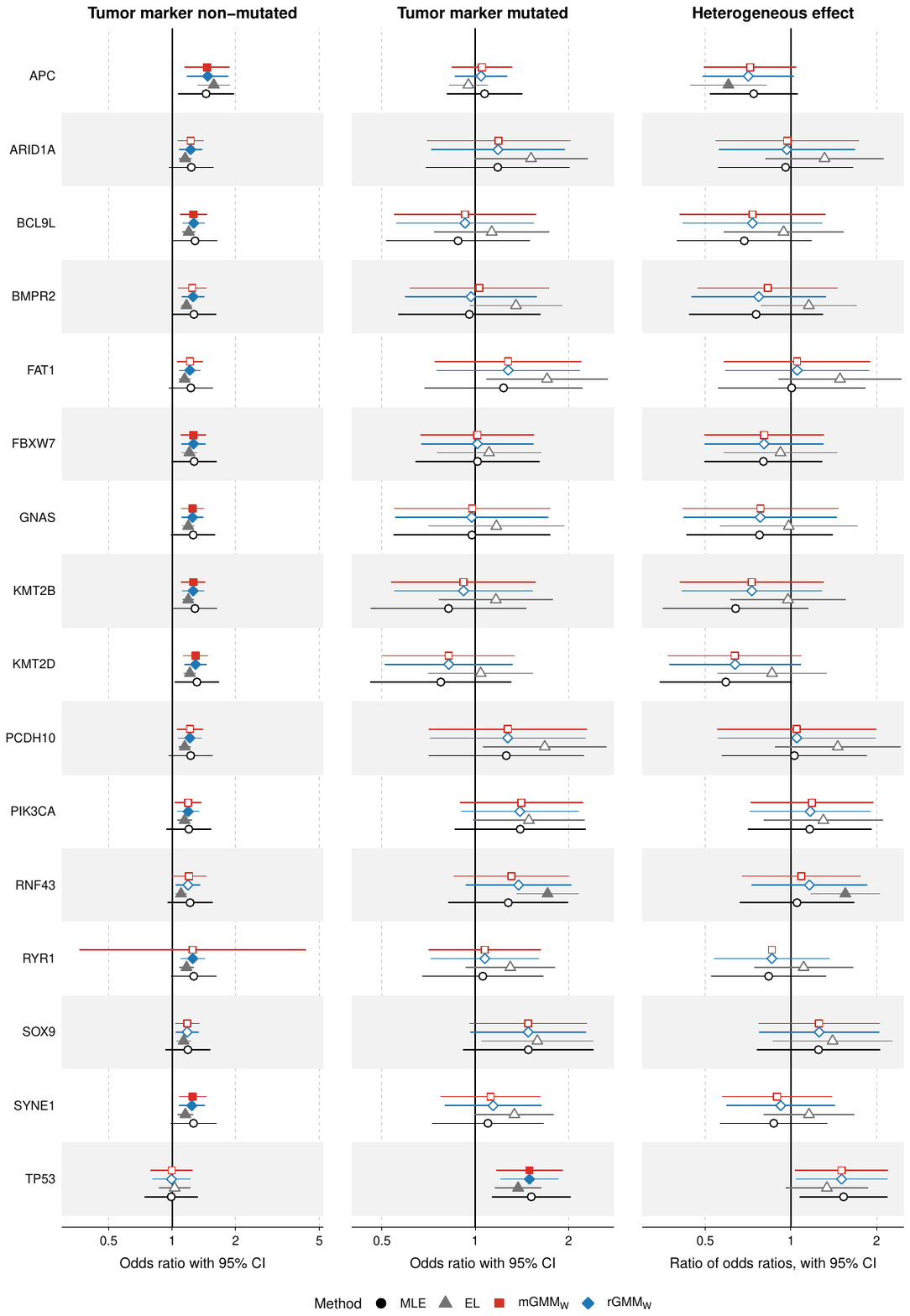}
    \caption{Forest plot of ORs of smoking on novel tumor markers using external established marker BRAF. $\text{mGMM}_{\text{W}}$ and $\text{rGMM}_{\text{W}}$ are proposed GMM approach using the huber-type weight. The nonsignificant results are displayed as hollow points using Bonferroni correction to account for multiple testing, with significance threshold 0.05/16 $\approx$ 0.003. The ratio of ORs is the ratio of the OR of smoking on the risk of developing CRC with mutated marker and non-mutated marker.}
    \label{fig:forest_plot_BRAF}
\end{figure}

% Requires \usepackage{booktabs}. SE highlighting is disabled; all SE values are printed as plain numbers.
% Requires \usepackage{booktabs}. SE highlighting is disabled; all SE values are printed as plain numbers.
\begin{table}[!htbp]
\centering
\caption{Log(OR) estimates, bootstrap SEs, and p-values for the association of smoking with the novel tumor markers using the established tumor marker CIMP}
\label{tab:smoking_novel_markers_mle_el_ppmgmm_pprgmm}
\resizebox{\textwidth}{!}{%
\begin{tabular}{llrrrrrrrrrrrr}
\toprule
 & & \multicolumn{3}{c}{MLE} & \multicolumn{3}{c}{EL} & \multicolumn{3}{c}{$\text{mGMM}_{\text{W}}$} & \multicolumn{3}{c}{$\text{rGMM}_{\text{W}}$} \\
\cmidrule(lr){3-5} \cmidrule(lr){6-8} \cmidrule(lr){9-11} \cmidrule(lr){12-14}
Tumor marker & Parameter & log(OR) & SE & p-value & log(OR) & SE & p-value & log(OR) & SE & p-value & log(OR) & SE & p-value \\
\midrule
APC & $\beta_{1,1}$ & 0.370 & 0.154 & 0.016 & 0.455 & 0.089 & <0.001 & 0.382 & 0.116 & 0.001 & 0.382 & 0.110 & <0.001 \\
 & $\beta_{2,1}$ & 0.069 & 0.142 & 0.626 & -0.051 & 0.072 & 0.480 & 0.014 & 0.103 & 0.891 & 0.014 & 0.095 & 0.885 \\
 & $\beta_{2,1}-\beta_{1,1}$ & -0.301 & 0.180 & 0.095 & -0.506 & 0.157 & 0.001 & -0.367 & 0.179 & 0.040 & -0.368 & 0.179 & 0.040 \\
\midrule
ARID1A & $\beta_{1,1}$ & 0.209 & 0.122 & 0.085 & 0.142 & 0.033 & <0.001 & 0.176 & 0.062 & 0.005 & 0.176 & 0.060 & 0.003 \\
 & $\beta_{2,1}$ & 0.167 & 0.271 & 0.538 & 0.414 & 0.214 & 0.053 & 0.204 & 0.264 & 0.439 & 0.204 & 0.257 & 0.426 \\
 & $\beta_{2,1}-\beta_{1,1}$ & -0.043 & 0.277 & 0.877 & 0.272 & 0.243 & 0.263 & 0.028 & 0.285 & 0.922 & 0.028 & 0.282 & 0.920 \\
\midrule
BCL9L & $\beta_{1,1}$ & 0.250 & 0.122 & 0.040 & 0.181 & 0.034 & <0.001 & 0.218 & 0.073 & 0.003 & 0.216 & 0.063 & <0.001 \\
 & $\beta_{2,1}$ & -0.128 & 0.271 & 0.638 & 0.122 & 0.217 & 0.574 & -0.087 & 0.259 & 0.737 & -0.088 & 0.250 & 0.725 \\
 & $\beta_{2,1}-\beta_{1,1}$ & -0.377 & 0.277 & 0.174 & -0.059 & 0.246 & 0.810 & -0.305 & 0.284 & 0.283 & -0.304 & 0.281 & 0.280 \\
\midrule
BMPR2 & $\beta_{1,1}$ & 0.238 & 0.122 & 0.050 & 0.157 & 0.029 & <0.001 & 0.194 & 0.071 & 0.006 & 0.195 & 0.061 & 0.001 \\
 & $\beta_{2,1}$ & -0.044 & 0.268 & 0.871 & 0.302 & 0.174 & 0.083 & 0.064 & 0.245 & 0.795 & 0.064 & 0.226 & 0.775 \\
 & $\beta_{2,1}-\beta_{1,1}$ & -0.282 & 0.275 & 0.305 & 0.145 & 0.197 & 0.461 & -0.131 & 0.263 & 0.620 & -0.130 & 0.252 & 0.606 \\
\midrule
FAT1 & $\beta_{1,1}$ & 0.204 & 0.121 & 0.091 & 0.136 & 0.030 & <0.001 & 0.168 & 0.070 & 0.017 & 0.167 & 0.059 & 0.005 \\
 & $\beta_{2,1}$ & 0.210 & 0.298 & 0.480 & 0.533 & 0.229 & 0.020 & 0.297 & 0.275 & 0.280 & 0.297 & 0.264 & 0.260 \\
 & $\beta_{2,1}-\beta_{1,1}$ & 0.006 & 0.303 & 0.984 & 0.397 & 0.253 & 0.117 & 0.129 & 0.295 & 0.663 & 0.130 & 0.287 & 0.651 \\
\midrule
FBXW7 & $\beta_{1,1}$ & 0.240 & 0.124 & 0.052 & 0.188 & 0.041 & <0.001 & 0.209 & 0.621 & 0.736 & 0.210 & 0.066 & 0.001 \\
 & $\beta_{2,1}$ & 0.017 & 0.234 & 0.941 & 0.102 & 0.196 & 0.601 & 0.020 & 0.659 & 0.976 & 0.021 & 0.210 & 0.922 \\
 & $\beta_{2,1}-\beta_{1,1}$ & -0.223 & 0.242 & 0.357 & -0.085 & 0.233 & 0.714 & -0.189 & 0.244 & 0.438 & -0.189 & 0.245 & 0.439 \\
\midrule
GNAS & $\beta_{1,1}$ & 0.229 & 0.121 & 0.058 & 0.176 & 0.032 & <0.001 & 0.200 & 0.073 & 0.007 & 0.198 & 0.061 & 0.001 \\
 & $\beta_{2,1}$ & -0.025 & 0.296 & 0.933 & 0.157 & 0.256 & 0.539 & 0.008 & 0.290 & 0.977 & 0.007 & 0.284 & 0.981 \\
 & $\beta_{2,1}-\beta_{1,1}$ & -0.254 & 0.301 & 0.400 & -0.018 & 0.283 & 0.948 & -0.191 & 0.313 & 0.541 & -0.192 & 0.311 & 0.538 \\
\midrule
KMT2B & $\beta_{1,1}$ & 0.249 & 0.121 & 0.039 & 0.176 & 0.029 & <0.001 & 0.209 & 0.135 & 0.123 & 0.211 & 0.060 & <0.001 \\
 & $\beta_{2,1}$ & -0.199 & 0.294 & 0.499 & 0.153 & 0.215 & 0.477 & -0.101 & 0.278 & 0.717 & -0.105 & 0.254 & 0.680 \\
 & $\beta_{2,1}-\beta_{1,1}$ & -0.448 & 0.299 & 0.135 & -0.024 & 0.237 & 0.921 & -0.310 & 0.290 & 0.285 & -0.316 & 0.280 & 0.260 \\
\midrule
KMT2D & $\beta_{1,1}$ & 0.270 & 0.122 & 0.027 & 0.193 & 0.033 & <0.001 & 0.231 & 0.072 & 0.001 & 0.231 & 0.060 & <0.001 \\
 & $\beta_{2,1}$ & -0.256 & 0.266 & 0.334 & 0.040 & 0.197 & 0.840 & -0.184 & 0.245 & 0.452 & -0.183 & 0.235 & 0.436 \\
 & $\beta_{2,1}-\beta_{1,1}$ & -0.527 & 0.272 & 0.053 & -0.153 & 0.224 & 0.494 & -0.415 & 0.270 & 0.124 & -0.413 & 0.263 & 0.116 \\
\midrule
PCDH10 & $\beta_{1,1}$ & 0.202 & 0.121 & 0.095 & 0.137 & 0.031 & <0.001 & 0.165 & 0.065 & 0.011 & 0.165 & 0.061 & 0.006 \\
 & $\beta_{2,1}$ & 0.230 & 0.293 & 0.433 & 0.516 & 0.233 & 0.027 & 0.309 & 0.293 & 0.290 & 0.311 & 0.286 & 0.277 \\
 & $\beta_{2,1}-\beta_{1,1}$ & 0.028 & 0.298 & 0.925 & 0.379 & 0.258 & 0.142 & 0.144 & 0.319 & 0.651 & 0.146 & 0.314 & 0.642 \\
\midrule
PIK3CA & $\beta_{1,1}$ & 0.182 & 0.123 & 0.137 & 0.136 & 0.039 & <0.001 & 0.151 & 0.088 & 0.087 & 0.152 & 0.062 & 0.014 \\
 & $\beta_{2,1}$ & 0.334 & 0.247 & 0.176 & 0.398 & 0.210 & 0.058 & 0.340 & 0.230 & 0.140 & 0.340 & 0.221 & 0.123 \\
 & $\beta_{2,1}-\beta_{1,1}$ & 0.152 & 0.254 & 0.550 & 0.262 & 0.245 & 0.284 & 0.189 & 0.253 & 0.453 & 0.188 & 0.250 & 0.451 \\
\midrule
RNF43 & $\beta_{1,1}$ & 0.196 & 0.124 & 0.114 & 0.097 & 0.030 & 0.001 & 0.144 & 0.088 & 0.104 & 0.141 & 0.067 & 0.034 \\
 & $\beta_{2,1}$ & 0.245 & 0.226 & 0.278 & 0.537 & 0.117 & <0.001 & 0.356 & 0.187 & 0.057 & 0.358 & 0.179 & 0.046 \\
 & $\beta_{2,1}-\beta_{1,1}$ & 0.049 & 0.235 & 0.836 & 0.440 & 0.141 & 0.002 & 0.212 & 0.216 & 0.325 & 0.217 & 0.211 & 0.303 \\
\midrule
RYR1 & $\beta_{1,1}$ & 0.236 & 0.124 & 0.057 & 0.157 & 0.039 & <0.001 & 0.192 & 0.077 & 0.013 & 0.195 & 0.065 & 0.003 \\
 & $\beta_{2,1}$ & 0.056 & 0.228 & 0.806 & 0.260 & 0.168 & 0.123 & 0.105 & 0.200 & 0.600 & 0.105 & 0.196 & 0.591 \\
 & $\beta_{2,1}-\beta_{1,1}$ & -0.180 & 0.236 & 0.447 & 0.103 & 0.203 & 0.612 & -0.087 & 0.232 & 0.707 & -0.090 & 0.230 & 0.695 \\
\midrule
SOX9 & $\beta_{1,1}$ & 0.171 & 0.123 & 0.163 & 0.125 & 0.039 & 0.002 & 0.141 & 0.072 & 0.051 & 0.141 & 0.060 & 0.019 \\
 & $\beta_{2,1}$ & 0.394 & 0.246 & 0.109 & 0.461 & 0.210 & 0.028 & 0.397 & 0.223 & 0.075 & 0.398 & 0.217 & 0.066 \\
 & $\beta_{2,1}-\beta_{1,1}$ & 0.223 & 0.253 & 0.379 & 0.337 & 0.245 & 0.169 & 0.256 & 0.245 & 0.297 & 0.257 & 0.243 & 0.291 \\
\midrule
SYNE1 & $\beta_{1,1}$ & 0.233 & 0.126 & 0.063 & 0.144 & 0.042 & 0.001 & 0.190 & 0.084 & 0.024 & 0.191 & 0.070 & 0.006 \\
 & $\beta_{2,1}$ & 0.094 & 0.210 & 0.654 & 0.290 & 0.148 & 0.050 & 0.136 & 0.195 & 0.486 & 0.136 & 0.183 & 0.455 \\
 & $\beta_{2,1}-\beta_{1,1}$ & -0.139 & 0.220 & 0.527 & 0.146 & 0.186 & 0.433 & -0.054 & 0.222 & 0.807 & -0.054 & 0.219 & 0.804 \\
\midrule
TP53 & $\beta_{1,1}$ & -0.010 & 0.147 & 0.946 & 0.027 & 0.086 & 0.753 & -0.014 & 0.109 & 0.900 & -0.014 & 0.102 & 0.894 \\
 & $\beta_{2,1}$ & 0.416 & 0.148 & 0.005 & 0.318 & 0.087 & <0.001 & 0.370 & 0.118 & 0.002 & 0.371 & 0.109 & <0.001 \\
 & $\beta_{2,1}-\beta_{1,1}$ & 0.426 & 0.180 & 0.018 & 0.291 & 0.169 & 0.085 & 0.384 & 0.185 & 0.038 & 0.385 & 0.184 & 0.036 \\
\midrule
\bottomrule
\end{tabular}%
}
\end{table}

\begin{figure}[!htbp]
    \centering
    \hspace{-2cm}
    \includegraphics[scale=0.7]{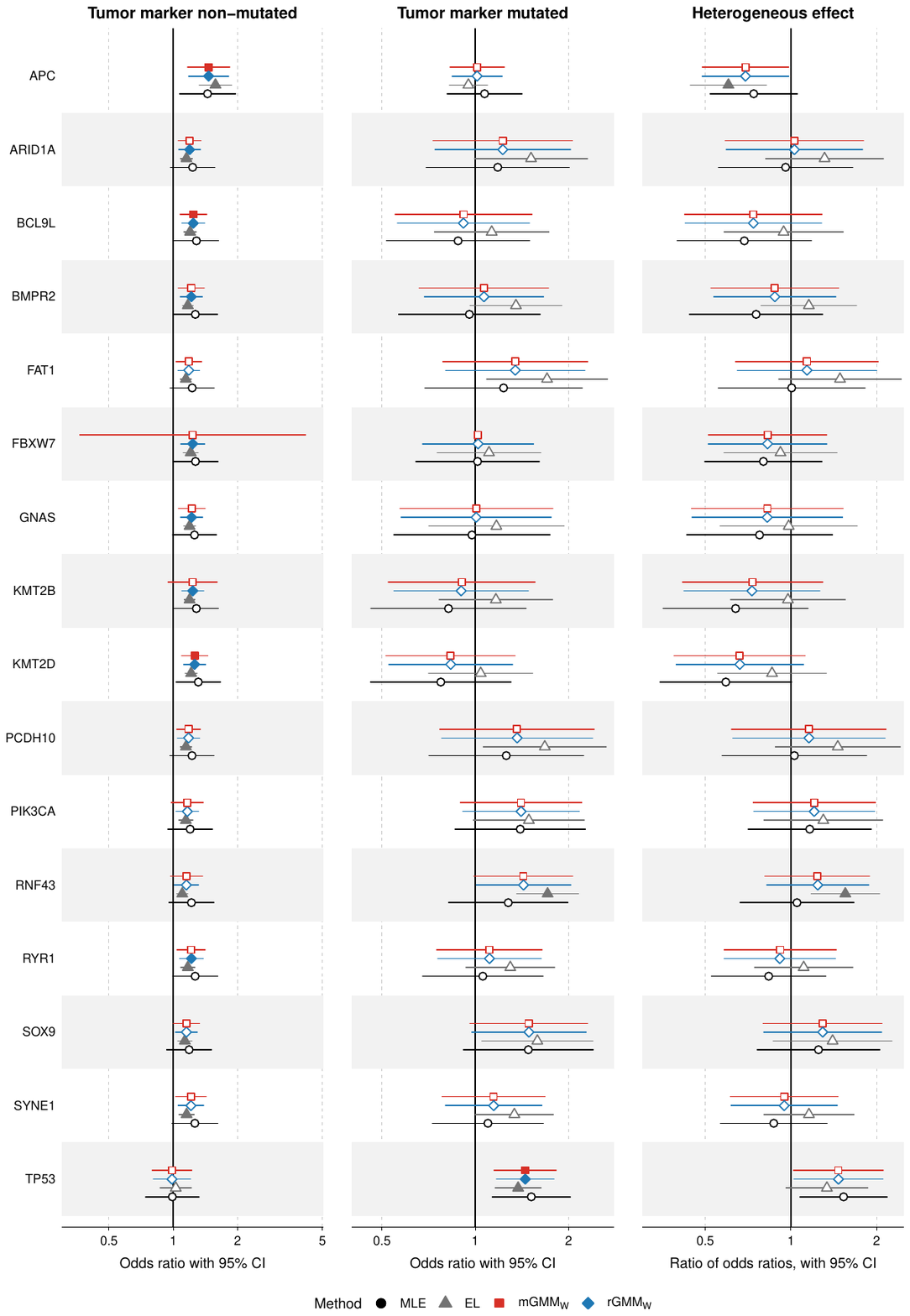}
    \caption{Forest plot of ORs of smoking on novel tumor markers using external established marker CIMP. $\text{mGMM}_{\text{W}}$ and $\text{rGMM}_{\text{W}}$ are proposed GMM approach using the huber-type weight. The nonsignificant results are displayed as hollow points using Bonferroni correction to account for multiple testing, with significance threshold 0.05/16 $\approx$ 0.003. The ratio of ORs is the ratio of the OR of smoking on the risk of developing CRC with mutated marker and non-mutated marker.}
    \label{fig:forest_plot_CIMP}
\end{figure}

% Requires \usepackage{booktabs}. SE highlighting is disabled; all SE values are printed as plain numbers.
% Requires \usepackage{booktabs}. SE highlighting is disabled; all SE values are printed as plain numbers.
\begin{table}[!htbp]
\centering
\caption{Log(OR) estimates, bootstrap SEs, and p-values for the association of smoking with the novel tumor markers using the established marker KRAS}
\label{tab:smoking_novel_markers_mle_el_ppmgmm_pprgmm}
\resizebox{\textwidth}{!}{%
\begin{tabular}{llrrrrrrrrrrrr}
\toprule
 & & \multicolumn{3}{c}{MLE} & \multicolumn{3}{c}{EL} & \multicolumn{3}{c}{$\text{mGMM}_{\text{W}}$} & \multicolumn{3}{c}{$\text{rGMM}_{\text{W}}$} \\
\cmidrule(lr){3-5} \cmidrule(lr){6-8} \cmidrule(lr){9-11} \cmidrule(lr){12-14}
Tumor marker & Parameter & log(OR) & SE & p-value & log(OR) & SE & p-value & log(OR) & SE & p-value & log(OR) & SE & p-value \\
\midrule
APC & $\beta_{1,1}$ & 0.370 & 0.154 & 0.016 & 0.455 & 0.089 & <0.001 & 0.326 & 0.124 & 0.009 & 0.327 & 0.120 & 0.006 \\
 & $\beta_{2,1}$ & 0.069 & 0.142 & 0.626 & -0.051 & 0.072 & 0.480 & 0.011 & 0.097 & 0.910 & 0.012 & 0.095 & 0.901 \\
 & $\beta_{2,1}-\beta_{1,1}$ & -0.301 & 0.180 & 0.095 & -0.506 & 0.157 & 0.001 & -0.315 & 0.188 & 0.093 & -0.315 & 0.186 & 0.091 \\
\midrule
ARID1A & $\beta_{1,1}$ & 0.209 & 0.122 & 0.085 & 0.142 & 0.033 & <0.001 & 0.155 & 0.079 & 0.050 & 0.155 & 0.064 & 0.015 \\
 & $\beta_{2,1}$ & 0.167 & 0.271 & 0.538 & 0.414 & 0.214 & 0.053 & 0.141 & 0.266 & 0.596 & 0.144 & 0.262 & 0.584 \\
 & $\beta_{2,1}-\beta_{1,1}$ & -0.043 & 0.277 & 0.877 & 0.272 & 0.243 & 0.263 & -0.015 & 0.287 & 0.960 & -0.011 & 0.287 & 0.969 \\
\midrule
BCL9L & $\beta_{1,1}$ & 0.250 & 0.122 & 0.040 & 0.181 & 0.034 & <0.001 & 0.194 & 0.065 & 0.003 & 0.198 & 0.062 & 0.001 \\
 & $\beta_{2,1}$ & -0.128 & 0.271 & 0.638 & 0.122 & 0.217 & 0.574 & -0.150 & 0.263 & 0.569 & -0.171 & 0.262 & 0.514 \\
 & $\beta_{2,1}-\beta_{1,1}$ & -0.377 & 0.277 & 0.174 & -0.059 & 0.246 & 0.810 & -0.344 & 0.287 & 0.231 & -0.369 & 0.287 & 0.199 \\
\midrule
BMPR2 & $\beta_{1,1}$ & 0.238 & 0.122 & 0.050 & 0.157 & 0.029 & <0.001 & 0.184 & 0.070 & 0.009 & 0.182 & 0.066 & 0.005 \\
 & $\beta_{2,1}$ & -0.044 & 0.268 & 0.871 & 0.302 & 0.174 & 0.083 & -0.056 & 0.255 & 0.826 & -0.057 & 0.251 & 0.822 \\
 & $\beta_{2,1}-\beta_{1,1}$ & -0.282 & 0.275 & 0.305 & 0.145 & 0.197 & 0.461 & -0.240 & 0.278 & 0.388 & -0.239 & 0.277 & 0.388 \\
\midrule
FAT1 & $\beta_{1,1}$ & 0.204 & 0.121 & 0.091 & 0.136 & 0.030 & <0.001 & 0.153 & 0.062 & 0.014 & 0.151 & 0.059 & 0.011 \\
 & $\beta_{2,1}$ & 0.210 & 0.298 & 0.480 & 0.533 & 0.229 & 0.020 & 0.187 & 0.282 & 0.508 & 0.186 & 0.282 & 0.510 \\
 & $\beta_{2,1}-\beta_{1,1}$ & 0.006 & 0.303 & 0.984 & 0.397 & 0.253 & 0.117 & 0.034 & 0.305 & 0.911 & 0.035 & 0.304 & 0.908 \\
\midrule
FBXW7 & $\beta_{1,1}$ & 0.240 & 0.124 & 0.052 & 0.188 & 0.041 & <0.001 & 0.182 & 0.630 & 0.772 & 0.183 & 0.064 & 0.004 \\
 & $\beta_{2,1}$ & 0.017 & 0.234 & 0.941 & 0.102 & 0.196 & 0.601 & -0.001 & 0.688 & 0.998 & -0.000 & 0.215 & 0.999 \\
 & $\beta_{2,1}-\beta_{1,1}$ & -0.223 & 0.242 & 0.357 & -0.085 & 0.233 & 0.714 & -0.184 & 0.817 & 0.822 & -0.183 & 0.245 & 0.454 \\
\midrule
GNAS & $\beta_{1,1}$ & 0.229 & 0.121 & 0.058 & 0.176 & 0.032 & <0.001 & 0.174 & 0.063 & 0.006 & 0.174 & 0.061 & 0.004 \\
 & $\beta_{2,1}$ & -0.025 & 0.296 & 0.933 & 0.157 & 0.256 & 0.539 & -0.034 & 0.293 & 0.907 & -0.040 & 0.292 & 0.892 \\
 & $\beta_{2,1}-\beta_{1,1}$ & -0.254 & 0.301 & 0.400 & -0.018 & 0.283 & 0.948 & -0.208 & 0.316 & 0.509 & -0.214 & 0.315 & 0.497 \\
\midrule
KMT2B & $\beta_{1,1}$ & 0.249 & 0.121 & 0.039 & 0.176 & 0.029 & <0.001 & 0.195 & 0.064 & 0.002 & 0.194 & 0.061 & 0.002 \\
 & $\beta_{2,1}$ & -0.199 & 0.294 & 0.499 & 0.153 & 0.215 & 0.477 & -0.213 & 0.277 & 0.440 & -0.214 & 0.276 & 0.438 \\
 & $\beta_{2,1}-\beta_{1,1}$ & -0.448 & 0.299 & 0.135 & -0.024 & 0.237 & 0.921 & -0.409 & 0.299 & 0.171 & -0.408 & 0.298 & 0.170 \\
\midrule
KMT2D & $\beta_{1,1}$ & 0.270 & 0.122 & 0.027 & 0.193 & 0.033 & <0.001 & 0.214 & 0.066 & 0.001 & 0.214 & 0.064 & <0.001 \\
 & $\beta_{2,1}$ & -0.256 & 0.266 & 0.334 & 0.040 & 0.197 & 0.840 & -0.271 & 0.251 & 0.279 & -0.272 & 0.250 & 0.277 \\
 & $\beta_{2,1}-\beta_{1,1}$ & -0.527 & 0.272 & 0.053 & -0.153 & 0.224 & 0.494 & -0.486 & 0.274 & 0.077 & -0.486 & 0.274 & 0.077 \\
\midrule
PCDH10 & $\beta_{1,1}$ & 0.202 & 0.121 & 0.095 & 0.137 & 0.031 & <0.001 & 0.148 & 0.062 & 0.016 & 0.146 & 0.060 & 0.015 \\
 & $\beta_{2,1}$ & 0.230 & 0.293 & 0.433 & 0.516 & 0.233 & 0.027 & 0.221 & 0.300 & 0.461 & 0.222 & 0.299 & 0.459 \\
 & $\beta_{2,1}-\beta_{1,1}$ & 0.028 & 0.298 & 0.925 & 0.379 & 0.258 & 0.142 & 0.073 & 0.323 & 0.822 & 0.075 & 0.323 & 0.816 \\
\midrule
PIK3CA & $\beta_{1,1}$ & 0.182 & 0.123 & 0.137 & 0.136 & 0.039 & <0.001 & 0.126 & 0.067 & 0.060 & 0.126 & 0.065 & 0.055 \\
 & $\beta_{2,1}$ & 0.334 & 0.247 & 0.176 & 0.398 & 0.210 & 0.058 & 0.321 & 0.219 & 0.144 & 0.320 & 0.219 & 0.145 \\
 & $\beta_{2,1}-\beta_{1,1}$ & 0.152 & 0.254 & 0.550 & 0.262 & 0.245 & 0.284 & 0.195 & 0.250 & 0.436 & 0.194 & 0.250 & 0.437 \\
\midrule
RNF43 & $\beta_{1,1}$ & 0.196 & 0.124 & 0.114 & 0.097 & 0.030 & 0.001 & 0.141 & 0.074 & 0.056 & 0.143 & 0.070 & 0.041 \\
 & $\beta_{2,1}$ & 0.245 & 0.226 & 0.278 & 0.537 & 0.117 & <0.001 & 0.211 & 0.222 & 0.341 & 0.208 & 0.219 & 0.341 \\
 & $\beta_{2,1}-\beta_{1,1}$ & 0.049 & 0.235 & 0.836 & 0.440 & 0.141 & 0.002 & 0.070 & 0.254 & 0.783 & 0.066 & 0.254 & 0.796 \\
\midrule
RYR1 & $\beta_{1,1}$ & 0.236 & 0.124 & 0.057 & 0.157 & 0.039 & <0.001 & 0.184 & 0.068 & 0.007 & 0.184 & 0.067 & 0.006 \\
 & $\beta_{2,1}$ & 0.056 & 0.228 & 0.806 & 0.260 & 0.168 & 0.123 & 0.014 & 0.210 & 0.948 & 0.013 & 0.209 & 0.951 \\
 & $\beta_{2,1}-\beta_{1,1}$ & -0.180 & 0.236 & 0.447 & 0.103 & 0.203 & 0.612 & -0.171 & 0.241 & 0.480 & -0.171 & 0.241 & 0.479 \\
\midrule
SOX9 & $\beta_{1,1}$ & 0.171 & 0.123 & 0.163 & 0.125 & 0.039 & 0.002 & 0.115 & 0.069 & 0.096 & 0.115 & 0.065 & 0.079 \\
 & $\beta_{2,1}$ & 0.394 & 0.246 & 0.109 & 0.461 & 0.210 & 0.028 & 0.380 & 0.220 & 0.084 & 0.378 & 0.218 & 0.082 \\
 & $\beta_{2,1}-\beta_{1,1}$ & 0.223 & 0.253 & 0.379 & 0.337 & 0.245 & 0.169 & 0.265 & 0.246 & 0.281 & 0.264 & 0.246 & 0.283 \\
\midrule
SYNE1 & $\beta_{1,1}$ & 0.233 & 0.126 & 0.063 & 0.144 & 0.042 & 0.001 & 0.174 & 0.073 & 0.017 & 0.175 & 0.068 & 0.010 \\
 & $\beta_{2,1}$ & 0.094 & 0.210 & 0.654 & 0.290 & 0.148 & 0.050 & 0.077 & 0.192 & 0.688 & 0.074 & 0.190 & 0.697 \\
 & $\beta_{2,1}-\beta_{1,1}$ & -0.139 & 0.220 & 0.527 & 0.146 & 0.186 & 0.433 & -0.097 & 0.226 & 0.669 & -0.101 & 0.226 & 0.656 \\
\midrule
TP53 & $\beta_{1,1}$ & -0.010 & 0.147 & 0.946 & 0.027 & 0.086 & 0.753 & -0.064 & 0.110 & 0.563 & -0.063 & 0.108 & 0.563 \\
 & $\beta_{2,1}$ & 0.416 & 0.148 & 0.005 & 0.318 & 0.087 & <0.001 & 0.367 & 0.111 & <0.001 & 0.367 & 0.111 & <0.001 \\
 & $\beta_{2,1}-\beta_{1,1}$ & 0.426 & 0.180 & 0.018 & 0.291 & 0.169 & 0.085 & 0.430 & 0.188 & 0.022 & 0.430 & 0.188 & 0.022 \\
\midrule
\bottomrule
\end{tabular}%
}
\end{table}

\begin{figure}[!htbp]
    \centering
    \hspace{-2cm}
    \includegraphics[scale=0.7]{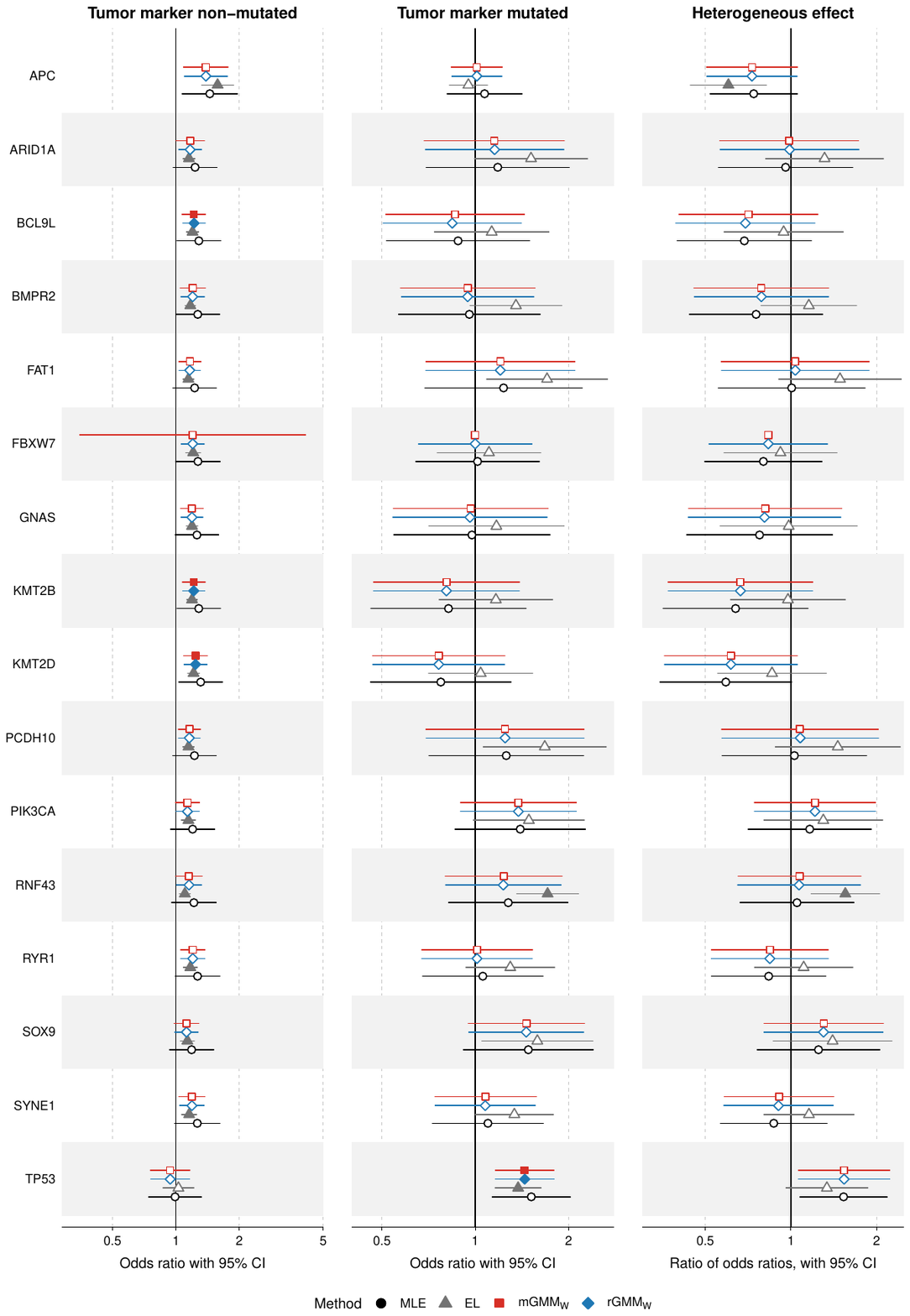}
    \caption{Forest plot of ORs of smoking on novel tumor markers using external established marker KRAS. $\text{mGMM}_{\text{W}}$ and $\text{rGMM}_{\text{W}}$ are proposed GMM approach using the huber-type weight. The nonsignificant results are displayed as hollow points using Bonferroni correction to account for multiple testing, with significance threshold 0.05/16 $\approx$ 0.003. The ratio of ORs is the ratio of the OR of smoking on the risk of developing CRC with mutated marker and non-mutated marker.}
    \label{fig:forest_plot_KRAS}
\end{figure}

This supplementary document provides detailed assumptions and proofs for the asymptotic results, together with additional empirical results, for the main manuscript.

\end{document}